\documentclass[showpacs,preprintnumbers,amsmath,amssymb,prd, superscriptaddress]{revtex4}
\usepackage{amssymb}
\usepackage{epsfig,amsmath,graphics}
\usepackage{epstopdf}
\usepackage{subfigure}

\newcommand{\OO}{\mathcal{O}}

\newcommand{\eff}{\text{eff}}
\newcommand{\keff}{k_\text{eff}}

\newcommand{\Rb}{\text{Rb}}

\newcommand{\SNR}{\text{SNR}}
\renewcommand{\v}[1]{\ensuremath{\mathbf{#1}}}
\newcommand{\ket}[1]{\ensuremath{\left|#1\right>}}
\newcommand{\bra}[1]{\ensuremath{\left<#1\right|}}
\newcommand{\braket}[2]{\ensuremath{\left<#1|#2\right>}}

\newcommand{\commutator}[2]{\ensuremath{\left[#1,#2\right]}}
\newcommand{\expectation}[1]{\ensuremath{\left<#1\right>}}
\newcommand{\abs}[1]{\ensuremath{\left|#1\right|}}

\def\be{\begin{equation}}
\def\ee{\end{equation}}

\newcommand{\OP}[1]{\ensuremath{\hat{#1}}}
\newcommand{\vOP}[1]{\ensuremath{\hat{\v{#1}}}}

\def\imagI{i}

\def\TotalState{\Psi}
\def\InternalState{A}
\def\ExternalState{\psi}
\def\CMState{\phi_{CM}}
\def\GalileanTrans{G}

\begin{document}

%\preprint{APS/123-QED}

\title{General Relativistic Effects in Atom Interferometry}

\author{Savas Dimopoulos}
%\email{savas@stanford.edu}
\affiliation{Department of Physics, Stanford University, Stanford, California 94305}

\author{Peter W. Graham}
%\email{pwgraham@stanford.edu}
\affiliation{SLAC, Stanford University, Menlo Park, California 94025}
\affiliation{Department of Physics, Stanford University, Stanford, California 94305}

\author{Jason M. Hogan}
%\email{hogan@stanford.edu}
\affiliation{Department of Physics, Stanford University, Stanford, California 94305}

\author{Mark A. Kasevich}
%\email{kasevich@stanford.edu}
\affiliation{Department of Physics, Stanford University, Stanford, California 94305}

\date{\today}% It is always \today, today,

\begin{abstract}
Atom interferometry is now reaching sufficient precision to motivate laboratory tests of general relativity. We begin by explaining the non-relativistic calculation of the phase shift in an atom interferometer and deriving its range of validity. From this, we develop a method for calculating the phase shift in general relativity.  Both the atoms and the light are treated relativistically and all coordinate dependencies are removed, thus revealing novel terms, cancellations, and new origins for previously calculated terms.  This formalism is then used to find the relativistic effects in an atom interferometer in a weak gravitational field for application to laboratory tests of general relativity. The potentially testable relativistic effects include the non-linear three-graviton coupling, the gravity of kinetic energy, and the falling of light. We propose specific experiments, one currently under construction, to measure each of these effects.  These experiments could provide a test of the principle of equivalence to 1 part in $10^{15}$ (300 times better than the present limit), and general relativity at the 10\% level, with many potential future improvements. We also consider applications to other metrics including the Lense-Thirring effect, the expansion of the universe, and preferred frame and location effects.
%Atom interferometry is now reaching sufficient precision to motivate laboratory tests of general relativity. We begin by explaining the non-relativistic calculation of the phase shift in an atom interferometer and deriving its range of validity. From this we develop a method for calculating the phase shift in general relativity. This formalism is then used to find the relativistic effects in an atom interferometer in a weak gravitational field for application to laboratory tests of general relativity. The potentially testable relativistic effects include the non-linear three-graviton coupling, the gravity of kinetic energy, and the falling of light. We propose experiments, one currently under construction, that could provide a test of the principle of equivalence to 1 part in $10^{15}$ (300 times better than the present limit), and general relativity at the 10\% level, with many potential future improvements. We also consider applications to other metrics including the Lense-Thirring effect, the expansion of the universe, and preferred frame and location effects.
\end{abstract}

\maketitle

\tableofcontents

\section{Introduction}
Atomic physics has made rapid progress recently, yielding several high precision techniques and experiments.  Atomic clocks can remain synchronous to 16-decimals \cite{Oskay} and have been used to search for the time variation of the fine structure constant \cite{Fischer:2004jt, Bize:2003bj, Peik:2004qn}.  In particular, atom interferometry has become a high precision tool that has been used successfully for a variety of applications including gyroscopes \cite{PhysRevLett.78.2046}, gradiometers \cite{PhysRevLett.81.971, Biedermann Thesis}, and gravimeters \cite{0026-1394-38-1-4, HolgerIsotropy}.  In light of this, it is interesting to consider the possibilities for using atom interferometry to test general relativity in the lab.  Indeed, the first atom interferometry experiment to push the limits on the Principle of Equivalence (PoE) \cite{Dimopoulos:2006nk} is currently under construction.

Experimental tests of general relativity (GR) have gone through two major phases.  The original tests of the perihelion precession and light bending were followed by a golden era from 1960 until today (see e.g. \cite{Will}).  These tests were in part motivated by alternatives to Einstein's theory, such as Brans-Dicke, designed to incorporate Mach's principle \cite{Brans:1961sx}.  More recently, the cosmological constant problem suggests that our understanding of general relativity is incomplete, motivating a number of proposals for modifying gravity at large distances \cite{Damour + Polyakov, DGP, Arkani-Hamed 2002}. In addition, possible alternatives to the dark matter hypothesis have led to theories where gravity changes at slow accelerations or galactic scales \cite{Arkani-Hamed Ghosts, Milgrom 1983, Bekenstein 2004}.

In a previous paper \cite{Dimopoulos:2006nk} we discussed the possibility of testing general relativity with atom interferometry.  We found that many relativistic effects will be large enough to be seen in the upcoming generation of experiments.  In this work we give the details of the framework for calculating the effects of general relativity in an atom interferometer.  We then apply this to an interferometer in the earth's gravitational field with the motivation of using the high precision of atom interferometry to test general relativity in a laboratory experiment.  The ability to find GR effects in an atom interferometer is more widely applicable as well.  In particular we consider other effects such as the Lense-Thirring effect.  Further, in \cite{Dimopoulos:2007cj, big gravity waves} we apply this technique to find the effect of a gravitational wave.

We will also discuss a few ideas for strategies to measure several of these GR effects in the lab.  We do not attempt to prove that such experiments are feasible, since this would require a very detailed analysis of the many relevant backgrounds.  Instead, we give a few arguments why the most important backgrounds may be controllable.  We wish mainly to motivate a more careful consideration of these experiments, given the interest in laboratory tests of general relativity.

An attempt was made to make the different sections of this paper as independent as possible.  Section \ref{Sec:AI} will discuss the experimental setup of an atom interferometer and give a description of the usual, nonrelativistic calculation of the final phase shift.  Section \ref{Sec:GR calc} will describe our method for finding the final phase shift in an atom interferometer in any space-time in general relativity.  Section \ref{Sec:earth calc} will specialize this discussion to the Schwarzschild metric for application to an atom interferometer in a weak gravitational field such as the earth's.  Section \ref{Sec:earth results} will give our results for the GR effects in an atom interferometer near the earth and discuss their physical origin in general relativity.  Section \ref{Sec:Measurement Strategies} will discuss a few ideas for measuring these GR terms in an actual experiment.  Section \ref{Sec:Other GR} will discuss other applications of this work including using atom interferometry to measure the Lense-Thirring effect and the expansion of the universe.

\section{Atom Interferometry}
\label{Sec:AI}
\subsection{Experimental Setup}
Atom interferometry can be used to measure accelerations precisely.  In an atom interferometer, the atom is forced to follow a superposition of two spatially and temporally separated free-fall paths. When recombined, the resulting interference pattern depends on the relative phase accumulated along the two paths.  This phase shift is exquisitely sensitive to inertial forces present during the interferometer sequence.  For example, the goal sensitivity of the next generation apparatus exceeds $10^{-14} \text{ m}/\text{s}^2$.\cite{Dimopoulos:2006nk}

A single measurement of acceleration in an atom interferometer consists of three steps: atom cloud preparation, interferometer pulse sequence, and detection.  In the first step, the cold atom cloud is prepared.  Using laser cooling and evaporative cooling techniques \cite{atomicsources}, a sub-microkelvin cloud of $~10^7$ $^{87}$\!Rb atoms is formed in $\sim 10 \text{ s}$.  Cold atom clouds are needed so that as many atoms as possible will travel along the desired trajectory and contribute to the signal.  In addition, many potential systematic errors (e.g., gravity gradients) are sensitive to the atom's initial conditions, so cooling can mitigate these unwanted effects.  Here, evaporative cooling is used since it yields the required tight control of the initial position and velocity of the cloud that cannot be achieved with laser cooling alone.  At the end of the cooling procedure, the final cloud number density is kept low enough so that atom-atom interactions within the cloud are negligible (Section \ref{Sec:Measurement Strategies} contains a quantitative discussion of this point).  This dilute ensemble of cold atoms is then launched with velocity $v_L$ by transferring momentum from laser light.  To avoid heating the cloud during launch, the photon recoil momenta are transferred to the atoms coherently, and spontaneous emission is avoided \cite{Phillips2002:JPhysB}.

\begin{figure}
\begin{center}
\includegraphics[width=250pt]{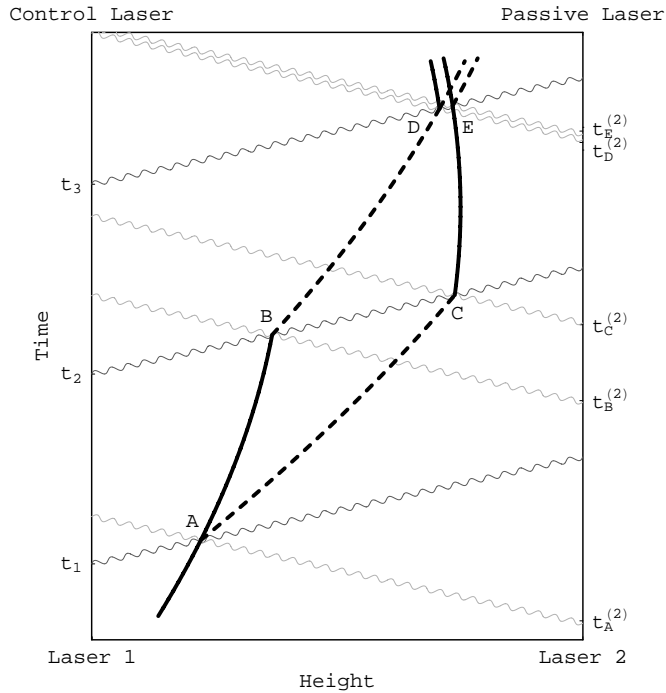}
\caption{ \label{Fig:AI-SingleInterferometer} A space-time diagram of a light pulse atom interferometer. The black lines indicate the motion of a single atom. Laser light used to manipulate the atom is incident from above (light gray) and below (dark gray) and travels along null geodesics.  Here the lasers' world-lines are taken to be the two vertical lines on the left and right edges of the graph.}
\end{center}
\end{figure}

In the second phase of the measurement, the atoms follow free-fall trajectories and the interferometry is performed.  A sequence of laser pulses serve as beamsplitters and mirrors that coherently divide each atom's wavepacket and then later recombine it to produce the interference.  Figure \ref{Fig:AI-SingleInterferometer} is a space-time diagram illustrating this process for a single atom.  The atom beamsplitter is implemented using a stimulated two-photon  transition.  In this process, laser light incident from the right of Fig. \ref{Fig:AI-SingleInterferometer} with wavevector $\v{k_1}$ is initially absorbed by the atom.  Subsequently, laser light with wavevector $\v{k_2}$ incident from the left stimulates the emission of a $\v{k_2}$-photon from the atom, resulting in a net momentum transfer of $\v{k_{\eff}}=\v{k_2}-\v{k_1}\approx 2\v{k_2}$.  These two-photon atom optics are represented in Fig. \ref{Fig:AI-SingleInterferometer} by the intersection of two counter-propagating photon paths at each interaction node.

\begin{figure}
\begin{center}
\subfigure[ ]{\label{Fig:Raman}}
\includegraphics[width=150pt]{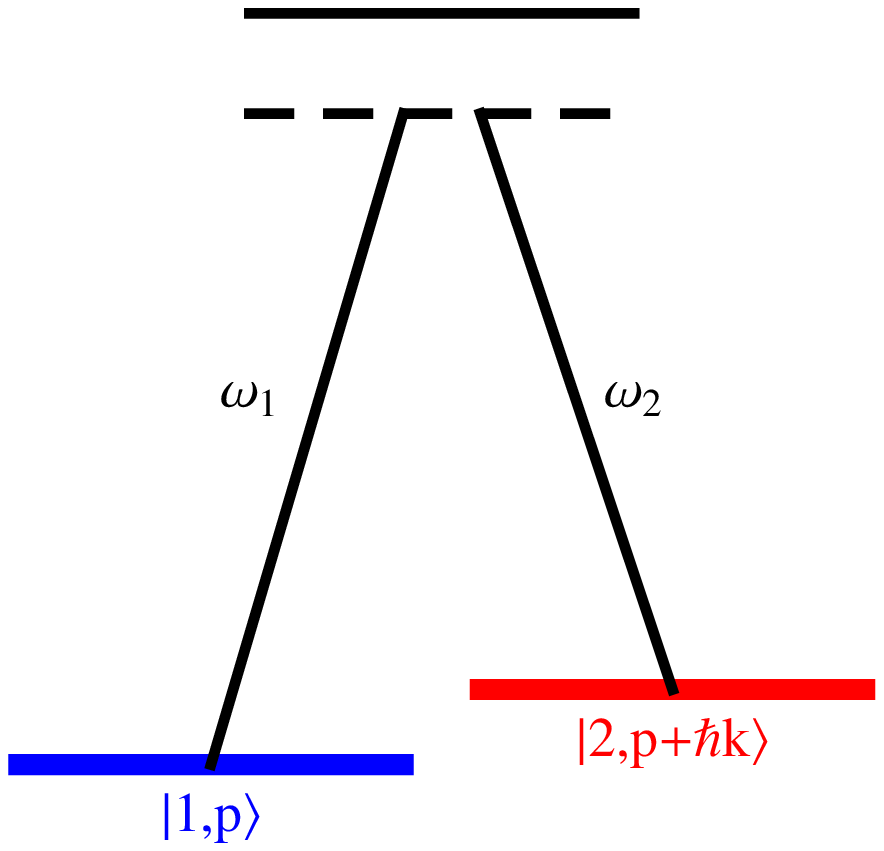}
\hspace{1 in}
\subfigure[ ]{\label{Fig:RabiPlot}}
\includegraphics[width=200pt]{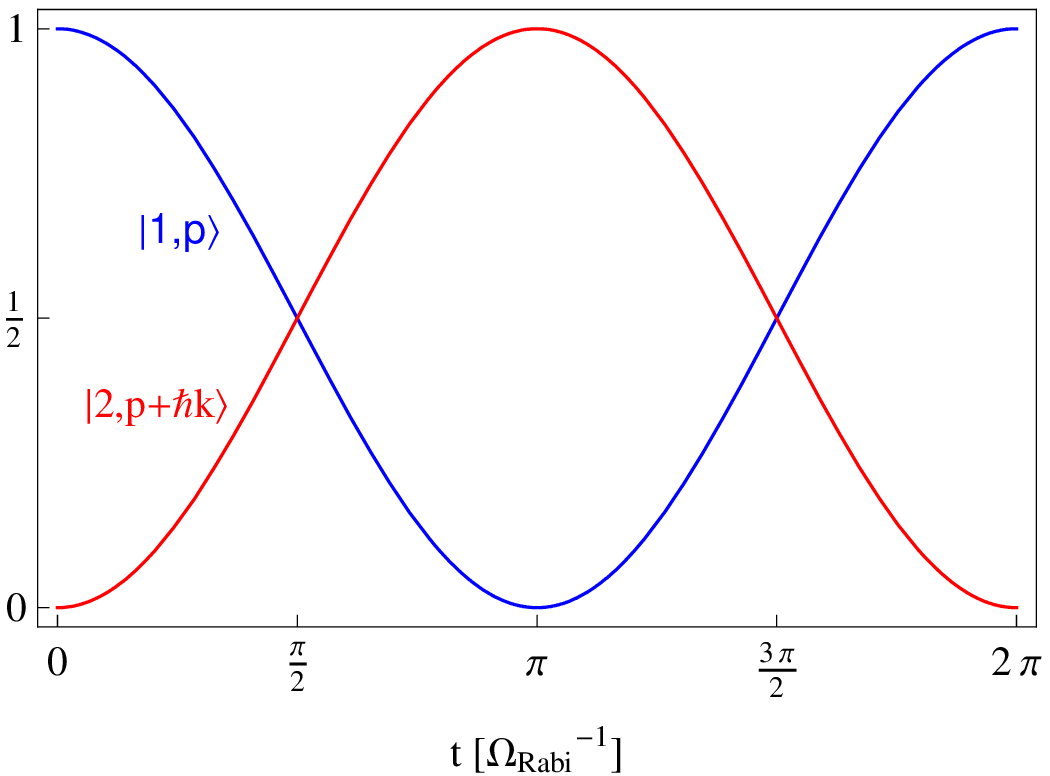}
\caption{(Color online) Figure \ref{Fig:Raman} shows an energy level diagram for a stimulated Raman transition between atomic states $\ket{1}$ and $\ket{2}$ through a virtual excited state using lasers of frequency $\omega_1$ and $\omega_2$.  Figure \ref{Fig:RabiPlot} shows the probability that the atom is in states $\ket{1}$ and $\ket{2}$ in the presence of these lasers as a function of the time the lasers are on.  A $\frac{\pi}{2}$ pulse is a beamsplitter since the atom ends up in a superposition of states $\ket{1}$ and $\ket{2}$ while a $\pi$ pulse is a mirror since the atom's state is changed completely.}
\end{center}
\end{figure}

\begin{figure}
\begin{center}
\includegraphics[width=250pt]{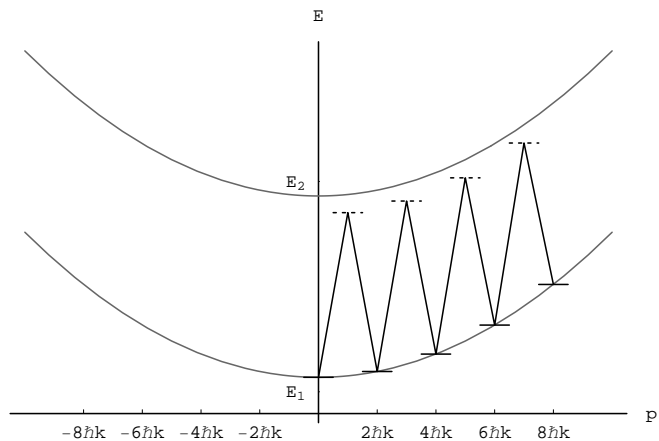}
\caption{ \label{Fig:BraggEnergyLevels} The atomic energy level diagram for a series of sequential two-photon Bragg transitions plotted as energy versus momentum.  The horizontal lines indicate the states through which the atom is transitioned.  The diagonal lines connecting the states represent the laser frequencies used in the transition.  The result of this transition is to give the atom a large momentum.}
\end{center}
\end{figure}

There are several schemes for exchanging momentum between the atoms and the lasers.  Figure \ref{Fig:Raman} shows the case of a Raman transition in which the initial and final states are different internal atomic energy levels.  The light fields entangle the internal and external degrees of freedom of the atom, resulting in an energy level change and a momentum kick.  As an alternative to this, it is also possible to use Bragg transitions in which momentum is transferred to the atom while the internal atomic energy level stays fixed (see Fig. \ref{Fig:BraggEnergyLevels}).  In both the Raman and Bragg scheme, the two lasers are far detuned from the optical transitions, resulting in a negligibly small occupancy of the excited state $\ket{e}$.  This avoids spontaneous emission from the short-lived excited state.  To satisfy the resonance condition for the desired two-photon process, the frequency difference between the two lasers is set equal to the atom's recoil kinetic energy (Bragg) plus any internal energy shift (Raman).  While the laser light is on, the atom undergoes Rabi oscillations between states $\ket{\v{p}}$ and $\ket{\v{p}+\v{k_{\eff}}}$ (see Fig. \ref{Fig:RabiPlot}). A beam splitter results when the laser pulse time is equal to a quarter of a Rabi period ($\frac{\pi}{2}$ pulse), and a mirror requires half a Rabi period ($\pi$ pulse).

After the initial beamsplitter, the atom is in a superposition of states which differ in velocity by $\v{k_{\eff}}/m$. The resulting spatial separation of the halves of the atom is proportional to the interferometer's sensitivity to acceleration along the direction of $\v{k_{\eff}}$.  In this work we consider the beamsplitter-mirror-beamsplitter $(\frac{\pi}{2}-\pi-\frac{\pi}{2})$ sequence \cite{PhysRevLett.67.181}, the simplest implementation of an accelerometer and the matter-wave analog of a Mach-Zender interferometer.

The third and final step of each acceleration measurement is atom detection.  At the end of the interferometer sequence, each atom is in a superposition of the two output velocity states, as shown by the diverging paths at the top of Fig. \ref{Fig:AI-SingleInterferometer}. Since these two states differ in velocity by $\sim k_{\eff}/m$, they spatially separate.  After an appropriate drift time, the two paths can be separately resolved, and the populations are then measured by absorption imaging. These two final velocity states are directly analogous to the two output ports of a Mach-Zender light interferometer after the final recombining beamsplitter.  As with a light interferometer, the probability that an atom will be found in a particular output port depends on the relative phase acquired along the two paths of the atom interferometer.

Recent atom interferometers have demonstrated sensor noise levels limited only by the quantum projection noise of the atoms (atom shot noise) \cite{clocks}.  Assuming a typical time--average atom flux of $n=10^{6}~\text{atoms/s}$, the resulting phase sensitivity is $\sim 1/\sqrt{n}=10^{-3}~\text{rad}/\sqrt{\text{Hz}}$ and the ultimate phase uncertainly is $\sim 10^{-6}~\text{rad}$ after $10^6~\text{s}$ of integration.  This noise performance can potentially be improved by using entangled states instead of uncorrelated atom ensembles \cite{spin_squeezing}.  For a suitably entangled source, the Heisenberg limit is $\SNR \sim n$, a factor of $\sqrt{n}$ improvement.  For $n\sim 10^{6}$ entangled atoms, the potential sensitivity improvement is $10^3$.  Recent progress using these techniques may soon make improvements in signal-to-noise ratio ($\SNR$) on the order of 10 to 100 realistic \cite{Tuchman_PRA}.  Of course, improvements in $\SNR$ may be easier to achieve simply with increased atom flux without using entanglement.

Another sensitivity improvement involves the use of more sophisticated atom optics.  The acceleration sensitivity of the interferometer is proportional to the effective momentum $\hbar \keff$ transferred to the atom during interactions with the laser.  Both the Bragg and Raman schemes described above rely on a two--photon process for which $\hbar \keff=2\hbar k$, but large momentum transfer (LMT) beamsplitters with up to $10\hbar k$ or perhaps $100\hbar k$ are possible \cite{HolgerLMT}.  Promising LMT beamsplitter candidates include optical lattice manipulations \cite{Phillips2002:JPhysB}, sequences of Raman pulses \cite{McGuirk} and adiabatic passage methods \cite{Chu}.  Figure \ref{Fig:BraggEnergyLevels} illustrates an example of an LMT process consisting of a series of sequential two--photon Bragg transitions as may be realized in an optical lattice.  As the atom accelerates, the resonance condition is maintained by increasing the frequency difference between the lasers.

Finally, we consider the expected acceleration sensitivity of the coming generation of atom interferometers. We are currently building a Rb interferometer that takes advantage of an $L\approx 10~\text{m}$ vacuum system which allows for a long interrogation time of $T=1.34~\text{s}$ \cite{Dimopoulos:2006nk, EP Varenna}. The phase response of the interferometer to an acceleration $g$ is $k g T^2$. Using $10\hbar k$ LMT beamsplitters and shot--noise limited detection of $10^6~\text{atoms}/\text{s}$ for this apparatus results in a sensitivity of $\sim \frac{1/\sqrt{10^6 \frac{\text{atoms}}{\text{s}} }}{(10 \frac{2 \pi}{780  ~\text{nm}})(1.34~\text{s})^2} \sim 7\times 10^{-13}~g/\sqrt{\text{Hz}}$ and a precision of $\sim 10^{-15}g$ after a day of integration. This estimate is based on realistic extrapolations from current performance levels, which are at $10^{-10}g$ \cite{Fixler}.

\subsection{Non-Relativistic Phase Shift Calculation}
\label{Sec:AI Calc}
%here do math to derive laser + sep + prop phases in the usual non-relativistic manner
%-discuss output ports, independence of phase shift on which one is used,...

In this section we review our non-relativistic method for calculating the phase difference between the two halves of the atom at the end of the atom interferometer.
These results are well-known \cite{CCT:1994, Bongs:2006}, but we are not aware of a complete, formal derivation of these rules in the literature.  Other equivalent formalisms for this calculation do exist (see, for example \cite{quantum_calculation, BorisDiamonds}); however, here we review our formalism because the relativistic calculation we discuss in Section \ref{Sec:GR calc} has the same structure.
%The general relativistic calculation method, discussed in Section \ref{Sec:GR calc}, is a generalization of this non-relativistic method.
For Section \ref{Sec:GR calc} it is necessary to understand the formulae for the phase difference (Section \ref{Sec: AI answer}).  The proof of these formulae as well as a discussion of their range of validity is given in Section \ref{Sec:AI proof} but is not necessary for the rest of the paper.

\subsubsection{Phase Shift Formulae}
\label{Sec: AI answer}
The main result we will show is that the total phase difference $\Delta \phi_\text{tot}$ between the two paths of an atom interferometer may be written as the sum of three easily calculated components:
\begin{equation}
\Delta \phi_\text{tot} = \Delta \phi_\text{propagation} + \Delta \phi_\text{separation} + \Delta \phi_\text{laser}.
\end{equation}
For this calculation and the rest of the paper we take $\hbar = c = 1$.

The propagation phase $\Delta \phi_\text{propagation}$ arises from the free--fall evolution of the atom between light pulses and is given by
\be\Delta \phi_\text{propagation} = \sum_{\text{upper}}\left(\int_{t\!_I}^{t\!_F}\! (L_c - E_i) dt\right) - \sum_{\text{lower}}\left(\int_{t\!_I}^{t\!_F}\! (L_c - E_i) dt\right)\label{PropagationPhaseIntro}\ee
where the sums are over all the path segments of the upper and lower arms of the interferometer, and $L_c$ is the classical Lagrangian evaluated along the classical trajectory of each path segment. In addition to the classical action, Eq. \eqref{PropagationPhaseIntro} includes a contribution from the internal atomic energy level $E_i$.  The initial and final times $t_I$ and $t_F$ for each path segment, as well as $L_c$ and $E_i$, all depend on the path segment.

The laser phase $\Delta \phi_\text{laser}$ comes from the interaction of the atom with the laser field used to manipulate the wavefunction at each of the beamsplitters and mirrors in the interferometer.  At each interaction point, the component of the state that changes momentum due to the light acquires the phase of the laser $\phi_L(t_0,\v{x}_c(t_0)) = \v{k}\cdot\v{x}_c(t_0)-\omega t_0+\phi$ evaluated at the classical point of the interaction:
\be\Delta \phi_\text{laser} = \left(\sum_j \pm\phi_L(t_j,\v{x}_u(t_j))\right)_\text{upper}-\left(\sum_j \pm\phi_L(t_j,\v{x}_l(t_j))\right)_\text{lower}\label{LaserPhaseIntro}\ee
The sums are over all the interaction points at the times $t_j$, and $\v{x}_u(t)$ and $\v{x}_l(t)$ are the classical trajectories of the upper and lower arm of the interferometer, respectively. The sign of each term depends on whether the atom gains $(+)$ or loses $(-)$ momentum as a result of the interaction.

The separation phase $\Delta\phi_\text{separation}$ arises when the classical trajectories of the two arms of the interferometer do not exactly intersect at the final beamsplitter (see Fig. \ref{Fig:SeparationPhase}).  For a separation between the upper and lower arms of $\v{\Delta x} = \v{x}_l - \v{x}_u$, the resulting phase shift is
\be\Delta\phi_\text{separation}= \v{\bar{p}}\cdot\v{\Delta x}\label{SeparationPhaseIntro}\ee
where $\v{\bar{p}}$ is the average classical canonical momentum of the atom after the final beamsplitter.  Of course, the separation phase is really an artifact of our semi-classical calculation method based on classical atom trajectories.  The observable is just the total phase shift at the end of the experiment.

\subsubsection{Proof}
\label{Sec:AI proof}
The interferometer calculation amounts to solving the Schrodinger equation with the following Hamiltonian:
\be \OP{H}_\text{tot} = \OP{H}_\text{a} + \OP{H}_\text{ext} + \OP{V}_\text{int}(\vOP{x})\ee
Here $\OP{H}_\text{a}$ is the internal atomic structure Hamiltonian, $\OP{H}_\text{ext}$ is the Hamiltonian for the atom's external degrees of freedom (center of mass position and momentum), and $\OP{V}_\text{int}(\vOP{x}) = -\vOP{\mu}\cdot\v{E}(\vOP{x})$ is the atom-light interaction, which we take to be the electric dipole Hamiltonian with $\vOP{\mu}$ the dipole moment operator.

The calculation is naturally divided into a series of light pulses during which $\OP{V}_\text{int}\neq 0$, and the segments between light pulses during which $\OP{V}_\text{int} = 0$ and the atom is in free--fall.  When the light is off, the atom's internal and external degrees of freedom are decoupled. The internal eigenstates satisfy
\be \imagI \partial_t \ket{\InternalState_i} = \OP{H}_\text{a} \ket{\InternalState_i} = E_i \ket{\InternalState_i}\label{InternalStateSchrodinger}\ee
and we write the solution as $\ket{\InternalState_i} = \ket{i}e^{-\imagI E_i (t-t_0)}$ with time-independent eigenstate \ket{i} and energy level $E_i$.

For the external state solution $\ket{\ExternalState}$, we initially consider $\OP{H}_\text{ext} = H(\vOP{x},\vOP{p})$ to be an arbitrary function of the external position and momentum operators:
\be \imagI \partial_t \ket{\ExternalState} = H(\vOP{x},\vOP{p}) \ket{\ExternalState}. \label{ExternalStateSchrodinger}\ee  It is now useful to introduce a Galilean transformation operator
\be \OP{\GalileanTrans}_c \equiv \OP{\GalileanTrans}(\v{x}_c,\v{p}_c,L_c) = e^{\imagI\int\! L_c dt}  e^{-\imagI \vOP{p} \cdot \v{x}_c}  e^{\imagI \v{p}_c\cdot\vOP{x}}\ee
which consists of momentum boost by $\v{p}_c$, a position translation by $\v{x}_c$, and a phase shift.  We choose to write
\be \ket{\ExternalState} = \OP{\GalileanTrans}_c \ket{\CMState}. \label{CMStateDef} \ee
We will show that for a large class of relevant Hamiltonians, if $\v{x}_c$, $\v{p}_c$, and $L_c$ are taken to be the classical position, momentum and Lagrangian, respectively, then $\ket{\CMState}$ is a wavepacket with $\left<\vOP{x}\right>=\left<\vOP{p}\right>=0$, and the dynamics of $\ket{\CMState}$ do not affect the phase shift result (i.e., $\ket{\CMState}$ is the center of mass frame wavefunction).  However, for now we maintain generality and just treat $\v{x}_c$, $\v{p}_c$, and $L_c$ as arbitrary functions of time.  Combining \eqref{ExternalStateSchrodinger} and \eqref{CMStateDef} results in
\begin{eqnarray}
  \imagI \partial_t \ket{\CMState} &=& \left\{ \OP{\GalileanTrans}_c^\dag H(\vOP{x},\vOP{p})\OP{\GalileanTrans}_c - \imagI \OP{\GalileanTrans}_c^\dag \partial_t\OP{\GalileanTrans}_c \right\} \ket{\CMState} \\
\nonumber
  &=& \left\{ H(\vOP{x} + \v{x}_c,\vOP{p} + \v{p}_c) + \v{\dot{p}}_c\cdot\vOP{x} - \left(\vOP{p} + \v{p}_c\right)\cdot\v{\dot{x}}_c + L_c\right\} \ket{\CMState}
\end{eqnarray}
where we used the following identities:
\begin{eqnarray}
  \OP{\GalileanTrans}_c^\dag\vOP{x}\OP{\GalileanTrans}_c &=& \vOP{x} + \v{x}_c\label{GalileanIdentities}\\
  \nonumber
  \OP{\GalileanTrans}_c^\dag\vOP{p}\OP{\GalileanTrans}_c &=& \vOP{p} + \v{p}_c \\
  \nonumber
  \OP{\GalileanTrans}_c^\dag H(\vOP{x},\vOP{p})\OP{\GalileanTrans}_c &=& H(\vOP{x} + \v{x}_c,\vOP{p} + \v{p}_c)
\end{eqnarray}
Next, we Taylor expand $H(\vOP{x} + \v{x}_c,\vOP{p} + \v{p}_c)$ about $\v{x}_c$ and $\v{p}_c$,
\be H(\vOP{x} + \v{x}_c,\vOP{p} + \v{p}_c) = H(\v{x}_c,\v{p}_c) + \nabla\!_{\vOP{x}} H(\v{x}_c,\v{p}_c)\cdot\vOP{x} + \nabla\!_{\vOP{p}} H(\v{x}_c,\v{p}_c)\cdot\vOP{p} + \OP{H}_2\ee
where $\OP{H}_2$ contains all terms that are second order or higher in $\vOP{x}$ and $\vOP{p}$.  (We will ultimately be allowed to neglect $\OP{H}_2$ in this calculation.)  Inserting this expansion and grouping terms yields
\be\imagI \partial_t \ket{\CMState} = \left\{
\big(H_c - \v{\dot{x}}_c\cdot\v{p}_c + L_c\big) +
\big(\nabla\!_{\v{x}_c}H_c + \v{\dot{p}}_c\big)\cdot\vOP{x} +
\big(\nabla\!_{\v{p}_c}H_c - \v{\dot{x}}_c\big)\cdot\vOP{p} +
\OP{H}_2
\right\} \ket{\CMState}\ee
where we have defined the classical Hamiltonian $H_c \equiv H(\v{x}_c,\v{p}_c)$. If we now let $\v{x}_c$, $\v{p}_c$, and $L_c$ satisfy Hamilton's equations,
\begin{eqnarray}
  \v{\dot{x}}_c &=& \nabla\!_{\v{p}_c}H_c \\
  \nonumber
  \v{\dot{p}}_c &=& - \nabla\!_{\v{x}_c}H_c\\
  \nonumber
  L_c &=& \v{\dot{x}}_c\cdot\v{p}_c - H_c
\end{eqnarray}
with $\v{p}_c\equiv\nabla\!_{\v{\dot{x}}_c}L_c$ the classical canonical momentum, then $\ket{\CMState}$ must satisfy
\be\imagI \partial_t \ket{\CMState} = \OP{H}_2 \ket{\CMState}\label{CMStateSchrodinger}\ee

Next we show that it is possible to choose \ket{\CMState} with $\expectation{\OP{x}}=\expectation{\OP{p}}=0$ for a certain class of $\OP{H}_2$, so that $\v{x}_c$ and $\v{p}_c$ completely describe the atom's classical center of mass trajectory.  This is known as the semi-classical limit. Starting from Ehrenfest's theorem for the expectation values of \ket{\CMState},
\begin{eqnarray}
  \partial_t \expectation{\OP{x}_i} &=& \imagI \expectation{\commutator{\OP{H}_2}{\OP{x}_i}} = \expectation{\partial_{\OP{p}_i}\OP{H}_2}\\
  \partial_t \expectation{\OP{p}_i} &=& \imagI \expectation{\commutator{\OP{H}_2}{\OP{p}_i}} = - \expectation{\partial_{\OP{x}_i}\OP{H}_2}
\end{eqnarray}
and expanding about $\expectation{\vOP{x}}$ and $\expectation{\vOP{p}}$,
\begin{eqnarray}
  \partial_t \expectation{\OP{x}_i} &=&   \left<\left.\partial_{\OP{p}_i}\OP{H}_2\right|_{\expectation{\vOP{x}},\expectation{\vOP{p}}} + \left.\partial_{\OP{p}_j}\partial_{\OP{p}_i}\OP{H}_2\right|_{\expectation{\vOP{x}},\expectation{\vOP{p}}} \left(\OP{p}_j - \expectation{\OP{p}_j}\right) + \left.\partial_{\OP{x}_j}\partial_{\OP{p}_i}\OP{H}_2\right|_{\expectation{\vOP{x}},\expectation{\vOP{p}}} \left(\OP{x}_j - \expectation{\OP{x}_j}\right) \right. \\\nonumber
  &&
  +\left.\frac{1}{2\,!}\left.\partial_{\OP{p}_i}\partial_{\OP{p}_j}\partial_{\OP{p}_k}\OP{H}_2\right|_{\expectation{\vOP{x}},\expectation{\vOP{p}}} \left(\OP{p}_j - \expectation{\OP{p}_j}\right)\left(\OP{p}_k - \expectation{\OP{p}_k}\right) + \cdots\right>\\
  \partial_t \expectation{\OP{p}_i} &=&   \left<\left.\partial_{\OP{x}_i}\OP{H}_2\right|_{\expectation{\vOP{x}},\expectation{\vOP{p}}} + \left.\partial_{\OP{x}_j}\partial_{\OP{x}_i}\OP{H}_2\right|_{\expectation{\vOP{x}},\expectation{\vOP{p}}} \left(\OP{x}_j - \expectation{\OP{x}_j}\right) + \left.\partial_{\OP{p}_j}\partial_{\OP{x}_i}\OP{H}_2\right|_{\expectation{\vOP{x}},\expectation{\vOP{p}}} \left(\OP{p}_j - \expectation{\OP{p}_j}\right)\right. \\\nonumber
  &&+\left.\frac{1}{2\,!}\left.\partial_{\OP{x}_k}\partial_{\OP{x}_j}\partial_{\OP{x}_i}\OP{H}_2\right|_{\expectation{\vOP{x}},\expectation{\vOP{p}}} \left(\OP{x}_j - \expectation{\OP{x}_j}\right)\left(\OP{x}_k - \expectation{\OP{x }_k}\right) + \cdots\right>
\end{eqnarray}
we find the following:
\begin{eqnarray}
  \partial_t \expectation{\OP{x}_i} &=& \left.\partial_{\OP{p}_i}\OP{H}_2\right|_{\expectation{\vOP{x}},\expectation{\vOP{p}}} +
  \frac{1}{2\,!}\left.\partial_{\OP{p}_k}\partial_{\OP{p}_j}\partial_{\OP{p}_i}\OP{H}_2\right|_{\expectation{\vOP{x}},\expectation{\vOP{p}}} \Delta p_{jk}^2 + \cdots \label{EhrenfestX}\\
  \partial_t \expectation{\OP{p}_i} &=& -\left.\partial_{\OP{x}_i}\OP{H}_2\right|_{\expectation{\vOP{x}},\expectation{\vOP{p}}} -
  \frac{1}{2\,!}\left.\partial_{\OP{x}_k}\partial_{\OP{x}_j}\partial_{\OP{x}_i}\OP{H}_2\right|_{\expectation{\vOP{x}},\expectation{\vOP{p}}} \Delta x_{jk}^2 + \cdots\label{EhrenfestP}
\end{eqnarray}
where $\Delta x_{jk}^2\equiv\expectation{\OP{x}_j\OP{x}_k}-\expectation{\OP{x}_j}\expectation{\OP{x}_k}$ and $\Delta p_{jk}^2\equiv\expectation{\OP{p}_j\OP{p}_k}-\expectation{\OP{p}_j}\expectation{\OP{p}_k}$ are measures of the wavepacket's width in phase space \footnote{In general, there will also be cross terms with phase space width such as $\expectation{\OP{x}_j\OP{p}_k} - \expectation{\OP{x}_j}\expectation{\OP{p}_k}$.}.  This shows that if $\OP{H}_2$ contains no terms higher than second order in $\vOP{x}$ and $\vOP{p}$, then Ehrenfest's theorem reduces to Hamilton's equations, and the expectation values follow the classical trajectories.  Furthermore, this implies that we can choose \ket{\CMState} to be the wavefunction in the atom's rest frame, since $\expectation{\OP{x}}=\expectation{\OP{p}}=0$ is a valid solution to Eqs. \eqref{EhrenfestX} and \eqref{EhrenfestP} so long as all derivatives of $\OP{H}_2$ higher than second order vanish.  In addition, even when this condition is not strictly met, it is often possible to ignore the non-classical corrections to the trajectory so long as the phase space widths $\Delta x_{jk}$ and $\Delta p_{jk}$ are small compared to the relevant derivatives of $\OP{H}_2$ (i.e., the semi-classical approximation).  For example, such corrections are present for an atom propagating in the non-uniform gravitational field $g$ of the earth for which $\partial_{\OP{r}}\partial_{\OP{r}}\partial_{\OP{r}}\OP{H}_2\sim \partial^2_r g$.  Assuming an atom wavepacket width $\Delta x\lesssim 1~\text{mm}$, the deviation from the classical trajectory is $\partial_t \expectation{\OP{p}}\sim (\partial^2_r g)\Delta x^2 \lesssim 10^{-20}g$, which is a negligibly small correction compared to the $\sim 10^{-15}g$ effects we are considering.

The complete solution for the external wavefunction requires a solution of Eq. \eqref{CMStateSchrodinger} for $\ket{\CMState}$, but this is non-trivial for general $\OP{H}_2$.  In the simplified case where $\OP{H}_2$ is second order in $\vOP{x}$ and $\vOP{p}$, the exact expression for the propagator is known \cite{ExactPropagator} and may be used to determine the phase acquired by $\ket{\CMState}$ (see also \cite{Audretsch:PRA47.5}).  However, this step is not necessary for our purpose, because for second order external Hamiltonians the operator $\OP{H}_2$ does not depend on either $\v{x}_c$ or $\v{p}_c$.  In this restricted case, the solution for the rest frame wavefunction $\ket{\CMState}$ does not depend on the atom's trajectory.  Therefore, any additional phase evolution in $\ket{\CMState}$ must be the same for both arms of the interferometer and so does not contribute to the phase difference.  This argument breaks down for more general $\OP{H}_2$, as does the semi-classical description of the atom's motion, but the corrections will depend on the width of $\ket{\CMState}$ in phase space as shown in Eqs. \eqref{EhrenfestX} and \eqref{EhrenfestP}.  We ignore all such wavepacket--structure induced phase shifts in this analysis by assuming that the relevant moments $\{\Delta x_{jk},\Delta p_{jk},\ldots\}$ are sufficiently small so that these corrections can be neglected.  As shown above for the non-uniform ($\partial^2_r g \neq 0$) gravitational field of the earth, this condition is easily met in many experimentally relevant situations.

Finally, we can write the complete solution for the free propagation between the light pulses:
\be\braket{\v{x}}{\ExternalState,\InternalState_i}=\bra{\v{x}}\OP{\GalileanTrans}_c\ket{\CMState}\ket{\InternalState_i}
= e^{\imagI\int_{t\!_I}^{t\!_F}\! L_c dt}  e^{\imagI  \v{p}_c\cdot\left(\v{x} - \v{x}_c \right)}\CMState\!\left(\v{x} - \v{x}_c \right)\ket{i}e^{-\imagI E_i (t_F-t_I)}\ee
We see that this result takes the form of a traveling wave with de Broglie wavelength set by $\v{p}_c$ multiplied by an envelope function $\CMState(\v{x})$, both of which move along the classical path $\v{x}_c$. Also, the wavepacket accumulates a propagation phase shift given by the classical action along this path, as well as an additional phase shift arising from the internal atomic energy:
\be\Delta \phi_\text{propagation} = \sum_{\text{upper}}\left(\int_{t\!_I}^{t\!_F}\! (L_c - E_i) dt\right) - \sum_{\text{lower}}\left(\int_{t\!_I}^{t\!_F}\! (L_c - E_i) dt\right)\label{PropagationPhase}\ee
where the sums are over all the path segments of the upper and lower arms of the interferometer, and $t_I$, $t_F$, $L_c$, and $E_i$ all depend on the path.

Next, we consider the time evolution while the light is on and $\OP{V}_\text{int}\neq 0$.  In this case, the atom's internal and external degrees of freedom are coupled by the electric dipole interaction, so we work in the interaction picture using the following state ansatz:
\be\ket{\TotalState}=\int\!d\v{p}\sum_i\,c_i(\v{p})\ket{\ExternalState_\v{p}}\ket{\InternalState_i}\label{InteractionPictureStateAnsatz}\ee
where we have used the momentum space representation of \ket{\CMState} and so $\ket{\ExternalState_\v{p}}\equiv\OP{\GalileanTrans}_c e^{-\imagI \OP{H}_2(t-t_0)}\ket{\v{p}}$. Inserting this state into the Schrodinger equation gives the interaction picture equations,
\begin{eqnarray}
  \imagI \partial_t \ket{\TotalState} &=& \imagI \int\!d\v{p}\sum_i\,\frac{\partial c_i(\v{p})}{\partial t}\ket{\ExternalState_\v{p}}\ket{\InternalState_i} + \OP{H}_\text{a}\ket{\TotalState}+\OP{H}_\text{ext}\ket{\TotalState}= \OP{H}_\text{tot}\ket{\TotalState}\\
  \Rightarrow \dot{c}_i(\v{p}) &\equiv& \frac{\partial c_i(\v{p})}{\partial t} = \frac{1}{\imagI} \int\!d\v{p'}\sum_j\,c_j(\v{p'})\bra{\InternalState_i}\bra{\ExternalState_\v{p}}\OP{V}_\text{int}(\vOP{x})\ket{\ExternalState_\v{p'}}\ket{\InternalState_j}\label{InteractionPictureEqs1}
\end{eqnarray}
where we used \eqref{InternalStateSchrodinger} and \eqref{ExternalStateSchrodinger} as well as the orthonormality of $\ket{\InternalState_i}$ and $\ket{\ExternalState_\v{p}}$.  The interaction matrix element can be further simplified by substituting in \ket{\ExternalState_\v{p}} and using identity \eqref{GalileanIdentities}:
\begin{eqnarray}
\bra{\ExternalState_\v{p}}\OP{V}_\text{int}(\vOP{x})\ket{\ExternalState_\v{p'}} &=& \bra{\v{p}}e^{\imagI \OP{H}_2(t-t_0)} \OP{V}_\text{int}(\vOP{x}+\v{x}_c) e^{-\imagI \OP{H}_2(t-t_0)}\ket{\v{p'}} \\\nonumber
&=& \bra{\v{p}}\OP{V}_\text{int}(\vOP{x}+\v{x}_c) \ket{\v{p'}} e^{\imagI\left(\frac{\v{p}^2}{2m}-\frac{\v{p'}^2}{2m}\right)(t-t_0)}
\end{eqnarray}
where we have made the simplifying approximation that $\OP{H}_2 \approx \frac{\vOP{p}^2}{2m}$.  This approximation works well as long as the light pulse time $\tau\equiv t-t_0$ is short compared to the time scale associated with the terms dropped from $\OP{H}_2$.  For example, for an atom in the gravitational field of earth, this approximation ignores the contribution $m (\partial_r g)\OP{x}^2$ from the gravity gradient, which for an atom of size $\Delta x \approx 1~\text{mm}$ leads to a frequency shift $\sim m (\partial_r g)\Delta x^2\sim 1~\text{mHz}$.  For a typical pulse time $\tau < 1~\text{ms}$, the resulting errors are $\lesssim 1~\mu\text{rad}$ and can usually be neglected.  Generally, in this analysis we will assume the short pulse (small $\tau$) limit and ignore all effects that depend on the finite length of the light pulse.  These systematic effects can sometimes be important, but they are calculated elsewhere\cite{AntoineFiniteTime}\cite{JansenThesis} and do not affect our main result for the largest general relativistic effects.

As mentioned before, we typically use a two photon process for the atom optics (i.e., Raman or Bragg) in order to avoid transferring population to the short-lived excited state.  However, from the point of view of the current analysis, these three-level systems can typically be reduced to effective two-level systems\cite{MolerRaman}\cite{ShoreBragg}. Since the resulting phase shift rules are identical,  we will assume a two-level atom coupled to a single laser frequency to simplify the analysis.  Assuming a single traveling wave excitation $\v{E}(\vOP{x})=\v{E}_0 \cos{\left(\v{k}\cdot\vOP{x}-\omega t+\phi\right)}$, Eq. \eqref{InteractionPictureEqs1} becomes
\be
\dot{c}_i(\v{p}) = \frac{1}{2\imagI} \int\!d\v{p'}\sum_j\,\Omega_{ij}\,c_j(\v{p'})
\bra{\v{p}}\left(e^{\imagI\left(\v{k}\cdot(\vOP{x}+\v{x}_c)-\omega t+\phi\right)}+h.c.\right)\ket{\v{p'}}e^{\imagI\int_{t\!_0}^{t}\!\omega_{ij}+\frac{\v{p}^2}{2m}-\frac{\v{p'}^2}{2m}dt}
\label{InteractionPictureEqs2}\ee
where the Rabi frequency is defined as $\Omega_{ij}\equiv\bra{i}(-\vOP{\mu}\cdot\v{E}_0)\ket{j}$ and $\omega_{ij}\equiv E_i - E_j$. Now we insert the identity         %{\ket{\v{p}}\rightarrow\ket{\v{p}+\v{k}}}
\be\v{k}\cdot(\vOP{x}+\v{x}_c)-\omega t + \phi= \underbrace{~\bigg{.}\v{k}\cdot\vOP{x}~}_\text{boost} + \underbrace{\bigg{.}\Big{(}\v{k}\cdot\v{x}_c(t_0)-\omega t_0+\phi\Big{)}}_\text{laser phase}+\underbrace{\int_{t_0}^{t}\!(\v{k}\cdot\v{\dot{x}}_c-\omega)dt}_\text{Doppler shift}\ee
into Eq. \eqref{InteractionPictureEqs2} and perform the integration over $\v{p'}$ using $\bra{\v{p}}e^{\pm\imagI\v{k}\cdot\vOP{x}}\ket{\v{p'}}=\braket{\v{p}}{\v{p'}\pm \v{k}}$:
\begin{eqnarray}
\dot{c}_i(\v{p}) = \frac{1}{2\imagI}\sum_j\,\Omega_{ij}\,\left\{c_j(\v{p}-\v{k})e^{\imagI \phi_L}
e^{\imagI\int_{t\!_0}^{t}\!(\omega_{ij}-\omega+\v{k}\cdot\v{\dot{x}}_c + \frac{\v{k}\cdot\v{p}}{m}-\frac{\v{k}^2}{2m})dt} +\right.\\\nonumber
\left. c_j(\v{p}+\v{k})e^{-\imagI \phi_L}
e^{-\imagI\int_{t\!_0}^{t}\!(-\omega_{ij}-\omega+\v{k}\cdot\v{\dot{x}}_c + \frac{\v{k}\cdot\v{p}}{m}+\frac{\v{k}^2}{2m})dt}\right\}
\label{InteractionPictureEqs3}\end{eqnarray}
where we define the laser phase at point $\{t_0,\v{x}_c(t_0)\}$ as $\phi_L\equiv \v{k}\cdot\v{x}_c(t_0)-\omega t_0+\phi$. Finally, we impose the two-level constraint $(i=1,2)$ and consider the coupling between $c_1(\v{p})$ and $c_2(\v{p}+\v{k})$:
\begin{eqnarray}
  \dot{c}_1(\v{p}) &=& \frac{1}{2\imagI}\Omega\,c_2(\v{p}+\v{k})e^{-\imagI \phi_L} e^{-\imagI\int_{t\!_0}^{t}\!\Delta(\v{p})dt}\label{InteractionPictureEqs4}\\\nonumber
  \dot{c}_2(\v{p}+\v{k}) &=& \frac{1}{2\imagI}\Omega^\ast c_1(\v{p})e^{\imagI \phi_L} e^{\imagI\int_{t\!_0}^{t}\!\Delta(\v{p})dt}
\end{eqnarray}
Here the detuning is $\Delta(\v{p})\equiv \omega_0-\omega+\v{k}\cdot(\v{\dot{x}}_c + \frac{\v{p}}{m})+\frac{\v{k}^2}{2m}$, the Rabi frequency is $\Omega \equiv \Omega_{12} = (\Omega_{21})^\ast$, and $\omega_0\equiv \omega_{21}>0$.  In arriving at Eqs. \eqref{InteractionPictureEqs4} we made the rotating wave approximation\cite{AllenEberly}, dropping terms that oscillate at $(\omega_0+\omega)$ compared to those oscillating at $(\omega_0-\omega)$. Also, $\Omega_{ii}=0$ since the $\ket{\InternalState_i}$ are eigenstates of parity and $\vOP{\mu}$ is odd.

The general solution to \eqref{InteractionPictureEqs4} is
\be
\left(\begin{array}{c}
        c_1(\v{p},t) \\
        c_2(\v{p}+\v{k},t)
      \end{array}
\right)=\left(
                \begin{array}{cc}
                  \Lambda_c(\v{p})e^{-\frac{\imagI}{2}\Delta(\v{p})\tau} & -\imagI\Lambda_s(\v{p})e^{-\frac{\imagI}{2}\Delta(\v{p})\tau}e^{-\imagI\phi_L} \\
                  -\imagI\Lambda_s^\ast(\v{p})e^{\frac{\imagI}{2}\Delta(\v{p})\tau}e^{\imagI\phi_L} & \Lambda_c^\ast(\v{p})e^{\frac{\imagI}{2}\Delta(\v{p})\tau} \\
                \end{array}
              \right)
\left(\begin{array}{c}
        c_1(\v{p},t_0) \\
        c_2(\v{p}+\v{k},t_0)
      \end{array}
\right)
\ee
\begin{eqnarray}
  \Lambda_c(\v{p}) &=& \cos{\left(\frac{1}{2}\sqrt{\Delta(\v{p})^2+\left|\Omega\right|^2}\,\tau\right)} + \imagI\frac{\Delta(\v{p})}{\sqrt{\Delta(\v{p})^2+\left|\Omega\right|^2}}\sin{\left(\frac{1}{2}\sqrt{\Delta(\v{p})^2+\left|\Omega\right|^2}\,\tau\right)}\\
  \Lambda_s(\v{p}) &=& \frac{\Omega}{\sqrt{\Delta(\v{p})^2+\left|\Omega\right|^2}}\sin{\left(\frac{1}{2}\sqrt{\Delta(\v{p})^2+\left|\Omega\right|^2}\,\tau\right)}
\end{eqnarray}
In integrating \eqref{InteractionPictureEqs4} we applied the short pulse limit in the sense of $\v{k}\cdot\v{\ddot{x}}_c \tau^2\ll 1$, ignoring changes of the atom's velocity during the pulse.  For an atom falling in the gravitational field of the earth, even for pulse times $\tau \sim 10~\mu\text{s}$ this term is $\sim k g \tau^2\sim 10^{-2}~\text{rad}$ which is non-negligible at our level of required precision.  However, for pedagogical reasons we ignore this error here.  Corrections due to the finite pulse time are calculated elsewhere \cite{Borde:FallingBeamsplitter,Audretsch:FallingBeamsplitter} and they do not affect our results to leading order.

For simplicity, from now on we assume the light pulses are on resonance: $\Delta(0)=0$.  We also take the short pulse limit in the sense of $\left|\Delta(\v{p})-\Delta(0)\right|\tau\ll 1$ so that we can ignore all detuning systematics.  This condition is automatically satisfied experimentally, since only the momentum states that fall within the Doppler width $\sim \tau^{-1}$ of the pulse will interact efficiently with the light.
\be
\left(\begin{array}{c}
        c_1(\v{p},t) \\
        c_2(\v{p}+\v{k},t)
      \end{array}
\right)=\left(
                \begin{array}{cc}
                  \Lambda_c & -\imagI\Lambda_s e^{-\imagI\phi_L} \\
                  -\imagI\Lambda_s^\ast e^{\imagI\phi_L} & \Lambda_c \\
                \end{array}
              \right)
\left(\begin{array}{c}
        c_1(\v{p},t_0) \\
        c_2(\v{p}+\v{k},t_0)
      \end{array}
\right)\qquad \begin{array}{c}
        \Lambda_c = \cos{\frac{\left|\Omega\right| \tau}{2}}\\
        \Lambda_s = \frac{\Omega}{\left|\Omega\right|}\sin{\frac{\left|\Omega\right| \tau}{2}}
      \end{array}
      \label{OnResonanceBeamsplitterMatrix}
\ee
In the case of a beamsplitter ($\frac{\pi}{2}$ pulse), we choose $\left|\Omega\right| \tau=\frac{\pi}{2}$, whereas for a mirror ($\pi$ pulse) we set $\left|\Omega\right| \tau=\pi$:
\be\Lambda_{\pi/2}=\left(
                             \begin{array}{cc}
                               \frac{1}{\sqrt{2}} & \frac{-\imagI}{\sqrt{2}}\,e^{-\imagI\phi_L} \\
                               \frac{-\imagI}{\sqrt{2}}\,e^{\imagI\phi_L} & \frac{1}{\sqrt{2}} \\
                             \end{array}
                           \right)
\qquad
\Lambda_{\pi}=\left(
                             \begin{array}{cc}
                               0 & -\imagI\,e^{-\imagI\phi_L} \\
                               -\imagI\,e^{\imagI\phi_L} & 0 \\
                             \end{array}
                           \right)\ee
These matrices encode the rules for the imprinting of the laser's phase on the atom: the component of the atom $c_1(\v{p},t_0)$ that gains momentum from the light (absorbs a photon) picks up a phase $+\phi_L$, and the component of the atom $c_2(\v{p}+\v{k},t_0)$ that loses momentum to the light (emits a photon) picks up a phase $-\phi_L$.  Symbolically,
\begin{eqnarray}
  \ket{\v{p}} &\rightarrow& \ket{\v{p}+\v{k}}e^{\imagI\phi_L} \label{laserPhaseRule1}\\
  \ket{\v{p}+\v{k}} &\rightarrow& \ket{\v{p}}e^{-\imagI\phi_L} \label{laserPhaseRule2}
\end{eqnarray}
As a result, the total laser phase shift is
\be\Delta \phi_\text{laser} = \left(\sum_j \pm\phi_L(t_j,\v{x}_u(t_j))\right)_\text{upper}-\left(\sum_j \pm\phi_L(t_j,\v{x}_l(t_j))\right)_\text{lower}\label{LaserPhase}\ee
where the sums are over all of the atom-laser interaction points $\{t_j,\v{x}_u(t_j)\}$ and $\{t_j,\v{x}_l(t_j)\}$ along the upper and lower arms, respectively, and the sign is determined by Eqs. \eqref{laserPhaseRule1}--\eqref{laserPhaseRule2}.

\begin{figure}
\begin{center}
\includegraphics[width=300pt]{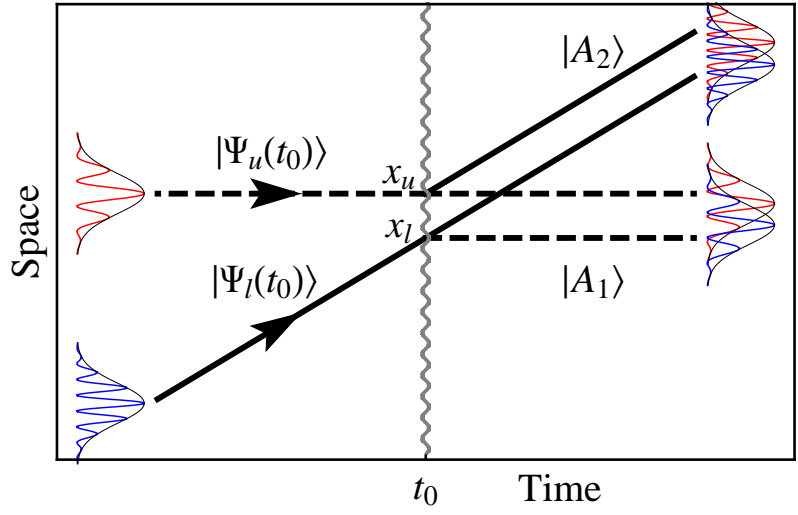}
\caption{ \label{Fig:SeparationPhase} (Color online) Separation Phase. This is a magnified view of the end of the interferometer which shows the upper and lower arms converging at the final beamsplitter at time $t_0$, and the resulting interference. The dashed and solid lines designate the components of the wavefunction in internal states $\ket{\InternalState_1}$ and $\ket{\InternalState_2}$, respectively. After the beamsplitter, each output port consists of a superposition of wavepackets from the upper and lower arm. Any offset $\v{\Delta x}= \v{x}_l - \v{x}_u$ between the centers of the wavepacket contributions to a given output port results in a separation phase shift.}
\end{center}
\end{figure}

The final contribution to $\Delta \phi_\text{tot}$ is the separation phase, $\Delta \phi_\text{separation}$. As shown in Fig. \ref{Fig:SeparationPhase}, this shift arises because the endpoints of the two arms of the interferometer need not coincide at the time of the final beamsplitter. To derive the expression for separation phase, we write the state of the atom at time $t=t_0+\tau$ just after the final beamsplitter pulse as
\be\ket{\TotalState(t)}=\ket{\TotalState_u(t)}+\ket{\TotalState_l(t)}\ee
where $\ket{\TotalState_u(t)}$ and $\ket{\TotalState_l(t)}$ are the components of the final state that originate from the upper and lower arms, respectively. Just before the final beamsplitter pulse is applied, we write the state of each arm as
\begin{eqnarray}
  \ket{\TotalState_u(t_0)} &=& \int\!d\v{p}\,c_1(\v{p},t_0)\OP{\GalileanTrans}_u(t_0)\ket{\v{p}}\ket{\InternalState_1}e^{\imagI \theta_u} \label{IC-UpperState}\\
  \ket{\TotalState_l(t_0)} &=& \int\!d\v{p}\,c_2(\v{p},t_0)\OP{\GalileanTrans}_l(t_0)\ket{\v{p}}\ket{\InternalState_2}e^{\imagI \theta_l}\label{IC-LowerState}
\end{eqnarray}
where $\OP{\GalileanTrans}_u\equiv \OP{\GalileanTrans}(\v{x}_u,\v{p}_u,L_u)$ and $\OP{\GalileanTrans}_l\equiv \OP{\GalileanTrans}(\v{x}_l,\v{p}_l,L_l)$ are the Galilean transformation operators for the upper and lower arm, respectively.  These operators translate each wavepacket in phase space to the appropriate position ($\v{x}_u$ or $\v{x}_l$) and momentum ($\v{p}_u$ or $\v{p}_l$).  Here we have assumed for clarity that prior to the final beamsplitter the upper and lower arms are in internal states $\ket{\InternalState_1}$ and $\ket{\InternalState_2}$ with amplitudes $c_1(\v{p},t_0)$ and $c_2(\v{p},t_0)$, respectively; identical results are obtained in the reversed case.  We have also explicitly factored out the dynamical phases $\theta_u$ and $\theta_l$ accumulated along the upper and lower arms, respectively, which contain by definition all contributions to laser phase and propagation phase acquired prior to the final beamsplitter.

We write the wavefunction components after the beamsplitter in the form of Eq. \eqref{InteractionPictureStateAnsatz}:
\begin{eqnarray}
  \ket{\TotalState_u(t)} &=& \int\!d\v{p}\sum_i\,c_i^{(u)}(\v{p},t)\OP{\GalileanTrans}_u\ket{\v{p}}\ket{\InternalState_i} \\
  \ket{\TotalState_l(t)} &=& \int\!d\v{p}\sum_i\,c_i^{(l)}(\v{p},t)\OP{\GalileanTrans}_l\ket{\v{p}}\ket{\InternalState_i}
\end{eqnarray}
where we invoked the short pulse limit so that $e^{-\imagI\OP{H}_2\tau}\approx 1$.  Next we time evolve the states using Eq. \eqref{OnResonanceBeamsplitterMatrix} assuming a perfect $\frac{\pi}{2}$ pulse and using the initial conditions given in Eqs. \eqref{IC-UpperState}--\eqref{IC-LowerState}: namely, $c_1^{(u)}(\v{p},t_0)=c_1(\v{p},t_0)e^{\imagI \theta_u}$ and $c_2^{(u)}(\v{p},t_0)=0$ for the upper arm and $c_1^{(l)}(\v{p},t_0)=0$ and $c_2^{(l)}(\v{p},t_0)=c_2(\v{p},t_0)e^{\imagI \theta_l}$ for the lower arm.
\begin{eqnarray}
  \ket{\TotalState_u(t)} &=& \int\!d\v{p}\,c_1(\v{p},t_0)\left\{ \frac{1}{\sqrt{2}}\OP{\GalileanTrans}_u\ket{\v{p}}\ket{\InternalState_1}+\frac{-\imagI}{\sqrt{2}}e^{\imagI\phi_L(\v{x}_u)}\OP{\GalileanTrans}_u\ket{\v{p}+\v{k}}\ket{\InternalState_2} \right\}e^{\imagI \theta_u} \\
  \ket{\TotalState_l(t)} &=& \int\!d\v{p}\,c_2(\v{p}+\v{k},t_0)\left\{ \frac{-\imagI}{\sqrt{2}}e^{-\imagI\phi_L(\v{x}_l)}\OP{\GalileanTrans}_l\ket{\v{p}}\ket{\InternalState_1}+\frac{1}{\sqrt{2}}\OP{\GalileanTrans}_l\ket{\v{p}+\v{k}}\ket{\InternalState_2} \right\}e^{\imagI \theta_l}
\end{eqnarray}
We now project into position space and perform the $\v{p}$ integrals,
\begin{eqnarray}
  \braket{\v{x}}{\TotalState_u(t)} &=& \frac{c_1(\v{x}-\v{x}_u,t_0)}{\sqrt{2}}\left\{ e^{\imagI\v{p}_u\cdot(\v{x}-\v{x}_u)}\ket{\InternalState_1}-\imagI e^{\imagI\phi_L(\v{x}_u)}e^{\imagI(\v{p}_u+\v{k})\cdot(\v{x}-\v{x}_u)}\ket{\InternalState_2} \right\}e^{\imagI \theta_u} \\
  \braket{\v{x}}{\TotalState_l(t)} &=& \frac{c_2(\v{x}-\v{x}_l,t_0)}{\sqrt{2}}\left\{-\imagI e^{-\imagI\phi_L(\v{x}_l)}e^{\imagI(\v{p}_l-\v{k})\cdot(\v{x}-\v{x}_l)}\ket{\InternalState_1}+ e^{\imagI\v{p}_l\cdot(\v{x}-\v{x}_l)}\ket{\InternalState_2} \right\}e^{\imagI \theta_l}
\end{eqnarray}
where we identified the Fourier transformed amplitudes using $c_i(\v{x}-\v{x}_c, t_0)=\int\!d\v{p}\,\braket{\v{x}-\v{x}_c}{\v{p}}c_i(\v{p},t_0)$.  The resulting interference pattern in position space is therefore
\begin{eqnarray}
\braket{\v{x}}{\TotalState(t)} &=& \braket{\v{x}}{\TotalState_u(t)}+\braket{\v{x}}{\TotalState_l(t)}\\\nonumber
&=& \frac{1}{\sqrt{2}}\ket{\InternalState_1}\left\{ c_1(\v{x}-\v{x}_u,t_0)e^{\imagI \theta_u}e^{\imagI\v{p}_u\cdot(\v{x}-\v{x}_u)} - \imagI\,c_2(\v{x}-\v{x}_l,t_0)e^{\imagI \theta_l}e^{-\imagI\phi_L(\v{x}_l)}e^{\imagI(\v{p}_l-\v{k})\cdot(\v{x}-\v{x}_l)} \right\}\\\nonumber
&+& \frac{1}{\sqrt{2}}\ket{\InternalState_2} \left\{c_2(\v{x}-\v{x}_l,t_0)e^{\imagI \theta_l}e^{\imagI\v{p}_l\cdot(\v{x}-\v{x}_l)} - \imagI\, c_1(\v{x}-\v{x}_u,t_0)e^{\imagI \theta_u}e^{\imagI\phi_L(\v{x}_u)}e^{\imagI(\v{p}_u+\v{k})\cdot(\v{x}-\v{x}_u)}   \right\}
\end{eqnarray}
The probability of finding the atom in either output port $\ket{\InternalState_1}$ or $\ket{\InternalState_2}$ is
\begin{eqnarray}
\abs{\bra{\InternalState_1}\braket{\v{x}}{\TotalState(t)}}^2 &=& \frac{\abs{c_1}^2+\abs{c_2}^2}{2}+\frac{1}{2}\left(\imagI\,c_1\,c_2^\ast\,e^{\imagI\Delta\phi_1}+c.c.\right)\\
\abs{\bra{\InternalState_2}\braket{\v{x}}{\TotalState(t)}}^2 &=& \frac{\abs{c_1}^2+\abs{c_2}^2}{2}-\frac{1}{2}\left(\imagI\,c_1\,c_2^\ast\,e^{\imagI\Delta\phi_2}+c.c.\right)
\end{eqnarray}
with $c_1\equiv c_1(\v{x}-\v{x}_u,t_0)$ and $c_2\equiv c_2(\v{x}-\v{x}_l,t_0)$. For the total phase shift we find
\begin{eqnarray}
\Delta\phi_1 &\equiv & \Big{\{}\theta_u+\v{p}_u\cdot(\v{x}-\v{x}_u)\Big{\}}-\Big{\{}\theta_l-\phi_L(\v{x}_l)+(\v{p}_l-\v{k})\cdot(\v{x}-\v{x}_l)\Big{\}}\label{DeltaPhi-1-raw}\\
&=&\underbrace{\theta_u-\Big{(}\theta_l-\phi_L(\v{x}_l)\Big{)}}_{\Delta\phi_\text{propagation,1}~+~\Delta\phi_\text{laser,1}}+\underbrace{\Big{.}\,\,\v{\bar{p}}_1\cdot\v{\Delta x}\,\,}_{\Delta\phi_\text{separation,1}}+\,\,\v{\Delta p}\cdot(\v{x}-\v{\bar{x}})\label{DeltaPhi-1}
\end{eqnarray}
and
\begin{eqnarray}
\Delta\phi_2 &\equiv & \Big{\{}\theta_u + \phi_L(\v{x}_u) + (\v{p}_u+\v{k})\cdot(\v{x}-\v{x}_u) \Big{\}}-
\Big{\{} \theta_l + \v{p}_l\cdot(\v{x}-\v{x}_l) \Big{\}}\label{DeltaPhi-2-raw}\\
&=&\underbrace{\Big{(}\theta_u + \phi_L(\v{x}_u)\Big{)}-\theta_l}_{\Delta\phi_\text{propagation,2}~+~\Delta\phi_\text{laser,2}}+\underbrace{\Big{.}\,\,\v{\bar{p}}_2\cdot\v{\Delta x}\,\,}_{\Delta\phi_\text{separation,2}}+\,\,\v{\Delta p}\cdot(\v{x}-\v{\bar{x}})\label{DeltaPhi-2}
\end{eqnarray}
where $\v{\bar{p}}_1=\frac{\v{p}_u + (\v{p}_l-\v{k})}{2}$ and $\v{\bar{p}}_2=\frac{(\v{p}_u+\v{k})+\v{p}_l}{2}$ are the average momenta in the \ket{\InternalState_1} (slow) and \ket{\InternalState_2} (fast) output ports, respectively.  In general, the separation phase is
\be\Delta\phi_\text{separation}= \v{\bar{p}}\cdot\v{\Delta x}\label{SeparationPhase}\ee
which depends on the separation $\v{\Delta x}\equiv \v{x}_l - \v{x}_u$ between the centers of the wavepackets from each arm as well as the average canonical momentum $\v{\bar{p}}$ in the output port.
We point out that even though the definitions \eqref{DeltaPhi-1-raw} and \eqref{DeltaPhi-2-raw} use the same sign convention as our previous expressions for laser \eqref{LaserPhase} and propagation \eqref{PropagationPhase} phase in the sense of $(~)_\text{upper}-(~)_\text{lower}$, the separation vector $\v{\Delta x}$ is defined as $(\v{x})_\text{lower}-(\v{x})_\text{upper}$.

Notice that the phase shift expressions \eqref{DeltaPhi-1} and \eqref{DeltaPhi-2} contain a position dependent piece $\v{\Delta p}\cdot(\v{x}-\v{\bar{x}})$, where $\v{\bar{x}}\equiv\frac{\v{x}_u + \v{x}_l}{2}$ and $\v{\Delta p}=(\v{p}_u+\v{k})-\v{p}_l=\v{p}_u - (\v{p}_l-\v{k})$, owing to the fact that the contributions from each arm may have different momenta after the last beamsplitter.  Typically this momentum difference is very small, so the resulting phase variation has a wavelength that is large compared to the spatial extent of the wavefunction.  Furthermore, this effect vanishes completely in the case of spatially averaged detection over a symmetric wavefunction.

Finally, we show that the total phase shifts $\Delta\phi_1$ and $\Delta\phi_2$ for the two output ports are actually equal, as required by conservation of probability.  According to Eqs. \eqref{DeltaPhi-1} and \eqref{DeltaPhi-2}, the contributions to the total phase differ in the following ways:
\begin{eqnarray}
\Big{(}\Delta\phi_\text{propagation,1}+\Delta\phi_\text{laser,1}\Big{)}-\Big{(}\Delta\phi_\text{propagation,2}+\Delta\phi_\text{laser,2}\Big{)}&=&\phi_L(\v{x}_l)-\phi_L(\v{x}_u)\qquad\\\nonumber
&=& \v{k}\cdot(\v{x}_l-\v{x}_u)=\v{k}\cdot\v{\Delta x}
\end{eqnarray}
\be\Delta\phi_\text{separation,1}-\Delta\phi_\text{separation,2}=\v{\bar{p}}_1\cdot\v{\Delta x} - \v{\bar{p}}_2\cdot\v{\Delta x} = -\v{k}\cdot\v{\Delta x}\ee
Together these results imply that $\Delta\phi_1=\Delta\phi_2$ and prove that the total interferometer phase shift $\Delta \phi_\text{tot}$ is independent of the output port.

The accuracy of the above formalism is dependent on the applicability of the aforementioned stationary phase approximation as well as the short pulse limit.  The stationary phase approximation breaks down when the external Hamiltonian varies rapidly compared to the phase space width of the atom wavepacket.  The short pulse limit requires that the atom's velocity not change appreciably during the duration of the atom-light interaction.  Both approximations are justified to a large degree for a typical light pulse atom interferometer, but in the most extreme high precision applications such as we consider here, important corrections are present.  However, we emphasize that these errors due to finite pulse duration and wavepacket size are well-known, previously established backgrounds.  Although they must be accounted for experimentally, these corrections do not affect the leading order general relativistic effects which we seek to calculate.

\section{General Relativistic Description of Atomic Interferometry}
\label{Sec:GR calc}

We are interested in the leading order effects of general relativity in an atom interferometer.  Section \ref{Sec:AI Calc} described our non-relativistic calculation.  Here we build on the results of that Section to create a formalism which treats the entire calculation in a relativistic manner.  It is very difficult to solve for the quantum mechanical evolution of the atom in a general metric background.  Thus, we will use the semi-classical approximation method outlined for the non-relativistic calculation.  This method can be used in relativity, with some minor modifications, since in general relativity the concept of potential is replaced by the least action principle.

In brief, the method is as follows.  Using the prescription in Section \ref{Sec: AI answer}, the free propagation of the atoms and the light in Figure \ref{Fig:AI-SingleInterferometer} is treated non-quantum mechanically.  Thus, both the laser pulses and the atoms are taken to move along geodesics of the space-time.  The description of the atom-light interaction is taken from non-relativistic quantum mechanics, but must be described in a covariant manner as will be discussed below.  Finally, the total resulting phase shift must be a coordinate invariant.  Further, to understand the result it is necessary to write it in terms of the physical variables of the problem as measured by an experiment, thus removing any coordinate dependence from the answer.

Our objective is to calculate the leading order general relativistic effects in order to to explore interesting and possibly measurable signals.  In order to simplify the calculation, many sub-leading order effects will be dropped including effects due to the finite pulse time of the lasers, AC Stark shifts, and the errors in the semi-classical approximation due to the finite size of the atom's wavefunction.  All these may give important backgrounds but they can be and have been calculated easily in the non-relativistic formalism.  We are interested in the largest effects that arise due to general relativity and so we can ignore all these effects.

We will consider an atom interferometer in a background space-time with metric $g_{\mu \nu}$.  The proper time for a particle in this space-time is then given by $d\tau^2 = g_{\mu \nu} dx^\mu dx^\nu$.  We will take $\hbar = c = 1$.

\subsection{Dynamics of the Interferometer}

%need full trajectories so need geodesics and their intersections
%must use physical variables (eg the laser is your clock, it sets your frequency reference)
%atom-laser interaction points done in LLF (ambiguity at O($v^3$))
%discuss finite wavefunction size effects and finite pulse time effects
%coordinate invariance?

The trajectories of the atoms and the laser pulses are found by solving the geodesic equation
\begin{equation}
\label{Eqn:Geodesic Eqn}
\frac{d^2 x^\mu}{d \tau^2} + \Gamma^\mu_{\alpha \beta} \frac{dx^\alpha}{d \tau} \frac{dx^\beta}{d \tau} = 0,
\end{equation}
where $\Gamma$ is the affine connection and Greek indices run 0 to 3.  In order to compute the leading order GR effects we will calculate the phase shift using the approximations explained in Section \ref{Sec:AI}, for which it is sufficient to find the motion of the center of the atomic wavefunction.  Of course, there are corrections to this semi-classical (or stationary phase) approximation due to the finite size of the atomic wavefunction which have been discussed in Section \ref{Sec:AI Calc}.  However, the leading order GR effects are just large enough to be experimentally measured and therefore these corrections to our GR results are negligible and will be ignored.  The interferometer is then defined by the initial space-time position and velocity of the atom before the first beamsplitter pulse, and by the starting positions of the three laser pulses which define the interferometer, the beamsplitter-mirror-beamsplitter sequence.  Once these are given, the rest of the interferometer is found by calculating the intersection of the geodesics as shown in Fig. \ref{Fig:AI-SingleInterferometer}.  We assume atomic transitions only occur when the atom is simultaneously in the presence of both laser beams.  We will assume that for every atom-light interaction point, the right laser is always turned on sufficiently far before the left laser so that the atom is already in the presence of the light from the right laser when the light from the left laser hits it.  Thus the atomic transitions will always occur when the light from the left laser reaches the atom.  This choice makes a small but potentially measurable effect that will be discussed below.  We will call the left laser the `control laser', because its timing determines the timing of the interferometer and the right laser the `passive laser'.

The intersection of the initial beamsplitter pulse (from the control laser) and the initial atomic trajectory defines point A.  After the first beamsplitter interaction, the half of the atom which is not affected by the light travels on to B along the same trajectory.  The half of the atom which is affected travels on a new trajectory originating from point A but with a new velocity which is a function of the incoming momenta of the atom and the light pulse as will be described below.  The two halves then travel to B and C respectively, which are defined by the intersections with the mirror pulse.  At B, the atom gets a kick from the light in the same way as before.  At C, it loses momentum via stimulated emission.  These halves then travel on to D and E respectively, where they interact with the final beamsplitter and interference is assumed to occur as described above.  The relativistic calculation of the final phase shift is described below.  Thus, given the initial conditions for the atom and laser pulses, the interferometer can be calculated in a fully covariant manner.

We must now give a coordinate invariant description of these initial conditions.  This means they must be written in terms of the physical variables measured by the experimentalist.  For all the calculations described here we will consider the laser to be at a fixed spatial coordinate location, $\vec{x}_L$.  As we will see, for all the metrics we consider, we will choose coordinates such that this is a suitable model for the laser.  The initial beamsplitter pulse then defines the start of the interferometer and should be considered to be at an arbitrary time coordinate, $t_1$ (unless the experimenter somehow has independent knowledge of the metric).  The mirror pulse is then taken to leave the laser at $\vec{x}_L$ a time $T$ later as measured on the laser's clock.  Similarly, a time $T$ later on the laser's clock the final beamsplitter pulse is emitted from the laser.  The laser's proper time is given by
\begin{equation}
T = \int  d \tau = \int_{t_1}^{t_2} \sqrt{g_{00}} dt = \int_{t_2}^{t_3} \sqrt{g_{00}} dt.
\end{equation}
Solving these two equations for $t_2$ and $t_3$ yields the time-coordinates at which the mirror and final beamsplitter pulses originate from the laser.  This then defines the three laser pulses in a coordinate invariant way.  The atom's initial position may be changed depending on the application and thus it is harder to give a single, relativistic description of it.  One natural way to define it is to take the atom to begin at the same position as the laser (and thus at a well-defined time on the laser's clock), and then to travel for a certain amount of time (again on the laser's clock) before the first beamsplitter pulse is emitted.  This defines the initial position in a relativistically invariant manner.  The atom's initial velocity can also be defined in several ways.  For example, it could be taken as the radar ranging velocity that the laser sees.  Where it is relevant, we will usually consider it to be the more experimentally realistic velocity that would result from getting some large number of momentum kicks from the laser.  Once the initial conditions for the atom interferometer have been defined in a coordinate invariant manner, we can calculate the entire interferometer sequence.

The atom-light interaction can most easily be thought about in a local Lorentz frame, $x'$, (LLF) of the atom (essentially Riemann normal coordinates).  This is a choice of coordinates such that the atom is at rest at the origin of these coordinates and space-time is locally flat near the atom.  Specifically, in the LLF the metric is locally flat with vanishing first derivatives at the position of the atom so that
\begin{equation}
\left. \Gamma^{\mu'}_{\alpha' \beta'} \right|_{x'=0} = 0
\end{equation} 
and near the origin (the position of the atom)
\begin{equation}
g_{\mu' \nu'} (x') = \eta_{\mu' \nu'} + \mathcal{O} ({x'}^2),
\end{equation}
where $\eta$ is the flat metric.
In these coordinates, the leading order effects of the interaction with the light on the atom are just the non-relativistic quantum mechanical rules given in Section \ref{Sec:AI}.  So in this frame the spatial momentum of the atom after the transition is equal to the sum of the spatial momenta of the atom before the transition and the incoming light which causes the transition, namely
\begin{equation}
\label{Eqn:atom-light momenta addition}
\left. m_\text{in} \frac{d{x'}^i_\text{atom}}{d\tau} \right|_\text{after} = \left. m_\text{fi} \frac{d{x'}^i_\text{atom}}{d\tau} \right|_\text{before} + \frac{d{x'}^i_\text{light}}{d\lambda} \quad \text{ (in LLF)}
\end{equation}
where $i=1,2,3$, $m_\text{in}$ and $m_\text{fi}$ are the rest masses of the atom before and after the atomic transition, and $\lambda$ is the affine parameter for the laser's null geodesic (the analogue of $\tau$ in Eq. \eqref{Eqn:Geodesic Eqn} but with different units).  Typically $m_\text{in} = m_\text{fi} \pm \omega_\text{a}$ where $\omega_\text{a}$ is the frequency difference between the initial and final (internal) atomic states that are coupled by the laser interaction \footnote{In the case of a two-photon Raman transition, $\omega_\text{a}$ is typically the hyperfine splitting of the ground state of an alkali atom ($\omega_\text{a}\approx 2\pi \times 6.8~\text{GHz}$ for $^{87}$Rb).  For a Bragg transition, $\omega_\text{a}=0$ since the initial and final internal states are the same.}. Note that in the case of a multi-photon transition (e.g., the two-photon transitions shown in Fig. \ref{Fig:AI-SingleInterferometer}) the photon momentum $\frac{d\vec{x}^i_\text{light}}{d\lambda}$ in Eq. \eqref{Eqn:atom-light momenta addition} must be replaced by the sum of the momenta of all photons that contribute to the transition.  Since the metric deviates slightly from flat, there are small corrections due to the tidal effects of gravity over the size $\Delta x$ of the atomic wavefunction which are $\mathcal{O}(R\Delta x^2)$, where $R$ represents the Riemann curvature tensor.  Thus any effects of gravity are suppressed by the size of the atomic wavefunction, so even the leading order Newtonian contributions to Riemann can be neglected during the atom-light interaction (see the discussion of the semi-classical approximation in Sec. \ref{Sec:AI proof}).  For a description of transforming coordinates to the LLF see \cite{Weinberg}.

To find these spatial momenta in the LLF ($x'$) it is necessary to know them in the main coordinate system ($x$) being used and then transform them to the LLF.  The atom's spatial momenta are simply determined from its geodesic and proper time coordinate $\tau$.  However, a light ray travels on the same geodesic no matter what its momentum is.  We must then know the correct affine parameterization, $x_\text{light}^\mu (\lambda)$, of the null geodesic such that $\frac{dx_\text{light}^i}{d\lambda}$ is actually the spatial momentum and not just proportional to it.  This is determined by the initial conditions for the laser pulse.  However, we cannot simply take the spatial momentum $\left. \frac{dx_\text{light}^i}{d\lambda} \right|_{x_L}$ of the laser pulse at emission to be equal to $k^i$, the laser's wavevector, because the coordinates $x^i$ do not necessarily have any physical meaning and so such a choice would be coordinate dependent.  We must instead write our answers only in terms of physical variables; in this case we must use the $k$ which an observer would measure the laser to have.  For definiteness we will assume the laser's frequency is measured by an observer at the same position $x_L$ as the laser and not moving with respect to it.  Then this defines
\begin{equation}
\left. \left( g_{\mu \nu} U^\mu \frac{dx_\text{light}^\nu}{d\lambda} \right) \right|_{x_L} \equiv \omega = k
\label{Eq: laser frequency reference}
\end{equation}
where $\omega$ is the frequency of the laser and $U^\mu = \frac{dx_\text{obs}^\mu}{d\tau}$ is the four-velocity of the observer.  This equation implicitly defines the $\lambda$ such that our observer sitting on the laser (in the laser's LLF) measures the emitted pulses to have the proper frequency $\omega$.  This pulse is then propagated to the atom in the main coordinate system of the problem ($x$).  When the light reaches the atom, its momentum is transformed to the atom's LLF ($x'$) in order to find the momentum transferred to the atom.  Eq. \eqref{Eq: laser frequency reference} is critical as it modifies the result for the GR corrections at leading order as we will see explicitly below.

Note that this rule (Eq. \eqref{Eqn:atom-light momenta addition}) of adding the momenta only applies in the LLF, and there is in fact some ambiguity about which LLF to use: the rest frame of the atom before or after the interaction.  This ambiguity implies relativistic corrections to the recoil velocity that are $\mathcal{O}(v_r^2)$, thus changing the atom's velocity at $\mathcal{O}(v_r^3)$.  This is far too small to be visible in the experiments we are considering; however if it becomes necessary to compute such corrections, the formalism given here would have to be adapted.

We have now completely determined the dynamics of the interferometer in a fully relativistic framework.  Importantly, the initial conditions have been given in terms of physical variables of the problem and so there is no coordinate dependence left.  All that remains is to determine the rules for calculating the final phase difference in general relativity.

\subsection{Relativistic Phase Shift Formulae}
\label{Sec: GR phase shift formulae}
%the GR generalization of each phase
%propagation is clear, just m dtau
%laser is laser's phase which is now fixed along null geodesic
%separation is pmu dxmu

The propagation phase is proportional to the integral of the Lagrangian, i.e., the action.  In general relativity, the action for a particle moving in a background space-time is the length of its world-line, $S=\int m d\tau$, where the mass $m$ of the particle is irrelevant for the equations of motion but is inserted so that the Lagrangian reduces to the normal non-relativistic Lagrangian in the appropriate limit.  Since $S=\int L dt$, the Lagrangian is $L = m \frac{d\tau}{dt}$.  To demonstrate that this reduces to the expected Lagrangian in a weak gravitational field, insert the Schwarzschild metric (Eq. \eqref{eqn:Schwarzschild metric}) and take a post-Newtonian expansion in $\phi$ and $\vec{v}^i \equiv \frac{d\vec{x}^i}{dt}$.  Then $L = m - \frac{1}{2} m \vec{v}^2 + m \phi + \mathcal{O}(v^4, \phi^2, v^2 \phi)$ as expected.  By analogy with the non-relativistic formula in Section \ref{Sec:AI}, the general relativistic action is the propagation phase
\begin{equation}
\label{eqn:GR prop phase}
\phi_\text{propagation} = \int L dt = \int m d\tau = \int p_\mu dx^\mu
\end{equation}
where $p^\mu \equiv m \frac{dx^\mu}{d\tau}$ is the particle's momentum.  The last equality follows from $p_\mu dx^\mu = m g_{\mu \nu}  \frac{dx^\mu}{d\tau}  dx^\nu = m \frac{d\tau}{d\tau} d\tau$.  Notice that this is the opposite sign convention from the non-relativistic expression for propagation phase from Section \ref{Sec:AI}.

The separation phase follows essentially from the formula outlined in Section \ref{Sec:AI} applied in the LLF.  The separation phase in the LLF is $\overline{E} \Delta t' - \vec{\overline{p}} \cdot \Delta \vec{x}'$.  We then employ the standard trick, to write the formula in this frame as a coordinate invariant 
\begin{equation}
\phi_\text{separation} = \int_E^D \overline{p}_{\mu'} d{x'}^{\mu'} \sim \overline{E} \Delta t' - \vec{\overline{p}} \cdot \Delta \vec{x}'
\end{equation}
where the integral is taken along the null geodesic connecting points E and D (the final beamsplitter pulse from laser 1).  Since this is a coordinate invariant and is true in the Local Lorentz Frame ($x'$), it is then valid in all frames.  Then in any coordinate system ($x$) the separation phase is
\begin{equation}
\label{Eqn: Separation Phase}
\phi_\text{separation} = \int_E^D \overline{p}_\mu dx^\mu.
\end{equation}
Here $\overline{p}^\mu$ is the average momentum of the two halves of the atom at points D and E
\begin{equation}
\overline{p}^\mu = \frac{1}{2} \left( m_\text{o} \left. \frac{dx^\mu}{d\tau} \right|_\text{D} + m_\text{o} \left. \frac{dx^\mu}{d\tau} \right|_\text{E} \right)
\end{equation}
and the momenta are evaluated at points D and E after the final beamsplitter pulse and in the same output port (either slow, $\ket{\InternalState_1}$, or fast, $\ket{\InternalState_2}$, whichever is being used to compute the final phase shift; see Fig. \ref{Fig:SeparationPhase}).  Here $m_\text{o}$ is the mass of the atom in the chosen output port.  The small coordinate ambiguities in this formula are negligible for every metric considered, as will be shown below.  As is clear from the formulas, the separation phase \eqref{Eqn: Separation Phase} can be thought of as the last piece of propagation phase \eqref{eqn:GR prop phase}.  The term $\overline{E} \Delta t$ is then roughly the phase acquired by bringing the half of the atom that transitioned earlier at the second beamsplitter up to the same time as the second half of the atom, so they can interfere.  We have chosen to define the separation phase along the null geodesic which is the final beamsplitter pulse because this is the first hypersurface on which interference can be considered to have occurred.  Of course, since quantum mechanics is linear, we can choose any later hypersurface, add the two halves of the atom's wavefunction and calculate a separation phase then (of course we would also need the correct rule for propagating the atoms forward after the final beamsplitter).  In other words, in order to find the total phase shift, the atom's wavefunction can be evaluated at any point in space-time after the final beamsplitter pulse.  For example, the number of atoms in each output port is often measured with a detection laser pulse.  This would correspond to a final null geodesic which is equivalently good for calculating the final phase shift.

As shown in Section \ref{Sec:AI}, the laser phase imparted to the atom during a beamsplitter or mirror pulse is the phase of the light at the interaction point.  Since this applies in the LLF and is also a coordinate invariant statement, it can be applied in any frame.  A null geodesic is a line of constant phase for the laser since e.g. it is the world-line in space-time that a crest of the laser pulse follows.  Thus the light's phase at an interaction point is the same as its phase at the time of emission of that light pulse from the laser.  For example, the laser phase of the pulse from laser 1, the control laser, at point A is just the phase of laser 1 at time $t_1$ in Fig. \ref{Fig:AI-SingleInterferometer}.  The total laser phase shift is then the sum of the laser phases from each laser over all the interaction points.  For example, in the slow output port, the total laser phase is
\begin{equation}
\label{Eqn: laser phase}
\Delta \phi^\text{slow}_\text{laser} = \phi_{L1} (t_1) - 2 \phi_{L1} (t_2) + \phi_{L1} (t_3) - \phi_{L2} (t_A^{(2)}) +  \phi_{L2} (t_B^{(2)}) +  \phi_{L2} (t_C^{(2)}) -  \phi_{L2} (t_D^{(2)})
\end{equation}
where $\phi_{Li} (t)$ means the phase of laser $i$ at time $t$.  Because we have assumed that laser 2 is always on \footnote{In fact, all we have assumed is that laser 2 is turned on well before laser 1, so that the interaction points are determined by the timing of laser 1.}, the transition points are defined by the intersection of the atom's geodesic with the light pulse from laser 1.  Of course, the interaction takes some small amount of time, but we assume that the phase of the laser at the beginning of the interaction is the one imparted because the corrections to this approximation are typically small and do not affect the leading order GR result (see the discussion of the short pulse limit in Section \ref{Sec:AI proof}).  Thus the phase which laser 2 imprints on the atom is the phase of the light from laser 2 (the passive laser) which is passing the atom at the first instant of the interaction as set by laser 1 (the control laser).  Finding this phase requires determining the time that this part of the light left laser 2, which we label $t_A^{(2)}$, $t_B^{(2)}$, etc.  We will usually assume that the phase of laser $i$ is just its frequency (as defined by Eq. \eqref{Eq: laser frequency reference}) times its proper time $\phi_{Li} (t) = \omega_i \tau_{Li}(t)$.  For a time independent metric this is also $\phi_{Li} (t) = \omega_i \sqrt{g_{00}} t$.  So in this case, the contributions to laser phase from laser 1, the control laser, completely cancel.

We now have rules for finding the final phase shift in an atom interferometer,
\begin{equation}
\label{Eqn:Total phase}
\Delta \phi_\text{tot} = \Delta \phi_\text{propagation} + \Delta \phi_\text{separation} + \Delta \phi_\text{laser}
\end{equation}
in general relativity.

It can be seen that this formula is independent of the output port used to calculate the phase.  The propagation phase does not depend on output port.  Laser phase depends on output port since in the slow output port it includes the phase of the laser at point D, while in the fast output port it includes the phase of the laser at point E.  As can be seen from Figure \ref{Fig:AI-SingleInterferometer}, this is a difference of $\sim \omega \left(t^{(2)}_E - t^{(2)}_D \right)$ where $\omega$ is the frequency of laser 2.  The separation phase also differs between ports because the momentum used in Eq. \eqref{Eqn: Separation Phase} is the average momentum of the relevant component of the atom's state after the final beamsplitter pulse.  In the fast output port this momentum is $\sim m v_r \approx k_\eff$ larger than in the slow output port.  So the difference in separation phase between the fast and the slow output port is $\sim \keff (x_E - x_D) \approx 2 \omega (x_E - x_D)$, since $\keff \approx 2 \omega$ for a two-photon transition.  But $x_E - x_D \approx t_E -t_D$ since points D and E lie on a null geodesic, and as a result we find that $x_E - x_D \approx \frac{1}{2} \left(t^{(2)}_E - t^{(2)}_D \right)$.  Therefore, the difference in the separation phase between the two output ports is exactly compensated by the change in the laser phase, and Eq. \eqref{Eqn:Total phase} is independent of which output port is used.

%{\bf maybe output port independence? maybe more on finite pulse time and cloud size effects?}

\section{GR Effects in the Earth's Gravitational Field}
\label{Sec:Schwarzschild}
The methods of the previous section can be used to find the effects of general relativity in an atom interferometer in the earth's gravitational field.  The space-time can be modeled with the Schwarzschild metric
\begin{equation}
\label{eqn:Schwarzschild metric}
ds^2 = \left(1+2 \phi \right) dt^2 - \frac{1}{1+2 \phi} dr^2 - r^2 d\Omega^2
\end{equation}
where $\phi = - \frac{GM}{r}$ is the gravitational potential.  For simplicity, in this section the rotation of the earth is neglected.  It will not modify the GR effects given here, and the possibility of measuring relativistic effects associated with that rotation will be considered in Section \ref{Sec:Lense-Thirring}.  Of course, this rotation will contribute non-relativistic terms that can be backgrounds, which can easily be calculated using simpler, non-relativistic methods.  Because the earth's gravitational field is weak, $\phi \sim 10^{-9}$ at the surface, we can take a post-Newtonian expansion in $\phi$.  In order to study effects beyond GR as well, we will calculate the phase shift in the parameterized post-Newtonian (PPN) expansion of the Schwarzschild metric
\begin{equation}
\label{Eqn: PPN metric}
ds^2 = (1 + 2 \phi + 2 \beta \phi^2) dt^2 - (1-2 \gamma \phi) dr^2 - r^2 d\Omega^2.
\end{equation}
Here $\beta$ and $\gamma$ parameterize modifications of general relativity and $\beta = \gamma = 1$ gives normal GR.  For the results presented in this section we will generally be considering this PPN expansion of the Schwarzschild metric, though there is little difference in a weak gravitational field.  Section \ref{Sec:earth calc} contains the details of the phase shift calculation for this metric.  In Section \ref{Sec:earth results} we present the results of this calculation and explain their physical significance.  It can be read without the previous Section \ref{Sec:earth calc}.

\subsection{Interferometer Calculation in the Schwarzschild Metric}
\label{Sec:earth calc}
%show details of calculating phase shift in (PPN expanded?) schwarzschild metric for application to earth based experiments
%discuss approximation techniques (pseries) and show geodesics
%say why we can take the laser to be at a fixed coordinate
%physical definition of v_L
%perhaps why LLF doesn't actually make a difference?

The geodesic equation \eqref{Eqn:Geodesic Eqn} for the metric \eqref{Eqn: PPN metric} cannot in general be solved exactly.  We approximate the solution using a power series solution in $\tau$.  By varying the order of the series, we ensure that we use a sufficiently high order to include all measurably large terms in the final phase shift.  For simplicity, we present first the approximate solutions using the metric \eqref{eqn:Schwarzschild metric} for a radial geodesic:
\begin{eqnarray}
r(\tau) & = & r_0+ {v_r}_0 \tau -\frac{\eta}{2} \left( \partial_r \phi (r_0)\right) \tau ^2-\frac{\eta}{6} {v_r}_0 \left(\partial^2_r \phi (r_0)\right) \tau ^3+ \mathcal{O}(\tau^4) \\
t(\tau) & = & t_0+\frac{\sqrt{{v^2_r}_0+\eta +2 \eta  \phi (r_0)} }{1+2 \phi (r_0)} \tau -\frac{{v_r}_0 \sqrt{{v^2_r}_0+\eta +2 \eta  \phi (r_0)} \partial_r \phi (r_0)}{(1+2 \phi (r_0))^2} \tau^2 + \\ & &
\frac{\sqrt{{v^2_r}_0+\eta +2 \eta  \phi (r_0)} \left(\left(4 {v^2_r}_0+\eta +2 \eta  \phi (r_0)\right) \partial_r \phi (r_0)^2-{v^2_r}_0 (1+2 \phi (r_0)) \partial^2_r \phi (r_0)\right) \tau ^3}{3 (1+2 \phi (r_0))^3}+ \mathcal{O}(\tau^4) \nonumber
\end{eqnarray}
where $\eta = g_{\mu\nu} \frac{dx^\mu}{d\tau} \frac{dx^\nu}{d\tau}$ is 0 for null geodesics and 1 for time-like geodesics.  Note that these are roughly the normal parabolic paths with some relativistic corrections.  Also, light rays do `bend' under gravity, but in these coordinates that effect shows up in the equation for $t(\tau)$ only.  In these solutions, the potential $\phi$ has effectively been expanded around $r_0$, making this approximation better the closer the paths are to $r_0$ or, roughly, the smaller $\tau$ is.  The scale this is to be compared to is the radius of the earth, $R_E$, as this determines the size of the higher $r$ derivatives of $\phi$.  Since the atom travels a distance much smaller than $R_E$, this approximation works very well here.  The paths in the PPN metric \eqref{Eqn: PPN metric} are
\begin{eqnarray}
r(\tau) & = & r_0+{v_r}_0 \tau +\frac{\left(-{v^2_r}_0 (-1+\gamma )+\eta +2 \left({v^2_r}_0 (\beta -2 \gamma )+\beta  \eta \right) \phi
(r_0)-6 {v^2_r}_0 \beta  \gamma  \phi (r_0)^2\right) \partial_r \phi (r_0) }{2 (-1+2 \gamma  \phi (r_0)) \left(1+2 \phi (r_0)+2
\beta  \phi (r_0)^2\right)} \tau ^2 +
\nonumber \\ & &
\frac{1}{6} {v_r}_0 \left(\frac{2 \gamma  \left({v^2_r}_0 (-1+2 \gamma )-\eta -2 \left({v^2_r}_0 (\beta -3 \gamma )+\beta  \eta \right)
\phi (r_0)+8 {v^2_r}_0 \beta  \gamma  \phi (r_0)^2\right) \partial_r \phi (r_0)^2}{(1-2 \gamma  \phi (r_0))^2 \left(1+2 \phi (r_0)+2
\beta  \phi (r_0)^2\right)}+\right.
\nonumber \\ & &
\frac{2 \left(-{v^2_r}_0-\eta +2 {v^2_r}_0 \gamma  \phi (r_0)\right) \left(-2+\beta +\gamma -6 (\beta -\gamma ) \phi (r_0)-6 \beta
 (\beta -3 \gamma ) \phi (r_0)^2+16 \beta ^2 \gamma  \phi (r_0)^3\right) \partial_r \phi (r_0)^2}{(1-2 \gamma  \phi (r_0))^2 \left(1+2
\phi (r_0)+2 \beta  \phi (r_0)^2\right)^2}+
\nonumber \\ & &
\left.\frac{{v^2_r}_0 \gamma  \partial^2_r \phi (r_0)}{1-2 \gamma  \phi (r_0)}+\frac{(1+2 \beta  \phi (r_0)) \left({v^2_r}_0+\eta -2
{v^2_r}_0 \gamma  \phi (r_0)\right) \partial^2_r \phi (r_0)}{(-1+2 \gamma  \phi (r_0)) \left(1+2 \phi (r_0)+2 \beta  \phi (r_0)^2\right)}\right)
\tau ^3+ \mathcal{O}(\tau^4)
\end{eqnarray}
\begin{eqnarray}
t(\tau) & = & t_0+\sqrt{\frac{{v^2_r}_0+\eta -2 {v^2_r}_0 \gamma  \phi (r_0)}{1+2 \phi (r_0)+2 \beta  \phi (r_0)^2}} \tau
-\frac{{v_r}_0 (1+2 \beta  \phi (r_0)) \sqrt{\frac{{v^2_r}_0+\eta -2 {v^2_r}_0 \gamma  \phi (r_0)}{1+2 \phi (r_0)+2 \beta
 \phi (r_0)^2}} \partial_r \phi (r_0) }{1+2 \phi (r_0)+2 \beta  \phi (r_0)^2} \tau ^2 -
 \nonumber \\ & &
\left(\left(\sqrt{\frac{{v^2_r}_0+\eta -2 {v^2_r}_0 \gamma  \phi (r_0)}{1+2 \phi (r_0)+2 \beta  \phi (r_0)^2}} \left(\left(-{v^2_r}_0
(-5+2 \beta +\gamma )+\eta +2 \left({v^2_r}_0 (-6 \gamma +\beta  (8+\gamma ))+2 \beta  \eta \right) \phi (r_0)+\right.\right.\right.\right.
\nonumber \\ & &
\left.2 \beta  \left({v^2_r}_0 (8 \beta -19 \gamma )+2 \beta  \eta \right) \phi (r_0)^2-36 {v^2_r}_0 \beta ^2 \gamma  \phi (r_0)^3\right)
\partial_r \phi (r_0)^2+
\nonumber \\ & &
\left.\left.\left.\left.{v^2_r}_0 (-1+2 \gamma  \phi (r_0)) \left(1+2 (1+\beta ) \phi (r_0)+6 \beta  \phi (r_0)^2+4 \beta ^2
\phi (r_0)^3\right) \partial^2_r \phi (r_0)\right)\right) \tau ^3\right)\right/
\nonumber \\ & &
\left(3 \left((-1+2 \gamma  \phi (r_0)) \left(1+2 \phi (r_0)+2 \beta  \phi (r_0)^2\right)^2\right)\right)+ \mathcal{O}(\tau^4)
\end{eqnarray}
These geodesics are calculated on a computer with a symbolic algebra package and so can easily be found to higher orders.  We present here the results up to third order to illustrate the method without overcomplicating the equations.

Theoretically, the intersections points of the laser pulses with the atom geodesics can now be found as discussed above.  However, these geodesics are too complicated to solve exactly so we must make two approximations.  First, the equations are solved self-consistently only to the order in $\tau$ to which the entire calculation is done.  In practice this means, for example, inverting the series to find $\tau(t)$ which can then be plugged in to $x(\tau)$ to give $x(t)$ for the atom.  An analogous procedure is used to find $x(t)$ for the light.  The atom and light coordinate trajectories are then equated and solved perturbatively order by order in $t$.  Second, we must still expand in the variables which are small in order to simplify the results sufficiently so that they remain tractable.  The correct way to do this expansion that keeps only relevant terms and avoids an ``order of limits'' problem is to Taylor expand in all the variables simultaneously, taking into account their relative sizes.  Specifically, let $\epsilon$ signify $10^{-1}$ and plug in $\epsilon^{9} \phi$ for $\phi$ everywhere (since $\phi \sim 10^{-9}$), and similarly for the other dimensionless variables: $v_L \sim 10^{-7}$, $\frac{\keff}{m} \sim v_r \sim 10^{-10}$, $\frac{\omega_a}{m} \sim 10^{-15}$, $\frac{R_E}{T} \sim 10^{-2}$, and $mT \sim 10^{26}$.  Since the corrections to the parabolic paths are small, we have a few large terms at low orders which give easily solved equations and then very many small corrections.  So we can Taylor expand all our results in $\epsilon$ and keep only terms which are large enough to possibly affect the final answer.  For the intersection points, we usually keep any terms which are $\gtrsim 10^{-30} R_E$.  We vary this order to make sure we haven't neglected any relevant terms in the final phase shift.  The entire calculation is done on a computer using a symbolic algebra package.

We take each laser to be at a fixed coordinate location because the lasers are assumed to be fixed to the earth.  Note that this is not a geodesic.  A fixed radial coordinate implies a fixed physical position since this is a static, time-independent metric.  Effects such as time variations in the earth's gravitational field or oscillations of the laser platform which cause the laser's effective coordinate location (or the whole metric) to vary are very small and so will affect the leading order GR signal at an unmeasurable level.  Of course, depending on the phenomenological characteristics of the signal being searched for, such effects crossed into the zeroth order Newtonian signal could be relevant backgrounds.  They can then be calculated simply in a non-relativistic fashion.  For now we ignore them since we are interested in calculating the effects of general relativity.  In Section \ref{Sec:Measurement Strategies} we consider measuring these GR signals and there we discuss the relevant backgrounds.

The rest of the calculation will be illustrated using metric \eqref{Eqn: PPN metric}.  The interferometer is defined, as shown above, by the initial positions and momenta of the laser pulses and the atom.  For simplicity, and because it illustrates all the effects we will be interested in, we make the following choices.  The lasers will be at positions $r= r_{L_1}$ and $r_{L_2}$ with frequencies $\omega_1$ and $\omega_2$, with laser 2 above laser 1.  We will take the usual definitions $\keff = \omega_1 + \omega_2$ and $\omega_\eff = \omega_1 - \omega_2$.  The pulses from laser 1 will originate at $t_1 = 0$, $t_2 = \frac{T}{\sqrt{1 + 2 \phi(r_{L_1}) + 2 \beta \phi(r_{L_1})^2 }}$, and $t_3 = 2 t_2$.  Following the prescription given in Eq. \eqref{Eq: laser frequency reference} for referencing the light momenta, an observer at the laser has coordinate velocity
\begin{equation}
U^\mu = \frac{dx_\text{obs}^\mu}{d\tau} = \left( \frac{dt_\text{obs}}{d\tau}, \frac{dx_\text{obs}}{d\tau} \right) = \left( \frac{1}{\sqrt{1 + 2 \phi(r_{L_1}) + 2 \beta \phi(r_{L_1})^2 }}, 0 \right)
\end{equation}
giving an initial light momentum for the pulses from laser 1 of
\begin{equation}
\frac{dx_\text{light}^\mu}{d\lambda}=\left( \frac{dt_\text{light}}{d\lambda}, \frac{dx_\text{light}}{d\lambda} \right) = \omega_1 \left( 1+\phi (r_{L_1})+\frac{3}{2} \phi (r_{L_1})^2 -\beta  \phi (r_{L_1})^2,1-\gamma  \phi (r_{L_1})+\frac{3}{2} \gamma ^2 \phi (r_{L_1})^2\right).
\end{equation}
This is close to $\left( \omega_1, \omega_1 \right)$ but with small GR corrections.  These corrections will modify the GR effects in the final answer at leading order and so they must be included.

We will take the atom to be initially at $r= r_{L_1}$ at $t=0$.  For now we leave all expressions in terms of the unphysical coordinate launch velocity, $v_L = \frac{dr}{dt}$.  We show below that this makes no important difference to the final phase shift and it keeps the expressions simpler.  The geodesics and intersections can now all be found as explained in Section \ref{Sec:GR calc}.

As one example, we describe the calculation of intersection point C.  The initial velocity of the fast half of the atom at point A is found by adding the light momentum to the atom's initial momentum as described above.  For the atom-light interaction at A, the coordinate transformations to the LLF take a vector $V^\mu$ to
\begin{equation}
V^{\mu'}_\text{LLF} = b^{\mu'}_\nu V^\nu
\end{equation}
with
\begin{eqnarray}
b^0_0 & = & \sqrt{\left(1+v_L^2+2 v_L^2 \gamma  \phi (r_{L_1})\right) \left(1-2 \phi (r_{L_1})+2 \beta  \phi (r_{L_1})^2\right)}
\\
b^0_1 & = & -v_L (1+2 \gamma  \phi (r_{L_1}))
\\
b^1_0 & = & v_L \sqrt{(1+2 \gamma  \phi (r_{L_1})) \left(1-2 \phi (r_{L_1})+2 \beta  \phi (r_{L_1})^2\right)}
\\
b^1_1 & = & -\sqrt{(1+2 \gamma  \phi (r_{L_1})) \left(1+v_L^2+2 v_L^2 \gamma  \phi (r_{L_1})\right)}
\end{eqnarray}
The velocity of the half of the atom going from A to C in the main coordinate system (metric \eqref{Eqn: PPN metric}) is then
\begin{equation}
\left. \frac{dr}{d\tau} \right|_A = v_L+\frac{k_\eff}{m}+\frac{k^2_\eff v_L}{2 m^2}+\frac{k_\eff}{m} v_L^2-\frac{k_\eff}{m} \frac{\omega_a}{m}-v_L \frac{\omega_\eff}{m}-\frac{k_\eff}{m} \gamma  \partial_r \phi (r_{L_1})+\frac{3}{2} \frac{k_\eff}{m} \gamma ^2 \partial_r \phi (r_{L_1})^2
\end{equation}
which is roughly $v_L + v_r$ with relativistic corrections.  Intersecting the atom and light geodesics using the approximations described above gives point C.  The expressions for the coordinates are very long and so are given in Appendix \ref{Sec:Results Appendix}.  Of course, these are roughly just $r_C \approx r_{L_1} + \left(v_L + \frac{k_\eff}{m} \right) T$ and $t_C \approx T$ as they would be in the non-relativistic case.

Computing the laser phase requires finding the times $t_A^{(2)}$, $t_B^{(2)}$, etc (see \eqref{Eqn: laser phase}) by dropping null geodesics from these intersection points to the second laser at $r_{L_2}$.

The entire calculation is done on a computer using a symbolic algebra package (Mathematica) so all Taylor series orders, initial conditions and such can be changed easily.  The calculation was written for a general metric so the same code is used to calculate the phase shift for several different metrics including \eqref{eqn:Schwarzschild metric} and \eqref{Eqn: PPN metric}.

%{\bf show that separation phase coordinate ambiguities are irrelevant}

%reference equations from general GR discussion for each example formula here? show separation and other phases?

\subsection{General Relativistic Effects and Interpretation}
\label{Sec:earth results}
%give our phase shift lists and explain some terms
%summarize our variable names? (m, omega, vL,...)
%say where kgT2 comes from for 2 and 3 level atom
%explain about the kgT2v2 terms (why no gamma, how get wrong without light bending...)
%discuss effect of changing whether left or right laser defines transitions
%maybe output port independence, maybe coordinate independence (ie do calc for several coordinates)?

We present here the results of the calculation of the phase shift in an atom interferometer placed in a weak gravitational field such as the earth's.  The physical origins of the important terms in the phase shift will be discussed, focusing on the GR terms and their interpretation.

To summarize the variables we are using, $T$ is the interrogation time between pulses on the laser's clock, $\omega_i$ is the frequency of laser $i$, $\keff \equiv \omega_1 + \omega_2$, $\omega_\eff \equiv \omega_1 - \omega_2$, $\omega_a$ is the frequency of the atomic transition between states 1 and 2 of the atom (see Figure \ref{Fig:Raman}), $m$ is the rest mass of the atom in the lower ground state, $v_L$ is the atom's launch velocity in the radial (vertical) direction, $r_{L_i}$ is the position of laser $i$, $\phi(r)$ is the local gravitational potential (for a perfect Schwarzschild metric $\phi(r) = - \frac{GM}{r}$), $g = - \nabla \phi$, $\beta$ and $\gamma$ are PPN parameters in the metric, and $c=\hbar = 1$.  We present the results for the final phase shift for metric \eqref{Eqn: PPN metric} in Table \ref{Tab: phases}.  The phase shift has been expanded into a sum of terms and we have grouped terms that have the same scalings with experimental control parameters.  Table \ref{Tab: phases} also displays the results of a non-relativistic (NR) calculation for the phase shift in a gravitational potential (see Section \ref{Sec:AI} for a description of the NR calculation).  Note that we have kept the laser frequencies the same for all three pulses, though in a real experiment these would have to be tuned to keep the transitions on resonance in order to compensate for the Doppler shift due to the atom's acceleration under gravity \footnote{Without such compensation, the transfer efficiency of the beamsplitter and mirror pulses would be prohibitively small.}.

\begin{table}
\begin{center}
\begin{math}
\begin{array}{|l|c|c|c|c|}
\hline
& \text{GR Phase Shift} & \text{Size (rad)} & \text{Interpretation} & \text{NR Phase Shift}\\
\hline
1. & -k_{\text{eff}} g T^2  & 3.\times 10^8 & \text{Newtonian gravity} & -k_{\text{eff}} g T^2\\
2. & -k_{\text{eff}}(\partial_r g)  v_L T^3 & -2.\times 10^3 & \text{1st gradient}& -k_{\text{eff}}(\partial_r g)  v_L T^3\\
3. & -\frac{7}{12}k_{\text{eff}} (\partial_r g) g T^4  & 9.\times 10^2 & & -\frac{7}{12}k_{\text{eff}} (\partial_r g) g T^4\\
4. & -3 k_{\text{eff}} g^2 T^3  & -4.\times 10^1 & \text{finite speed of light and} & \\
5. & -3 k_{\text{eff}} g  v_L T^2 & 4.\times 10^1 & \text{Doppler shift corrections}& \\
6. & -\frac{k_{\text{eff}}^2}{2 m}(\partial_r g) T^3  & -7.\times 10^{\text{-1}} & \text{1st gradient recoil} & -\frac{k_{\text{eff}}^2}{2 m}(\partial_r g) T^3\\
7. &  \left(\omega_{\text{eff}}-\omega_a\right) g T^2& -4.\times 10^{-1} & \text{detuning}& \\
8. & (2-2 \beta -\gamma) k_{\text{eff}} g \phi T^2  & -2.\times 10^{\text{-1}} & \text{GR (non-linearity)}& \\
9. & -\frac{3 k_{\text{eff}}^2}{2 m} g T^2 & 2.\times 10^{\text{-2}} & & \\
10. & -\frac{7}{12} k_{\text{eff}}  v_L^2 (\partial_r^2 g) T^4 & 8.\times 10^{\text{-3}} & \text{2nd gradient}& -\frac{7}{12} k_{\text{eff}}  v_L^2 (\partial_r^2 g) T^4\\
11. & -\frac{35}{4} k_{\text{eff}} (\partial_r g) g v_L T^4 & 6.\times 10^{\text{-4}} && \\
12. & -4 k_{\text{eff}}(\partial_r g)  v_L^2 T^3 & -3.\times 10^{\text{-4}} && \\
13. & 2 \omega _a g^2 T^3  & 2.\times 10^{\text{-4}} && \\
14. & 2 \omega _a g v_L T^2  & -2.\times 10^{\text{-4}} && \\
15. & -\frac{7 k_{\text{eff}}^2 }{12 m} v_L (\partial_r^2 g) T^4  & 7.\times 10^{\text{-6}} & \text{2nd gradient recoil} & -\frac{7 k_{\text{eff}}^2 }{12 m} v_L (\partial_r^2 g) T^4\\
16. & -12 k_{\text{eff}} g^2 v_L T^3  & -7.\times 10^{\text{-6}} & & \\
17. & -7 k_{\text{eff}} g^3 T^4  & 4.\times 10^{\text{-6}} && \\
18. & -5 k_{\text{eff}} g v_L^2 T^2  & 3.\times 10^{\text{-6}} & \text{GR (velocity-dependent force)}& \\
19. & (2-2 \beta -\gamma) k_{\text{eff}} \partial_r (g\phi) v_L T^3  & 2.\times 10^{\text{-6}} & \text{GR 1st gradient}& \\
20. & \frac{7}{12}(4-4 \beta -3\gamma)k_{\text{eff}} \phi  (\partial_r g) g T^4   & -2.\times 10^{\text{-6}} & \text{GR}& \\
21. & \left(\omega _{\text{eff}}-\omega _a\right)(\partial_r g) v_L T^3  & 2.\times 10^{\text{-6}} && \\
22. & \frac{7}{12} \left(\omega _{\text{eff}}-\omega _a\right)(\partial_r g) g T^4  & -1.\times 10^{\text{-6}} && \\
23. & -\frac{7}{12} (2-2 \beta -\gamma) k_{\text{eff}} g^3 T^4   & -3.\times 10^{\text{-7}} & \text{GR}& \\
24. & -\frac{7 k_{\text{eff}}^2}{2 m}(\partial_r g) v_L T^3  & -2.\times 10^{\text{-7}} && \\
25. & -\frac{27 k_{\text{eff}}^2}{8 m}(\partial_r g) g T^4  & 2.\times 10^{\text{-7}} && \\
26. & \frac{k_{\text{eff}} \omega _a}{m}g T^2  & -1.\times 10^{\text{-7}} && \\
27. & 6 (2-2 \beta -\gamma) k_{\text{eff}} \phi  g^2 T^3   & 5.\times 10^{\text{-8}} & \text{GR}& \\
28. & 3 \left(\omega _{\text{eff}}-\omega _a\right) g^2 T^3  & 4.\times 10^{\text{-8}} && \\
29. & 3 \left(\omega _{\text{eff}}-\omega _a\right)g v_L T^2  & -4.\times 10^{\text{-8}} && \\
30. & 6 (1-\beta) k_{\text{eff}} \phi  g v_L T^2  & ~3.\times 10^{\text{-8}} & \text{GR}& \\
\hline
\end{array}
\end{math}
\caption{\label{Tab: phases} A list of all the terms above a certain size in the phase shift from the full GR calculation for metric \eqref{Eqn: PPN metric}, along with their numerical size in radians and an interpretation.  The NR phase shift column shows the results of a completely non-relativistic phase shift calculation for comparison. The sizes of the terms assume the initial design, sensitive to accelerations $\sim 10^{-15} g$, which has $L = 9 \text{ m}$, $T = 1.3 \text{ s}$, $v_L = 13 \frac{\text{m}}{\text{s}}$, $\keff = 2 \frac{2 \pi}{780 \text{ nm}}$, $\omega_a = 6.8 \text{ GHz}$, and $m = 81 \text{ GeV}$ (for $^{87}$Rb). All detuning terms assume $\omega_\eff - \omega_a = 1 \text{~kHz}$.  Note that there is some ambiguity in how some of the terms are grouped since by definition $g = - \partial_r \phi$.}
\end{center}
\end{table}

To understand the GR effects underlying some of these phases, recall that, roughly, the atom interferometer is sensitive to accelerations.  The following discussion will be highly coordinate dependent and not rigorous, but its only purpose is to gain some intuition for the GR effects we find.  Combining the geodesic equations \eqref{Eqn:Geodesic Eqn} for the spatial $\vec{x}^i$ ($i=1,2,3$) and $t$, the coordinate acceleration of an atom in the frame of Eq. \eqref{Eqn: PPN metric} is
\begin{equation}
\label{Eqn: force}
\frac{d \vec{v}}{dt} = - \vec{\nabla} ( \phi + (\beta + \gamma) \phi^2 ) + \gamma (3 (\vec{v} \cdot \hat{r})^2 - 2 \vec{v}^2) \vec{\nabla} \phi + 2 \vec{v} (\vec{v} \cdot \vec{\nabla} \phi)
\end{equation}
with $\vec{v} = \frac{d\vec{x}}{dt}$ for this equation only.  The acceleration is approximately that from Newtonian gravity, $- \vec{\nabla} \phi$, but with leading order GR corrections.  These corrections fall into two classes, both of which will interest us.

The $\nabla \phi^2$ terms are related to the non-linear nature of gravity, the fact that a gravitational field seems to source itself in general relativity.  This could also be called the non-Abelian nature of gravity since gravitational energy gravitates through a three-graviton vertex. To see that this is the origin of the $\nabla \phi^2$ terms, note that, because of these terms, the divergence of the gravitational field given in Eq. \eqref{Eqn: force} is nonzero (here by gravitational field we mean $\vec{g} \equiv \frac{d \vec{v}}{dt}$ from Eq. \eqref{Eqn: force}).  Just as for an electric field, a nonzero divergence of the gravitational field implies a local source density (in general relativity this means a local energy density) that is proportional to that divergence.  So Eq. \eqref{Eqn: force} implies that there is a local energy density in free space proportional to $\nabla \cdot \vec{g} \propto \nabla \cdot \nabla \phi^2 = 2 (\nabla \phi)^2$.  But note that to leading order $\nabla \phi \approx \vec{g}$ so that $\nabla \cdot \vec{g} \propto \vec{g}^2$.  In other words, the local energy density is proportional to the field squared, exactly as expected from the electric field analogy.  This energy is then the source of the $\nabla \phi^2$ terms.  The non-linearity of gravity is parametrized in the standard way by the PPN parameter $\beta$.

The other terms in Eq. \eqref{Eqn: force} proportional to $\vec{v}^2 \nabla \phi$ are velocity dependent forces.  These terms are related to the gravitation of the atom's kinetic energy (or the kinetic energy of the source mass in the frame where the atom is stationary and the source is moving), since all energy, not just mass, gravitates in general relativity.

The non-linear GR corrections in Eq. \eqref{Eqn: force} are smaller than Newtonian gravity by a factor of $\phi \sim 10^{-9}$, while the velocity dependent force terms are smaller by $v^2 \sim 10^{-15}$ for the atom velocities we are considering. We will see that the non-linear terms can only be measured through a gradient of the force produced and so are reduced by an additional factor of $\frac{10 \text{m}}{R_\text{earth}} \approx 10^{-6}$ for a 10m long experiment.  Both effects are then $\sim 10^{-15} g$.

%put in gravitomagnetism?

% para explaining many terms briefly (gravity, 1st grad, doppler shifts, detuning effects from omega gT2 and omega_a g T^2
These effects can be seen in the total phase shift in the interferometer.  Table \ref{Tab: phases} presents the answer for the total phase shift as found by the relativistic calculation outlined above.  It lists all the terms in the total phase shift large enough to be measured by the initial apparatus.  Effectively, the local gravitational acceleration is expressed as a Taylor series in the height above the Earth's surface.  The first phase shift in Table \ref{Tab: phases} represents the effect of the leading order (constant) piece of the local acceleration while the 2nd and 10th terms are the next gradients in the Taylor series.  Notice that even the second gradient of the gravitational field is relevant for this interferometer.  The terms in this list that have been measured agree with the results of previous experiments.  The largest two phase shifts due to the first two terms in the Taylor expansion of the local $g$ field were known and measured several times (e.g. \cite{PhysRevLett.67.181, 0026-1394-38-1-4}).  The 4th and 5th terms arise from the second order Doppler shift of the laser's frequency as seen by the moving atom.  These Doppler shift and finite speed of light corrections (terms 4 and 5) were known and measured to cancel each other in a `symmetric' interferometer \cite{Peters Thesis}.  The 7th term is proportional to the two-photon detuning between the difference in the two lasers' frequencies, $\omega_\eff$, and the resonant frequency of the atomic transition, $\omega_a$.  In any practical experiment this detuning is kept quite small and this term will be negligible.  The terms proportional to $\omega_a$ and $\omega_\eff$ almost cancel since the lasers' frequencies are usually chosen to be on resonance with the atomic transition and so could never have been measured given the precision of previous interferometers.  These terms were not previously known because their calculation requires a fully relativistic calculation.  The recoil shift $\frac{k^2_\eff}{m} T^3 \partial_r g$ was known and measured \cite{Chu recoil, Chu recoil2}.

The 8th, 18th through 20th, 23rd, 27th, and 30th terms arise only from GR and are not present in the results of our Newtonian calculation.  The 8th and 19th terms arise in part from the non-linear nature of gravity.  This is clear since they look like the analogue of $kgT^2$ and the 1st gradient terms but with $g$ replaced by the part of the acceleration coming from the non-linearity of gravity, $g \phi$ in Eq. \eqref{Eqn: force}.  Similarly the 18th term arises in part from the velocity dependent forces in Eq. \eqref{Eqn: force} since it appears to be an acceleration $\sim g v_L^2$.  Note that of course the acceleration from these velocity dependent forces is actually proportional to the integrated effects along the entire trajectory of the atom.  However, this is obscured by the expansion we are taking, and so we just see the largest term, $\propto v_L^2$, with terms proportional to the other velocities $v_r$, $gT$ and so on, farther down the list.  In fact, for every term $\propto v_L$, we expect and see a term with $v_L$ replaced by $gT$ also on the list, since the velocity changes over the course of the interferometer by roughly this amount.

We now address the fact that we have left the phase shift in Table \ref{Tab: phases} in terms of the unphysical (coordinate-dependent) launch velocity.  This is the only coordinate dependent variable in Table \ref{Tab: phases}; all others ($\keff$, $T$, etc.) have coordinate invariant definitions.  Ultimately, in any real experiment the experimenter determines how to measure the launch velocity, and this gives the physical, coordinate-invariant definition.  On this point, different experiments will surely vary, so here we assume a simple prescription but leave Table \ref{Tab: phases} in the general form in terms of $v_L$, which should allow any prescription to be applied.  We assume that the atom is launched by the lasers using $n_L$ photon kicks (Raman or Bragg transitions).  The experimental definition of the physical launch velocity will then be $v_p = n_L \frac{k_\eff}{m}$.  Note that $n_L$, $k_\eff$, and $m$ all have physical, coordinate-indepedent definitions.  Repeating the normal atom-light interaction calculation (Eq. \eqref{Eqn:atom-light momenta addition} and ensuing discussion) $n_L$ times then gives the relation
\begin{equation}
\label{Eqn:Physical velocity}
v_L \approx v_p \left(1 - \gamma \phi - \frac{\omega_a}{m} \left(1 - \gamma \phi  \right) - \frac{n_L -1}{2} \frac{\omega_\eff}{m} \left(1 - \gamma \phi  \right)  \right)
\end{equation}
with higher order terms dropped.  It is not surprising that there are higher order GR corrections when the coordinate launch velocity is written in terms of a physically measurable parameter.  Here we are only interested in this if it changes the GR effects we seek to measure, for example by changing the dependence on the PPN parameters.  It is clear that substituting Eq. \eqref{Eqn:Physical velocity} into the phase shift in Table \ref{Tab: phases} will not affect the two GR terms we are most interested in, 8 and 18.  It can affect other GR terms at the level of $10^{-6} \text{ rad}$ and below, but it cannot remove totally the dependence on the PPN parameters of GR terms 8 and 18, and so it does not qualitatively change their interpretation.

\begin{table}
\begin{center}
\begin{tabular}{|l|c|c|c|c|c|c|c|}
\hline
& Parameter & Total Phase Shift & Propagation Phase & Separation Phase & Laser Phase &  \\
& Dependence & Coefficient & Coefficient & Coefficient & Coefficient & Size (rad) \\
\hline
1. & $k_\eff T^3 (\partial_r g)$ & 0 & $1$ & $-1$ & 0 & $4 \times 10^{10}$ \\
2. & $k_{\eff} g T^2$ & $- 1$ & $- 1$ & $1$ & $- 1$ & $3 \times 10^8$ \\
3. & $\omega_\eff g T^2$ & $1$ & $2$ & $-2$ & $1$ & $3 \times 10^3$ \\
4. & $\omega_a g T^2$ & $- 1$ & $- 1$ & $0$ & $0$ & $3 \times 10^3$\\
5. & $k_{\eff} (\partial_r g) T^3 v_L$ & $- 1$ & $2$ & $-2$ & $- 1$ & $2 \times 10^3$ \\
6. & $k_\eff  (\partial_r g)\phi T^3$ & 0 & $2\gamma+2\beta-2$ & $-2\gamma-2\beta+2$ & 0 & $3 \times 10^{1}$ \\
7. & $k_{\eff} g T^2 v_L$ & $-3$ & $-5$ & $5$ & $-3$ & $1 \times 10^1$  \\
8. & $k_{\eff} g \phi T^2$ & $2 - 2 \beta - \gamma$ & $2 - 2 \beta - \gamma$ & $-2+2 \beta+\gamma$ & $2 - 2 \beta - \gamma$ & $2 \times 10^{-1}$ \\
9. & $k_{\eff} g^2 T^3 v_L$ & $-12$ & $-17$ & $17$ & $-12$ & $7 \times 10^{-6}$ \\
10. & $ k_{\eff} \partial_r (g \phi) T^3 v_L$ & $2 -2 \beta - \gamma$ & $-4+4 \beta+2 \gamma$ & $4-4 \beta - 2\gamma$ & $2 -2 \beta - \gamma$ & $2 \times 10^{-6}$ \\
11. & $k_{\eff} g T^2 v_L^2$ & $-5$ & $-9$ & $9$ & $-5$ & $5 \times 10^{-7}$ \\
\hline
\end{tabular}
\caption{\label{Tab: phases origin} A breakdown of some of the terms in the phase shift, Table \ref{Tab: phases}, list by origin.  The sizes are given for the coefficient of the term in the total phase shift.  Note that there is some ambiguity in dividing terms 9 and 10 since by definition $g = - \partial_r \phi$.}
\end{center}
\end{table}

The origin of some of the largest terms in the phase shift list highlights important differences between a relativistic calculation and a non-relativistic one.  From Table \ref{Tab: phases origin} we see that frequently the contributions to a given term in the phase shift lift from the propagation and separation phases cancel, and so the term can be considered to come from laser phase.  There is even a term in the propagation and separation phases that is larger than $k g T^2$ which cancels.  Note that this term is $k_\eff c T^3 (\partial_r g)$ if we do not take $c=1$.  Many of the terms in propagation and separation phase can be considered to arise from the fact that, in a relativistic calculation, the endpoints of the interferometer (points D and E in Fig. \ref{Fig:AI-SingleInterferometer}) are not simultaneous.  This has a large effect because the phase of the atom evolves at a rate proportional to its mass $m$, as follows from the separation phase formula \eqref{Eqn: Separation Phase}.  In other words, the Compton wavelength measures separation in time just as the de Broglie wavelength does in space.

It is interesting to consider an atom interferometer with only a single laser driving the atomic transitions directly between two levels, instead of the normal two-photon transition through a virtual intermediate level.  In this case, there is no passive laser and the laser phase is zero, as mentioned in Section \ref{Sec: GR phase shift formulae}.  This means that the $k_{\eff} g T^2$ term is removed as are most of the $k_\eff$ terms.  However we must now have the two atomic levels spaced by an energy which is roughly the frequency of a laser, so $\omega_a \sim k$, instead of $\omega_\eff$.  After doing such a single laser calculation we find that the largest term in the phase shift is $\omega_a g T^2$, which is roughly the same size as $kgT^2$.  This phase shift now comes from propagation phase instead of laser phase, as would be guessed from Table \ref{Tab: phases origin}.  This term arises because the rest mass of the atom is different in the two atomic states.  This means that the dominant phase shift does not depend on the laser frequency and is instead set by the intrinsic structure of the atom.  In the normal non-relativistic calculation there is only the $k_\eff g T^2$ term which comes from laser phase (not from propagation phase), and there is no $\omega_a g T^2$ term at all.  So the non-relativistic single laser calculation gets the origin of the major part of the phase shift wrong, and it gives an answer which is off by the amount the laser frequency $\omega$ is detuned from resonance $\omega_a$.

The term $-5 k_{\eff} g T^2 v_L^2$ (term 18) receives contributions from the velocity-dependent forces in Eq. \eqref{Eqn: force}, but its coefficient is independent of $\gamma$.  There are two canceling contributions to this term coming from the $\gamma$ terms in the force law for the atom and the photon.  Note that the $\gamma$ terms in the equation of motion for the light are not suppressed by any small velocity factors (since $c=1$), so they are just as large as the `normal' Newtonian gravity term (this is the origin of the famous factor of 2 in the equation for the bending of light by the sun).  This phase shift term thus measures both the effect of gravity on light and the velocity-dependent force on the atom.  If we put a different parameter, $\delta_\text{light}$, in front of the $\phi$ in the component $g_{00}$ of the metric governing the motion of the light and redo the entire calculation, this term becomes $(4+ \delta_\text{light} + \gamma_\text{light} - \gamma_\text{atom}) k_\eff g T^2 v_L^2$, where the $\gamma$'s are the PPN parameters in the metrics for the light and the atom.  This term then tests a matter-light principle of equivalence, namely that they both feel the same metric.  A new force may well couple to light and matter in a different manner than gravity and so may be testable in this way.

\section{Measurement Strategies}
\label{Sec:Measurement Strategies}
%here possible experiments and sensitivities
%backgrounds - rotations, B-fields, ...
%control parameters

In this section we consider several ideas for measuring GR or beyond GR effects using atom interferometry.  Possible measurement strategies are discussed for testing the Principle of Equivalence (PoE) and measuring the effects of the velocity dependent forces arising from GR, the non-linear nature of gravity, and the bending of light in a gravitational field.  We have discussed some of these ideas in a previous paper \cite{Dimopoulos:2006nk} but here we give a few more thoughts.  Our main motivation is to describe some phenomenological differences between the GR effects we have calculated and classical, non-relativistic effects in order to show that these GR effects are not coordinate artifacts, and are possibly distinguishable from Newtonian gravity and other backgrounds.  We do not claim to have proven that every conceivable background is under control.  Instead we only wish to argue that it may be possible to test GR in the lab using this technology, and so it is well worth trying to design experiments to do so.

In general we consider a $\sim 10 \text{ m}$ long interferometer with $T \sim 1 \text{ s}$ and $\sim 10^6$ atoms cooled and launched per shot, resulting in a final phase sensitivity of $\sim 10^{-6}~\text{rad}$ after about $10^6$ shots.  This implies sensitivity above atom shot noise of $10^{-15} g$ where $g$ is the acceleration due to gravity on earth.  As we have seen, this is enough to start measuring GR effects in the lab.  Of course, there are many possible improvements to this technology, and the potentially achievable sensitivities (at least above shot noise) were discussed in \cite{Dimopoulos:2006nk}.

\subsection{Principle of Equivalence}
\begin{figure}
\begin{center}
\includegraphics[width=300 pt]{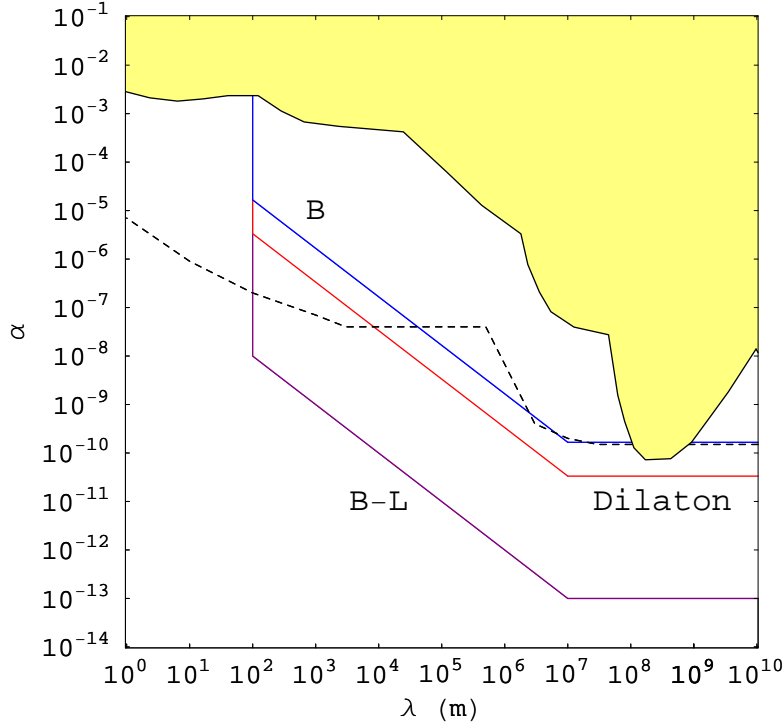}
\caption{ \label{Fig:EP limits} (Color online) Projected limits to be set by the Principle of Equivalence measurement.  Limits are shown (labeled solid lines) for several possible new Yukawa forces arising from a coupling to Baryon number (B), Baryon minus Lepton number (B-L) and the dilaton.  These assume a uniform earth density.  Previous limits are shown in solid (yellow) shading \cite{Adelberger:2003zx}.  The dashed line is the current limit on a force coupling to B from an Equivalence Principle experiment using a realistic earth model \cite{Schlamminger:2007ht}.}
\end{center}
\end{figure}

Atomic interferometers have been used to test the Principle of Equivalence (PoE) \cite{Hansch EP exp, 0026-1394-38-1-4}.  We intend to perform a similar test to $10^{-15}$ accuracy using a Rb apparatus currently under construction (see Section \ref{Sec:AI}) \cite{EP Varenna}.   The basic idea is to co-locate two atomic clouds of different species or isotope (e.g., $^{85}\Rb$ and $^{87}\Rb$) and run simultaneous atom interferometers using the same laser pulses on these two clouds (a simultaneous dual species fountain has been demonstrated \cite{Marion}).  This test is essentially looking for the presence of new forces that act in a PoE violating way, namely that are not proportional simply to the mass of the atoms but instead depend on the atomic species or isotope.  It seems likely that any new force would be PoE violating since it is hard to design a force other than gravity that does not violate the PoE.  Indeed most examples of new forces in the literature do violate the PoE.  This experiment, being done on the surface of the earth, has maximal sensitivity to PoE violating forces with a range greater than the earth's radius.  An infinite range force would also be detectable in this manner.  As the range, $\lambda$, of the new force decreases below the earth's radius, the sensitivity of the experiment decreases as $\alpha \sim 10^{-15} \left( \frac{R_E}{\lambda} \right)$ (with the overly simplistic assumption of uniform earth density) since the volume sourcing the new Yukawa force goes as $\sim \lambda^3$.  Notice that this does give sensitivity to forces with shorter ranges, down to $\sim 10 \text{ m}$, the rough size of the experiment and the distance over which it is not clear what the local mass distribution is.  Figure \ref{Fig:EP limits} shows projected limits from this experiment on a new force that couples to Baryon number (B) and Baryon minus Lepton number (B - L), and arising from a light dilaton \cite{Kaplan:2000hh}.  The large suppression factors come from the fact that $^{85}\Rb$ and $^{87}\Rb$ have similar couplings to these new forces and also roughly equal masses.  Also shown in Figure \ref{Fig:EP limits} are the current experimental limits on a force coupling to B coming from an equivalence principle experiment \cite{Schlamminger:2007ht}.  This line shows the difference that a realistic earth model makes for the limits.

%gravity gradients for PoE controlled by diff measurement, keep atoms within 10nm.
One possible background to the PoE measurement arises from the fact that the earth's gravitational field is not constant but in fact has a gradient $\sim \frac{g}{R_e}$, where $R_e$ is the radius of the earth.  Thus an increase in height of $\delta h \sim 10 \text{ nm}$ changes the acceleration by $\sim g \frac{\delta h}{R_e} \approx 10^{-15} g$, which is at the desired sensitivity.  The primary way to reduce this is to use a differential measurement between two simultaneous atom interferometers run with the same laser pulses.  In the PoE measurement it is necessary to co-locate the two isotopes in the initial atomic trap to $\lesssim 10~\text{nm}$ in order to avoid this systematic.  The differential sag in the combined gravity plus magnetic trap depends only on the isotope masses since we can trap $^{85}\Rb$ and $^{87}\Rb$ using states with the same magnetic moment.  For our designed magnetic trap curvature, the expected differential sag is $\sim 10~\text{nm}$. Further, this differential position can be measured by characterizing the magnetic trap in situ, which can be used to reduce the systematic phase error below $10^{-15}g$.

%One way this can be achieved is by stabilizing the trap position at $100 \text{ nm}$ (achievable with current traps).  Then difference in the position of the isotopes, e.g. $^{85}\Rb$ and $^{87}\Rb$, will be stable to better than $10 \text{ nm}$ because their mass to magnetic moment ratios (determining their position in the trap) only differ by $\approx 5 \%$. {\bf Jason check this paragraph too}

%control grav wrinkles because your experiment naturally low-passes gravity below 10m scale but only variations that matter are on 10nm scale because of diff measurement.  and they're naturally quite small to begin with.

%say how do velocity measurement, differential in launch velocity cancels laser phase noise, vibrations, and even gravity (kgT2).  Still get a background from grav gradients, doppler shift but these are linear in v and have different T scalings so can probably pick out GR term at order 1.  Also, coriolis/centrifugal and magnetic fields.  Could keep length fixed and scale up velocity so that you see all the same wrinkles but your signal doesn't change.  Every other background varies with v but signal is constant so just do a 1 parameter fit.  Also, the other backgrounds fall as v rises so just take huge launch velocities.  Even wrinkles goes down with v (see general formula).

\subsection{Velocity Dependent Forces}
In order to measure the GR effects of the `velocity dependent forces' and the gravitational effect on the laser light, we must pick out the phase shift term with the $v^2_L$ scaling.  The basic idea is that it is very difficult for classical gravity to mimic the effect of a true velocity dependent force.  Again we will consider a differential measurement between simultaneous atom interferometers since this aids greatly in the control of many backgrounds.  Here we consider two atom interferometers with differing launch velocities in order to select the velocity dependent term we are looking for.  This will naturally cancel many phase shifts, including the leading order one from gravity, $k_\eff g T^2$.  Of course, there are still several possible background terms that are larger than the signal and scale with launch velocity.  These include terms coming from the earth's gravity gradient, $k_\eff (\partial_r g) T^3 v_L$, the effect of the doppler shift of the laser as seen by the moving atom, $k_\eff g T^2 v_L$, and the effect of the Earth's rotation, $\keff v_L \Omega^2 T^3$, where $\Omega$ is the component of the Earth's rotation rate perpendicular to the launch velocity \cite{Bongs:2006}.  Magnetic fields do not induce velocity dependent phase shifts when the internal atomic state remains unchanged, as in a Bragg beamsplitter.  All of these background terms, although much larger than our signal, scale differently with $v_L$ and $T$ than our signal.  There are no terms which scale as $v_L^2 T^2$; this is a unique sign of GR.  Varying these around the typical values ($v_L = 13 \frac{\text{m}}{\text{s}}$ and $T=1.3~\text{s}$) then allows the GR term to be picked out from the backgrounds with a sensitivity limited only by the atom shot noise.  It is crucial for this fit that $v_L$ can be known experimentally very precisely (better than the ratio of the background to the signal).  This is possible since the launch velocity is precisely linked to laser frequencies (see the discussion of physical velocity and Eq. \eqref{Eqn:Physical velocity}) which can be known extremely well.

Additionally, in case backgrounds do become a problem, it is possible to reduce the measured size of the background terms even before this fit.  Because the GR term scales as $(v_L T)^2$, there is no loss in the signal by going to the regime where the launch velocity is large.  In this regime, the atom's velocity is roughly constant over the length $L$ of the interferometer and $L = v_L T$.  If $v_L$ and $T$ are then always scaled inversely so that $L$ is fixed, then the signal does not change but all the background terms do.  Further, by taking $v_L$ large, all the background terms are suppressed by at least one power of $v_L$ because they all have more powers of $T$ than of $v_L$.  For example, the gravity gradient term becomes $k_\eff (\partial_r g) \frac{L^3}{v_L^2}$.  Thus, the sizes of the phase shifts from these backgrounds can be directly reduced even before data analysis.

%many digits of perpendicularity come from precise length measuring (optical interferometry) and geometric constructions.  shimming can also help.

\subsection{Non-Linearity of Gravity}
\label{Sec: Non-linearity of gravity}
To measure the terms which arise from the non-linearity of gravity, it may help to run three simultaneous atomic gradiometers along three mutually orthogonal axes in a `divergence configuration'.  Such a configuration effectively measures the divergence of the local gravitational field, which must be zero in Newtonian gravity outside the source mass.  This should then allow the non-linear GR effect, $k_\eff g \phi T^2$, to be picked out.  In particular, the atoms can be launched along a single, large (e.g. 10 m) vertical axis.  Then three perpendicular atom interferometers can be run along this same axis using three perpendicular sets of lasers.  Thus the atoms can be split vertically or in either horizontal direction to make the three perpendicular atom interferometers.  Yet all three interferometers are in essentially the same position, separated only by the much smaller $v_r T \sim 1 \text{ cm}$.

One question which arises is the extent to which the three laser axes can be made mutually orthogonal, since they must be perpendicular to one part in $10^9$ (since on the earth $\phi \sim 10^{-9}$) in order to reduce the earth's gravity gradient below the GR signal.  Methods for measuring angles with nanoradian precision have already been demonstrated, albeit for angles much smaller than 90 degrees \cite{nanoradian}.  It may also be possible to geometrically construct laser axes which are perpendicular to very high accuracy by using the ability to accurately measure distances with a laser interferometer.

%Commercially available systems can already change angles at the {\bf nanoradian level cite???}.

%For example if you construct four points whose distances between nearest neighbors are all equal and whose diagonals are equal than it must be a square.  Thus the two diagonals will be perpendicular axes.  This is one possible way of converting a very accurate measurement of distance into a similarly accurate construction of perpendicular laser axes.  This is a proof of principle but there may be many more practical constructions.

It is also possible to ameliorate the requirement on the perpendicularity of the laser axes by reducing the local gravity gradient with an appropriately constructed local mass distribution.  We will show that it is possible to reduce the gravity gradient along all three perpendicular axes by $\OO(1)$ of its natural size on the earth.  With such a construction, the three gravity gradients can then be measured by the atom interferometer itself.  It should then be possible to make minor modifications to the mass distribution to cancel the gravity gradients with increasing precision, without an exact knowledge of the angles of the atom interferometers.  Every order of magnitude reduction in the size of the local gravity gradient reduces the requirement on perpendicularity of the laser axes by an order of magnitude.  Since the atom interferometer can measure gravity gradients very precisely, it may be possible to align the lasers to sufficient accuracy without a complicated alignment mechanism.

Now we must show that the earth's gravity gradients in all three directions (i.e. $\partial_x \vec{g}_x$, $\partial_y \vec{g}_y$, and $\partial_z \vec{g}_z$) can be cancelled to $\OO(1)$.  If the $z$-axis runs perpendicularly to the local surface of the earth at the point in question, then $\vec{g}_z \approx - \frac{G M_\text{earth}}{R^2_\text{earth}}$ \footnote{Of course there are small corrections due to the fact that the earth is not perfectly spherical.} is negative and the gradient $\partial_z \vec{g}_z \approx - 2 \frac{\vec{g}_z}{R_\text{earth}}$ is positive.  Adding more mass on the $z$-axis either completely above or completely below the atom interferometer will only add a positive quantity to $\partial_z \vec{g}_z$, thus increasing the gravity gradient.  Therefore we must add mass around the atom interferometer.  As a proof of principle, take a sphere of mass centered on the point in question with a cylindrical hole along the $z$-axis (the atom interferometer apparatus would be placed in this hole).  Assuming the radius of the cylinder is small compared with the radius of the sphere, the Newtonian gravitational field due to the sphere inside itself has a derivative with an opposite sign compared to the earth's gradient: $\partial_z \vec{g}_z = - \frac{4}{3} \pi G \rho_\text{sphere}$.  Since in vacuum the divergence is zero in Newtonian gravity ($\partial_x \vec{g}_x + \partial_y \vec{g}_y + \partial_z \vec{g}_z = 0$) and there is a rotational symmetry about the $z$-axis, the other two components of the sphere's gravity gradient are $\partial_x \vec{g}_x = \partial_y \vec{g}_y = - \frac{1}{2} \partial_z \vec{g}_z$.  With an appropriate choice of $\rho_\text{sphere}$ then, the earth's gravity gradient can be cancelled off.  Notice that the sphere cancelled the $z$-component of the earth's gradient while the cylindrical hole cancelled the $x$- and $y$-components, since in the infinite limit the cylinder only has $x$- and $y$-components of acceleration.

\begin{figure}
\begin{center}
\includegraphics[width=400 pt]{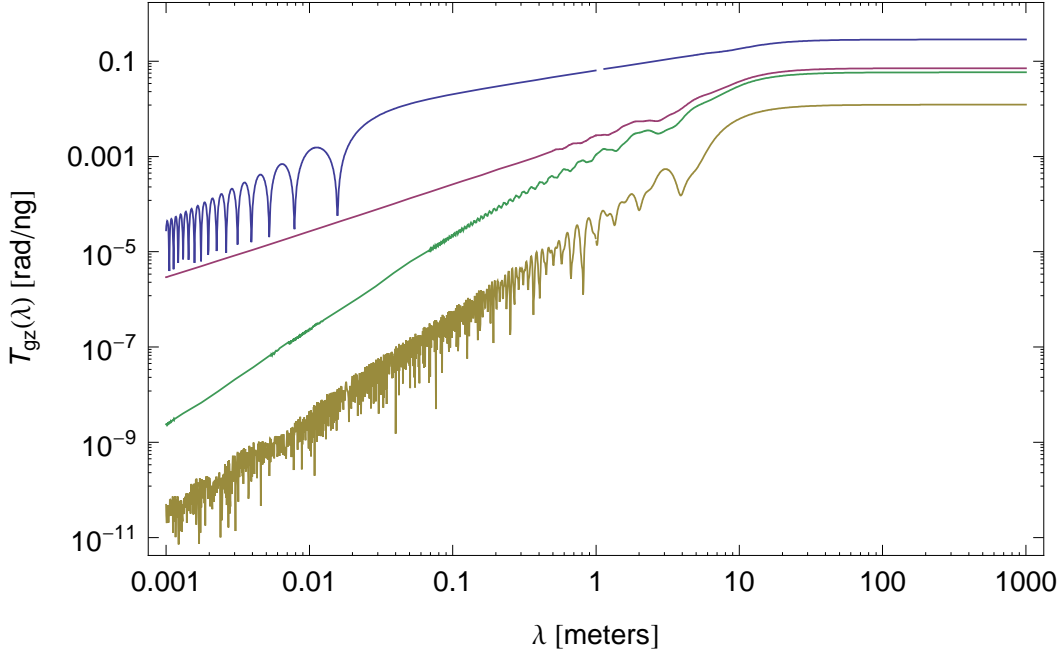}
\caption{ \label{Fig:gravity wrinkles} (Color online) The response function, $T_{gz}$, as defined in Eq. \eqref{Eq: gravity wrinkles}.  It gives the phase shift response of the atom interferometer to a Fourier component of the local gravitational field with wavelength $\lambda$ and amplitude $10^{-9} g_\text{earth}$.  All curves assume the example 10 m atom interferometer.  The top curve assumes it is run in a `symmetric' configuration (see text), the next lower curve assumes the atoms are dropped from rest at the top of the device and the following two curves assume a downward launch velocity of $1 \frac{\text{m}}{\text{s}}$ and $10 \frac{\text{m}}{\text{s}}$ respectively.}
\end{center}
\end{figure}

\subsection{General Backgrounds}
%B field paragraph: all E and M is background B
%give dominant effect of background B
%control with shielding and m=0 states
One possible background to any measurement made with an atom interferometer arises from the interaction of the atoms with ambient electromagnetic fields.  The electric fields present in any realistic setup are too small to give detectable phase shifts as they are easily screened.  Surface effects such as the Casimir interaction are negligible as the atoms are kept far from all surfaces.  Only ambient magnetic fields can give large enough phase shifts to be potential backgrounds.  An atom responds strongly to a background magnetic field, so we usually consider a magnetically shielded interferometer.  Ambient magnetic fields can be shielded down to the nT level (see for example \cite{nnbar B shield}) which leads to appropriately small phase shifts.  The atoms are prepared in a magnetically insensitive, $m=0$, state and so the energy shifts that arise are second order in the magnetic field, $\Delta E = \frac{1}{2}\alpha B^2$ where $\alpha$ is the second order Zeeman coefficient of the state. Since the internal levels of the atom can have different values of $\alpha$, magnetic phase shifts are generally smaller when the internal state is fixed, as is the case in an interferometer that uses Bragg transitions.  Systematic phase shifts can then only result from spatial variations in the field: $$\Delta \phi \approx  -\keff\frac{\alpha}{m} B_0 \frac{\partial \delta B}{\partial z} T^2$$
where $B=B_0+\delta B$, $B_0\sim 100~\text{nT}$ is the constant bias magnetic field and $\delta B$ is a small field perturbation. Variations of $\delta B \sim 1~\text{nT}$ over the length of the interferometer give negligible phase shifts for a $^{87}\text{Rb}$ apparatus.  The above formula generally holds for field perturbations that vary on length scales that are long compared to the interferometer arm splitting.  In the opposite limit, the interferometer averages over perturbations with wavelengths that are small compared to the overall interferometer length.  This spatial averaging behavior also occurs for short wavelength gravity perturbations, as we describe in more detail below.

Another potential background in the interferometer arises from atom-atom collisions within the atom cloud.  For $^{87}$\!Rb, the frequency shift of the atomic state $\ket{F=2, m_F=0}$ due to atom-atom collisions is
$$\delta\nu \approx (-0.9~\text{mHz})\left(\frac{n}{10^9~\text{cm}^{-3}}\right)\sqrt{\frac{\tau}{1~\mu\text{K}}}$$ for a cloud of number density $n$ and temperature $\tau$ \cite{GibbleNumberDensity,NumberDensityRbData}.  Unlike in atomic clocks, phase errors due to this effect in a $\frac{\pi}{2}$--$\pi$--$\frac{\pi}{2}$ pulse sequence atom interferometer implemented using Bragg atom optics are the result of unequal densities between the two arms.  Nominally, the upper and lower arms of the interferometer have the same atom number density, but an imperfect initial beamsplitter can cause an asymmetry between the arms, resulting in a phase shift
$$\delta\phi_\text{collision}=4\pi \delta\nu T \approx (1.1\times 10^{-2}~\text{rad})\left(\frac{ n}{10^9~\text{cm}^{-3}}\right)\sqrt{\frac{\tau}{1~\mu\text{K}}}\left(\frac{T}{1~\text{s}}\right) \left( \frac{\Delta n}{n} \right) $$
for a density difference $\Delta n$.  This represents an upper bound on the atom-atom phase shift, since in reality the cloud density decreases in time during the experiment due to ballistic expansion\footnote{This result does not apply to an interferometer that uses Raman atom optics, since in that case the time evolution of the density leads to the main effect.}.  Making the conservative assumption that the density difference can be controlled at the level of $\frac{\Delta n}{n}\sim 10^{-2}$ implies a phase error of $\sim 10^{-4}~\text{rad}$.  However, this systematic offset is not a concern for many of the experiments we consider since it is expected to cancel as a result of our differential measurement strategies.  This cancellation relies on the condition that the density does not depend on any of the other control parameters in the experiment, an assumption that must be verified experimentally\footnote{This cancellation does not occur in the case of the Equivalence Principle measurement, since the two isotopes have different atom-atom interaction strengths.}.

An additional tool that may be useful for reducing backgrounds and picking out the signal terms is the possibility of running the atom interferometer with different pulse sequences.  While the $\frac{\pi}{2}$-$\pi$-$\frac{\pi}{2}$ sequence is an accelerometer, more complicated sequences can be used which suppress accelerations and leave the gravity gradients, or vice versa.  In general, either $T^2$ or $T^3$ terms, or both, can be removed by a suitable choice of pulse sequence \cite{BorisDiamonds,Audretsch:PRA53.1}.  This can remove most of the relevant backgrounds to the velocity-dependent force measurement since they scale with higher powers of $T$.  These different pulse sequences improve the ability to pick out a term that scales in a particular way with the control parameters.

Finally, it is possible that small gravitational `anomalies' due to local masses may be a background.  In fact, only the small wavelength variations in the local $g$ field can be a relevant background for these experiments.  Perturbations to local $g$ at wavelengths larger than the rough size of the experiment (e.g. 10 m) are well described by the Taylor series expansion that we have assumed for the earth's field.  Since these perturbations are naturally small compared to the earth's field, the differential measurement strategies discussed above for the Principle of Equivalence, velocity dependent force, and non-linearity of gravity measurements will remove these long wavelength perturbations in exactly the same way as they removed the earth's field.  

%However, short wavelength perturbations to the local $g$ field can be relevant.

Short wavelength perturbations in the local gravitational field can be relevant.  Luckily, the atom interferometer naturally averages over these perturbations because of its spatial length.  We write the total phase shift due to gravity anomalies along the vertical ($z$) direction summed over all wavelengths $\lambda$ as
\be
\label{Eq: gravity wrinkles}
\Delta\phi_{g}=\int T_{gz}(\lambda) \widetilde{\delta g}_z(\lambda) d\lambda\ee
where the gravitational field at a position $z$ in the interferometer is $\delta g_z(z) = \int  \widetilde{\delta g}_z(\lambda) e^{\frac{i 2 \pi z}{\lambda}} d \lambda$, $\widetilde{\delta g}_z(\lambda)$ is the Fourier component of a gravity perturbation with wavelength $\lambda$, and $T_{gz}(\lambda)$ is the interferometer's gravity perturbation response function. Figure \ref{Fig:gravity wrinkles} shows the response of an atom interferometer in the example 10 m configuration to perturbations in the local $g$ field of wavelength $\lambda$.  The top curve is for the atom interferometer run in the `symmetric' configuration where the atoms are launched from the bottom of the interferometer region with exactly the right velocity to stop at the top, $v_L = g T$ and the first and last beamsplitter pulses happen when the atom is at the bottom on its way up and on its way down, respectively.  The next lower curve assumes the atom is dropped from rest at the top.  The next two curves assume the atom is launched downwards from the top with velocities $1 \frac{\text{m}}{\text{s}}$ and $10 \frac{\text{m}}{\text{s}}$, respectively.  As expected, the atom interferometer always averages down the perturbations on scales below its size, here $10 m$.  If the atom is launched so that its velocity is never zero then the suppression is much bigger at shorter lengths.  For the lower two curves, the phase falls off as $\lambda^2$, as compared with $\lambda$ for the case in which the atom comes to rest during the interferometer.  Roughly, the more uniform the atoms' velocity, the larger the reduction that comes from averaging over the small scale gravity perturbations.  This is a very useful tool for reducing backgrounds from local masses.  It also scales favorably with the length of the interferometer, since a longer interferometer suppresses larger distance scales, leading to an even greater suppression at short distances.

\begin{figure}
\begin{center}
\includegraphics[width=400 pt]{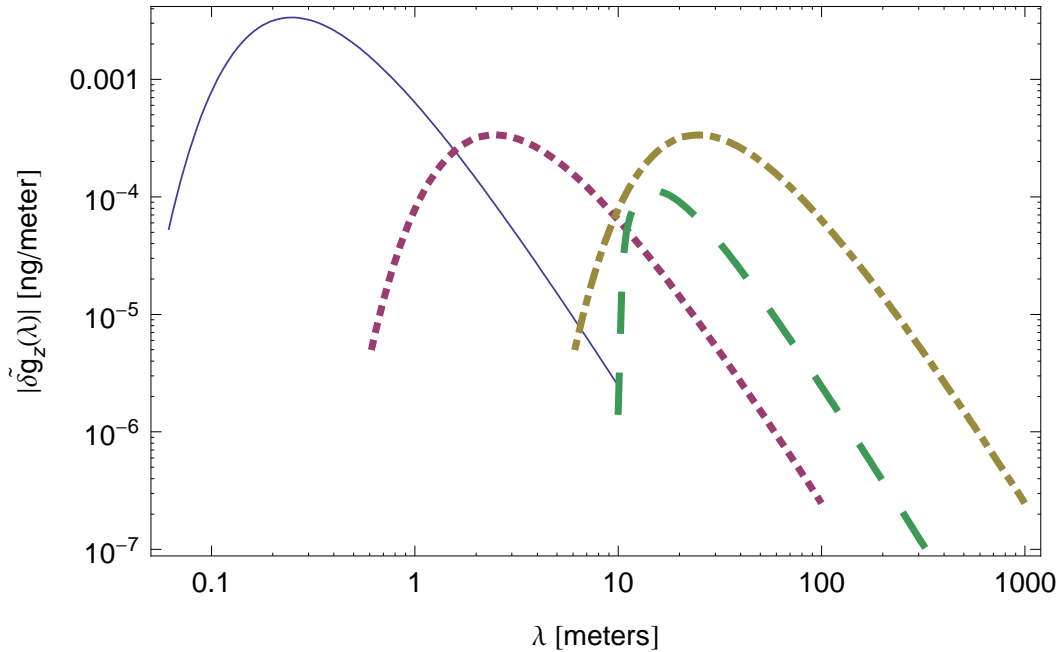}
\caption{ \label{Fig:gravity sources} (Color online) The magnitude of the power spectra of the local gravitational field, $\widetilde{\delta g}_z(\lambda)$ from Eq. \eqref{Eq: gravity wrinkles}, for several example sources.  The solid (blue) curve is a $10^{-2} \text{~kg}$ point source, $10 \text{~cm}$ from the center of the interferometer.  Similarly, the dotted (purple) curve is a $1 \text{~kg}$ source at $1 \text{~m}$ and the dash-dotted (yellow) curve is $1000 \text{~kg}$ at $10 \text{~m}$.  The long-dashed (green) curve is a thin $10 \text{~m}$ long rod of mass $10 \text{~kg}$, parallel to the interferometer, whose center is $1 \text{~m}$ from the interferometer.}
\end{center}
\end{figure}

The differential measurement strategies suppress the longer wavelength contributions.  In fact, at very long wavelengths there is no difference between what we have called a `perturbation' and the previously included part of earth's gravitational field.  The differential measurement strategies were designed to allow us to control systematics arising from the earth's gravitational field.  For the proposed Principle of Equivalence measurement, atom clouds of the two isotopes are separated by less than $1 \mu \text{m}$.  This provides a large suppression to longer wavelength gravitational perturbations.  Similarly, in the divergence strategy measurement (see Section \ref{Sec: Non-linearity of gravity}) the three atom interferometers are separated by $\sim 1 \text{ cm}$.  For wavelengths longer than this scale the divergence measurement looks like a true divergence (instead of a finite difference approximation to the derivative) and so the longer wavelength gravitational modes are suppressed.  As discussed above, the velocity dependent force measurement benefits greatly from the ability to launch the atoms with a large initial velocity to suppress all Newtonian gravitational effects without suppressing the GR signal.  This suppression of all wavelengths can be seen by comparing the lowest two curves in Figure \ref{Fig:gravity wrinkles}.  The lowest curve has a larger launch velocity and is therefore suppressed at all wavelengths as expected.

If we know the magnitude of the perturbation in the local gravitational field, we can compute the phase shift induced in an interferometer from Eq. \eqref{Eq: gravity wrinkles}.  This implies some constraints on the mass and distance of nearby objects that must be taken into account when designing an actual experiment.  Fig. \ref{Fig:gravity sources} displays $\widetilde{\delta g}_z(\lambda)$ for several example sources.  Each source is taken to be some perpendicular distance from the center of the atom interferometer.  We have always plotted the magnitude of the Fourier transforms in Figures \ref{Fig:gravity wrinkles} and \ref{Fig:gravity sources} (i.e. we have taken the sum in quadrature of the sine and cosine components).  An object that is farther away than its size looks like a point mass.  Its gravitational field has a Fourier transform, $\widetilde{\delta g}_z(\lambda)$, that is sharply peaked around its distance from the interferometer.  An object that is large compared to its distance (represented by the rod in Figure \ref{Fig:gravity sources}) produces a $\widetilde{\delta g}_z(\lambda)$ that is sharply peaked around the object's size.  This analysis allows us to predict the expected size of local gravitational perturbations as a function of wavelength.

The differential response curve allows us to compute systematic errors arising from the specific gravity environment of our interferometer.  Quantitative estimates of these effects requires knowledge of the local $\delta g_z(z)$, which may be obtained through a combination of modelling and characterization.  The atom interferometer itself can be used as a precision gravimeter for mapping $\delta g_z(z)$ in situ.  By varying the interrogation time $T$, a local gravity measurement can be made over a small spatial region.  Many such measurements in different positions can be made by varying the launch velocity and time of the initial beamsplitter, resulting in a measurement of $\delta g_z(z)$.  

The interferometer's response to the short wavelength modes can be quite suppressed as in Fig. \ref{Fig:gravity wrinkles}.  Additionally, the amplitudes of the gravitational perturbations can be kept small as in Fig. \ref{Fig:gravity sources}.  Further, depending, to some extent, on the nature of the gravitational source, the phase shifts due to these local gravitational anomalies are unlikely to vary at order one with the control parameters used in the experiment, or to vary at order one from shot to shot.  This leads to a further suppression, since a truly constant phase shift would not be a background to many of these proposed experiments.  Thus, it seems possible to reduce the background phase shifts due to local gravitational effects below the required $10^{-6} \text{~rad}$ level.

There are, of course, many details we have not addressed here that are important for a real experiment.  Here we have given our ideas for ways of distinguishing the main GR effects of the earth's gravitational field from the relevant backgrounds.  These effects are found to be possibly phenomenologically distinguishable from Newtonian gravity and other backgrounds, and are thus true GR (or beyond GR) effects, indescribable in a Newtonian model.  We have tried to motivate why these or similar experiments may be possible, but we have not proven that they will work in the full detail necessary for a real experiment.  We just wish to motivate future work.

\section{Other GR Effects?}
\label{Sec:Other GR}

\subsection{Hubble Expansion (Vacuum Energy)}
%Riemann is too small, violations of PoE are known to be small as well

The cosmological expansion of the universe can be described by the metric
\begin{equation}
\label{Eqn:Hubble metric}
ds^2 = dt^2 - e^{2 H t} d\vec{x}^2
\end{equation}
where $H$ is the Hubble constant.  Using the geodesic equation \eqref{Eqn:Geodesic Eqn}, a particle moving in this metric has an acceleration
\begin{equation}
\frac{d^2 \vec{x}}{d\tau^2} = 2 H \frac{d\vec{x}}{d\tau} \frac{dt}{d\tau} \approx 2 H v.
\end{equation}
For laboratory atomic velocities this would be $\sim 10^{-18} g$, large enough to be very interesting compared to the initial sensitivity of the interferometer under construction $\sim 10^{-15} g$.  We would like to consider whether it is then possible to measure the expansion rate of the universe in an atom interferometer.  Since the local neighborhood has collapsed gravitationally, the normal cosmological expansion of the metric is presumably not occurring inside the solar system.  However, a true cosmological constant would necessarily be present everywhere and must therefore affect the metric locally.  We will loosely model the local effects of the vacuum energy using metric \eqref{Eqn:Hubble metric} (to include the effects of local mass a McVittie metric \cite{McVittie} should really be used), but it must be remembered that $H$ will then presumably refer only to the contribution of the vacuum energy.

Unfortunately, this is just an artificial, coordinate acceleration, not the physical acceleration that would be measured by any experiment.  A calculation of the acceleration that would be measured in an actual experiment, either an observer using radar-ranging to determine the atom's position or using an atom interferometer, shows that the physically observable accelerations are of $\OO(H^2)$.

Another way to see an effect linear in $H$ is to consider two observers on geodesics of \eqref{Eqn:Hubble metric} which are at constant $\vec{x}$ positions.  The spatial distance (defined by integrating only the spatial terms in the metric at a fixed time) between these observers is $e^{H t} \Delta \vec{x}$ and so it would appear that there is a relative velocity of $\mathcal{O}(H \Delta x)$.  In fact a radar-ranging calculation of the distance or velocity shows essentially the same thing.  This is not a surprise, since this is in fact just the normal observation of Hubble's law for the expansion of the universe.  Such observers represent galaxes, which are measured to recede from each other at velocities of order $H$.  However we cannot create a similar laboratory version of this experiment in order to measure the Hubble constant (really vacuum energy) without starting the atom with an initial velocity of $\mathcal{O}(H)$, which defeats the purpose.  The galaxies naturally have such velocities, and so tend to mark the cosmological expansion, due to the action of the $\OO(H^2)$ acceleration acting for the age of the universe $H^{-1}$.

More generally, it seems there cannot be effects linear in $H$, at least in `normal' variables such as position, velocity, or acceleration, in any experiment we can set up.  Without access to observers in some special reference frame such as the galaxies provide, any experiment will be free-falling in the cosmological expansion.  By the Principle of Equivalence, such an experiment will then see a metric that is locally flat with corrections proportional to the Reimann curvature.  But this is $R \propto H^2$ (all components are either $0$, $- H^2$, or $\pm H^2 e^{2 H t}$).  This is very similar to transforming Eq. \eqref{Eqn:Hubble metric} to the static patch
\begin{equation}
\label{Eqn:Static Patch metric}
ds^2 = \left( 1- H^2 r^2 \right) dt^2 - \frac{1}{1- H^2 r^2} dr^2 - r^2 d\Omega^2.
\end{equation}
In these coordinates it is clear that all accelerations (all Christoffel symbols), the Reimann curvature and other similar quantities will be $\mathcal{O}(H^2)$.  Of course, the argument that in a LLF the answer will just be proportional to the curvature and therefore to $H^2$ misses the important point that the Reimann tensor (at least in either of the coordinates \eqref{Eqn:Hubble metric} or \eqref{Eqn:Static Patch metric}) still has coordinate dependence, and therefore cannot actually be observable.  This problem highlights the importance of working entirely in physical variables, as we stressed for the atom interferometer in Section \ref{Sec:GR calc}.

The expansion of the universe is occasionally proposed as an explanation for the Pioneer Anomaly \cite{pioneer} which is an anomalous acceleration of the Pioneer spacecrafts of order $H$.  Radar ranging the Pioneer is quite similar in spirit to the atom interferometer, which can be thought of as laser ranging the atom to find its acceleration.  Thus, similar arguments apply to this case as well, and the Pioneer anomaly cannot be explained within general relativity as being due to the cosmological expansion.

Intuitively, it is impossible to observe effects linear in Hubble in a local experiment because of the equivalence principle.  Essentially, everything in the experiment is `falling' together in the expansion of the universe.  Similarly, it is impossible to detect (at leading order) the acceleration toward the dark matter of the galaxy.  Only the gradient of this force is detectable, and this is too small to be measured.  However, violations of the equivalence principle could in principle lead to observable effects both for the expansion of the universe and dark matter.  These would probably be suppressed by a small factor which is the extent to which the equivalence principle is known to be valid.

\subsection{Lense-Thirring}
\label{Sec:Lense-Thirring}
%give rough estimate of size of effect, but need to pick out of coriolis so too small.
%maybe varying angle would work better?

The Lense-Thirring effect is a gravitomagnetic effect due to the rotation of a source mass.  It is difficult to measure and has been searched for in several experiments \cite{Lammerzahl:2001qr, Ciufolini:2006up, Iorio:2007nn, Nordtvedt:1988vt, AngoninWillaime:2003vp} but no undisputed, conclusive measurements exist yet to better than $\mathcal{O}(1)$.  Given the success of atom interferometers used as gyroscopes, it seems worthwhile to consider whether gravitomagnetic effects could be measured in an atom interferometer.  To understand the effect, the metric outside a spinning body can be written as \cite{Weinberg}
\begin{equation}
ds^2 = \left( -1 - 2 \phi - 2 \phi^2 -2 \psi \right) dt^2 + \left( 1 - 2 \phi \right) d\vec{x}^2 + 2 \vec{\zeta} \cdot d\vec{x} dt
\end{equation}

Using this in the geodesic equations \eqref{Eqn:Geodesic Eqn} gives the coordinate acceleration as
\begin{equation}
\frac{d\vec{v}}{dt} = -\nabla \left( \phi + 2 \phi^2 + \psi \right) - \frac{d\vec{\zeta}}{dt} + \vec{v} \times \left( \nabla \times \vec{\zeta} \right) + 3 \vec{v} \frac{d\phi}{dt} + 4 \vec{v} \left( \vec{v} \cdot \nabla \right) \phi - \vec{v}^2 \nabla \phi
\end{equation}
where $\vec{v} = \frac{d\vec{x}}{dt}$.  The terms proportional to $\vec{\zeta}$ give the gravitomagnetic terms we are interested in.  Outside a spinning, spherical body with angular momentum $\vec{J} \propto MR^2 \vec{\Omega}$
\begin{equation}
\vec{\zeta}(\vec{x}) = \frac{2 G}{r^3} \left( \vec{x} \times \vec{J} \right).
\end{equation}
The relevant acceleration of a test body caused by this effect is then
\begin{equation}
\label{Eqn:Lense-Thirring}
a_\text{LT} =  \vec{v} \times \left( \nabla \times \vec{\zeta} \right) \sim \frac{G M R_E^2 \Omega v}{r^3}.
\end{equation}
Near the earth's surface, using the launch velocity $v \sim 10^{-7}$, the highest this acceleration can be is $a_\text{LT} \sim 10^{-13} g$, which is certainly above the planned sensitivity of upcoming interferometers \cite{Dimopoulos:2006nk}.  Though we have not done a full GR calculation as outlined above, following the usual guess the phase shift is $k a_\text{LT} T^2 \sim 10^{-4} \text{rad}$.  Unfortunately, this phase shift scales very similarly with control parameters as the phase shift due to the Coriolis effect (assuming an earth-bound atom interferometer).  The Coriolis effect gives a phase shift
\begin{equation}
\vec{k} \cdot \vec{a_C} T^2 \sim \vec{k} \cdot \left( \vec{\Omega} \times \vec{v_L} \right) T^2.
\end{equation}
On the earth this acceleration is $\sim 10^{-4} g$.  Then, by \eqref{Eqn:Lense-Thirring}, the Lense-Thirring effect is roughly a factor of $\phi$ smaller than Coriolis in magnitude: $a_\text{LT} \sim \phi a_C \sim 10^{-9} a_C$.  Further, they scale the same way with the control parameters $k$, $v_L$, and $T$, although their directions and dependencies on the directions of the vectors involved, $\vec{v_L}$, $\vec{x}$, and $\vec{\Omega}$, are different.  This means the Lense-Thirring effect cannot be directly measured beneath the much larger Coriolis background.  However, the Coriolis effect is a kinematical effect and is thus qualitatively different from the dynamical Lense-Thirring effect which depends on the rotation of the source mass itself and not just of the laboratory in which the experiment is being performed.  For one thing, it means that the Lense-Thirring effect depends on the distance to the source mass and not just the angular velocity.  Our idea to measure the Lense-Thirring effect in an atom interferometer exploits this difference to isolate the effect from the much larger Coriolis background.

To subtract off the Coriolis background, we can use a differential measurement between two simultaneous interferometers that measure a different Lense-Thirring acceleration but the same Coriolis acceleration.  Our idea for accomplishing this is to have the two interferometers differ in only one control parameter, their height above the earth's surface.  This gives the same Coriolis force, up to the level that the other control parameters can be made equal between the two interferometers (note that time variations are not relevant here so long as the two interferometers remain identical in everything but height).  However there are constraints on how identical the two interferometers can be.  One of the most important is the need to make the launch velocities equal to high precision, since the Coriolis force scales directly with $v_L$.  One idea for doing this is a common launch of a single cloud that is subsequently split into two at differing heights.  This limits how far apart the interferometers can be.  We will take a height difference of $1 \text{ m}$ as an optimistic but not unreasonable guess.  Then the size the Lense-Thirring effect that can be measured is reduced by a gradient factor to $a_\text{LT} \frac{1 \text{ m}}{R_E} \sim 10^{-19}$.  This is roughly a factor of $10^4$ below the initial sensitivity we are considering.  As mentioned in Section \ref{Sec:AI}, there are many possibilities for improving this sensitivity by orders of magnitude.  However this number seemed challenging enough that we have not pursued this idea further (we are explicitly not considering tying an atom interferometer to a telescope in a Gravity Probe B (GPB)-like configuration, since that has already been proposed \cite{AngoninWillaime:2003vp}).  It is also possible that there are better ideas for isolating Lense-Thirring from the Coriolis backgrounds that would allow a measurement, since the Lense-Thirring effect is naturally quite large compared to the sensitivities of upcoming atom interferometers.  This is left to future work.

\subsection{Preferred Frame}
%discuss possibilities for measuring other PPN params.
%mostly preferred frame/location so same as accelerometer, look for daily, yearly modulation.  but this has already been done cite.
%mention possibilities for future work (do calcs to find out all the effects of other PPNs, varying angle for lense-thirring,...)

There are many possible modified theories of gravity beyond those parametrized by the PPN parameters $\beta$ and $\gamma$ in metric \eqref{Eqn: PPN metric}.  As one example, the PPN formalism includes eight other parameters that parametrize the possible metric modifications of general relativity \cite{Will}.  There are many other, non-metric, theories as well.  The full PPN metric \cite{Will} would still fall under the calculation method outlined above.

We have not performed the full calculation to see what other effects from the PPN metric would be present in an atom interferometer, but we can easily guess one.  Often, the new effects introduced by the full PPN metric can usefully be thought of as preferred frame or location effects, usually called violations of Lorentz invariance.  Such effects have been pursued before as modulations of the local acceleration in an accelerometer on the earth's surface with periods of a day, a year, and so on \cite{Will, Warburton, HolgerIsotropy}.  While the atom interferometer can yield an impressive increase in sensitivity over the accelerometers used to do the previous searches (and has already led to improved limits \cite{HolgerIsotropy}), the previous experiments were ultimately limited by geophysical uncertainties.  Thus, it is not clear that an increase in accelerometer sensitivity would lead to an improved ability to search for such Lorentz violating effects without an equally improved geophysical model.  Of course, the atom interferometer is much more than just an accelerometer, as attested to by the many control parameters and measurement strategies employing simultaneous differential measurements outlined above.  Further, there may be signals from these PPN and Lorentz violating effects that are more than just modulations of the local acceleration.  We cannot exclude the possibility that there are novel search techniques that would allow atom interferometers to provide stringent tests of Lorentz violating theories, but we leave such considerations to future work.

\section{Summary and Comparison}
Relativistic effects in interferometry have been discussed before in several contexts \cite{previous1, previous2, previous3, previous4, previous5}.  None of these discussed specific, viable experiments for the post-Newtonian relativistic effects we have considered.  Mostly, they focus on calculation methods instead of specific laboratory experiments.  These methods are not applicable to our experimental setup.

In \cite{previous2}, general discussion and motivation was given for considering the effects of general relativity on devices such as atom or neutron interferometers.  Relativistic calculations for a certain type of interferometer were given in \cite{previous3, previous4} and were used to give a rough estimate of the phase shift that might arise in a setup similar to neutron interferometry.  While it is possible that a similar estimate could be given for the atom interferometer, it would miss most of the important effects since their analysis does not take into account the laser pulses which actually form a light-pulse atom interferometer like the ones we consider here.  None of these calculations can be applied directly to the atom interferometer for several reasons, including the lack of description of the laser pulses and the difficulty of solving the necessary equations for the full atom interferometer sequence in a general metric background.  Essentially, these calculations only have what we call propagation phase.  Further, this phase is not calculated along the trajectories that are relevant for light-pulse atom interferometry.   The measurements suggested in these papers are also quite different from the our proposals.

Atom interferometry is considered in \cite{previous5} as a way to measure space-time curvature, which is essentially the leading order effect of (Newtonian) gravity.  We are interested in measuring the post-Newtonian corrections.  We are thus led to consider a specific setup in which the effect of the laser pulses and the platform on which the laser rests is crucial and must be taken into account.  The formalism given in \cite{previous1} does not give the final phase shift for the general relativistic effects we are interested in, though some of the effects are mentioned.  We cannot use this calculation because the equations become too difficult to solve when all the post-Newtonian terms are kept to the order which is necessary given the precision of the experiment.  Further, the effect of gravity on the laser pulses that serve as our beamsplitters and mirrors is not taken into account, and thus a relativistic prescription for the atom-light interaction is not included.  Thus, in order to have a fully relativistic, coordinate-invariant calculation, we use our semi-classical method for calculating the phase shift in the interferometer.  Additionally, in order to simplify the calculations, many of these previous papers worked in the linearized gravity approximation where the metric is expanded as $g_{\mu \nu} \approx \eta_{\mu \nu} + h_{\mu \nu}$.  This cannot yield a correct result for the non-linear effects in general relativity.

We build and expand upon this previous work by considering a specific experimental scenario for a light pulse atom interferometer, in which the GR effects can be calculated in a fully relativistic framework.  This includes important effects such as a relativistic treatment of the laser pulses forming the beamsplitters and mirrors which accounts for the influence of gravity on the propagation of the light, as well as changes to the phase shift formulas.  This framework is able to go beyond linearized gravity to reveal the effect of the non-linearity of gravity on an atomic interferometer.  This requires calculating phase shifts for terms in the Hamiltonian higher than quadratic order.  We also consider several specific experimental strategies for testing general relativity in an atom interferometer.
%Previous works paragraph here or maybe in beginning of calc section?
%previous works paragraph.  including need for finite speed of light.  also how quadratic hamiltonians misses many things including the non-linearity of gravity.
%Previous works on GR and interferometry \cite{previous} have not dealt with a specific, viable experiment or a full relativistic calculation.  Important effects, such as the influence of gravity on light, the corresponding changes to the separation and laser phases, and the non-linearity of gravity were not discussed.  The typical perturbation theory calculation, which integrates the linearized GR Lagrangian over the unperturbed Newtonian trajectories, does not give the correct coefficients or even the dependence on PPN parameters of the phase shifts in Table \ref{Tab: phases}.  For example, the aforementioned cancellation of the $\gamma$'s in $k_\eff g T^2 v_L^2$ would be missed by such a calculation.

%also maybe something here about how easy or likely it is to build such an experiment.  the likelihood and timescale for getting the necessary technology (eg. heisenberg statistics)?
%future work?  gravity waves.

%We have demonstrated 

\section*{Acknowledgments}
We would like to thank Eric Adelberger, Nima Arkani-Hamed, Raphael Bousso, Matthew Green, David Johnson, Nemanja Kaloper, Surjeet Rajendran, Jay Wacker, and Robert Wagoner.
PWG acknowledges the support of the Mellam Family Graduate Fellowship during a portion of this work.

\appendix

\section{Earth Field Calculation Results}
\label{Sec:Results Appendix}

The coordinates for point C using metric \eqref{Eqn: PPN metric} as discussed in Section \ref{Sec:earth calc} are
\begin{eqnarray}
r_C & = & \frac{1}{24} \left(24 r_{L_1}+24 \frac{k_\eff}{m} T+24 \frac{k^2_\eff}{m^2} T+12 \frac{k^3_\eff}{m^3} T+24 T v_L+48 \frac{k_\eff}{m} T v_L+\right. \nonumber \\ & &
48 \frac{k^2_\eff}{m^2} T v_L+24 T v_L^2+60 \frac{k_\eff}{m} T v_L^2+72 \frac{k^2_\eff}{m^2} T v_L^2+12 T v_L^3+48 \frac{k_\eff}{m}
T v_L^3-3 T v_L^5- \nonumber \\ & & 
24 \frac{k_\eff}{m} T \frac{\omega_a}{m}-48 \frac{k^2_\eff}{m^2} T \frac{\omega_a}{m}-48 \frac{k_\eff}{m} T v_L \frac{\omega_a}{m}-24 T v_L
\frac{\omega_\eff}{m}-48 \frac{k_\eff}{m} T v_L \frac{\omega_\eff}{m}- \nonumber \\ & & 
48 T v_L^2 \frac{\omega_\eff}{m}-6 T^2 \left(-2+2 v_L^2 (-4+\gamma )+2 \frac{k^2_\eff}{m^2} (-4+6 v_L (-2+\gamma )+\gamma )+v_L^3
(-7+4 \gamma )+\right. \nonumber \\ & & 
\left.\frac{k_\eff}{m} \left(4 v_L (-4+\gamma )+3 v_L^2 (-9+4 \gamma )+6 (-1+\frac{\omega_a}{m})\right)+6 v_L (-1+\frac{\omega_\eff}{m})\right) \partial_r \phi (r_{L_1})-
 \nonumber \\ & & 
4 T^3 \left(-3+v_L (-11+2 \beta +4 \gamma )+v_L^2 (-23+8 \beta +16 \gamma )+ \right. %
 \nonumber \\ & & 
\left. \frac{k_\eff}{m} (-11+2 \beta +4 \gamma +2 v_L (-23+8
\beta +16 \gamma ))\right) \partial_r \phi (r_{L_1})^2+
 \nonumber \\ & & 
4 \frac{k_\eff}{m} T^3 \partial^2_r \phi (r_{L_1})+16 \frac{k^2_\eff}{m^2} T^3 \partial^2_r \phi (r_{L_1})+4 T^3 v_L \partial^2_r \phi (r_{L_1})+32 \frac{k_\eff}{m}
T^3 v_L \partial^2_r \phi (r_{L_1})+
 \nonumber \\ & & 
92 \frac{k^2_\eff}{m^2} T^3 v_L \partial^2_r \phi (r_{L_1})+16 T^3 v_L^2 \partial^2_r \phi (r_{L_1})+94 \frac{k_\eff}{m} T^3 v_L^2 \partial^2_r \phi (r_{L_1})+30
T^3 v_L^3 \partial^2_r \phi (r_{L_1})+
 \nonumber \\ & & 
36 T^3 v_L^4 \partial^2_r \phi (r_{L_1})-12 \frac{k^2_\eff}{m^2} T^3 v_L \gamma  \partial^2_r \phi (r_{L_1})-12 \frac{k_\eff}{m} T^3 v_L^2 \gamma
 \partial^2_r \phi (r_{L_1})-
  \nonumber \\ & & 
4 T^3 v_L^3 \gamma  \partial^2_r \phi (r_{L_1})-12 T^3 v_L^4 \gamma  \partial^2_r \phi (r_{L_1})-4 \frac{k_\eff}{m} T^3 \frac{\omega_a}{m} \phi
''[r_{L_1}]-4 T^3 v_L \frac{\omega_\eff}{m} \partial^2_r \phi (r_{L_1})-
 \nonumber \\ & & 
32 T^3 v_L^2 \frac{\omega_\eff}{m} \partial^2_r \phi (r_{L_1})-4 T \partial_r \phi (r_{L_1}) \left(3 \left(2 \frac{k^2_\eff}{m^2}-2 v_L^2 (1+v_L)+\frac{k_\eff}{m}
\left(2-3 v_L^2\right)\right) \gamma +\right.
 \nonumber \\ & & 
3 T \left(2 (-1+\beta +\gamma )+3 v_L (-2+2 \beta +\gamma )+v_L^2 \left(-8+8 \beta -2 \gamma ^2\right)+ \right. %
 \nonumber \\ & & 
\left. 2 \frac{k_\eff}{m} \left(3 (-1+\beta +\gamma )+v_L \left(-8+8 \beta +4 \gamma -3 \gamma ^2\right)\right)\right) \partial_r \phi (r_{L_1})+
 \nonumber \\ & & 
T^2 \left(3 (-4+4 \beta +3 \gamma )-2 v_L \left(22+6 \beta  (-4+\gamma )-15 \gamma +8 \gamma ^2\right)\right) \partial_r \phi (r_{L_1})^2+
 \nonumber \\ & & 
\left.T^2 (\frac{k_\eff}{m} (-2+2 \beta +3 \gamma +16 v_L (-1+\beta +\gamma ))+2 v_L (-1+\beta +\gamma +2 v_L (-2+2 \beta +\gamma
))) \partial^2_r \phi (r_{L_1})\right)+
 \nonumber \\ & & 
2 T \partial_r \phi (r_{L_1})^2 \left(3 T \left(8-12 \beta -8 \gamma +8 \beta  \gamma +8 \gamma ^2+3 v_L \left(8+4 \beta  (-3+\gamma )-4 \gamma
+3 \gamma ^2\right)\right) \partial_r \phi (r_{L_1})+\right.
 \nonumber \\ & & 
\left.\left.2 \left(3 \left(3 \frac{k_\eff}{m}-v_L^2\right) \gamma ^2+2 T^2 v_L \left(2-3 \beta -2 \gamma +2 \beta  \gamma +2 \gamma ^2\right)
\partial^2_r \phi (r_{L_1})\right)\right)\right)
 \nonumber \\ & & 
+ \mathcal{O}(10^{-34} r_{L_1}) \nonumber
\end{eqnarray}
\begin{eqnarray}
t_C & = & T+T v_L+\frac{k_\eff}{m} T+\frac{k^2_\eff}{m^2} T+\frac{\frac{k^3_\eff}{m^3} T}{2}+2 \frac{k_\eff}{m} T v_L+2 \frac{k^2_\eff}{m^2} T v_L+T
v_L^2+\frac{5}{2} \frac{k_\eff}{m} T v_L^2+3 \frac{k^2_\eff}{m^2} T v_L^2+\frac{T v_L^3}{2}+ \nonumber \\ & & 
2 \frac{k_\eff}{m} T v_L^3-\frac{T v_L^5}{8}-\frac{k_\eff}{m} T \frac{\omega_a}{m}-2 \frac{k^2_\eff}{m^2} T \frac{\omega_a}{m}-2 \frac{k_\eff}{m}
T v_L \frac{\omega_a}{m}-T v_L \frac{\omega_\eff}{m}-2 \frac{k_\eff}{m} T v_L \frac{\omega_\eff}{m}- 2 T v_L^2 \frac{\omega_\eff}{m}+ \nonumber %\\ 
\end{eqnarray}
\begin{eqnarray}
& & 
\frac{k_\eff}{m} T \partial_r \phi (r_{L_1})+\frac{k^2_\eff}{m^2} T \partial_r \phi (r_{L_1})+T v_L
\partial_r \phi (r_{L_1})+2 \frac{k_\eff}{m} T v_L \partial_r \phi (r_{L_1}) + \frac{1}{2} \frac{k_\eff}{m} T^3 \partial_r \phi (r_{L_1}) \partial^2_r \phi (r_{L_1}) + \nonumber \\ & & 
T v_L^2 \partial_r \phi (r_{L_1})+\frac{5}{2} \frac{k_\eff}{m} T v_L^2 \partial_r \phi (r_{L_1})+\frac{1}{2} T v_L^3 \partial_r \phi (r_{L_1})+T
v_L \gamma  \partial_r \phi (r_{L_1})+2 \frac{k_\eff}{m} T v_L \gamma  \partial_r \phi (r_{L_1})+ \nonumber \\ & & 
2 T v_L^2 \gamma  \partial_r \phi (r_{L_1})+4 \frac{k_\eff}{m} T v_L^2 \gamma  \partial_r \phi (r_{L_1})+\frac{3}{2} T v_L^3 \gamma  \phi
[r_{L_1}]-\frac{k_\eff}{m} T \frac{\omega_a}{m} \partial_r \phi (r_{L_1})-\frac{k_\eff}{m} T \gamma  \frac{\omega_a}{m} \partial_r \phi (r_{L_1})- \nonumber \\ & & 
T v_L \frac{\omega_\eff}{m} \partial_r \phi (r_{L_1})-T v_L \gamma  \frac{\omega_\eff}{m} \partial_r \phi (r_{L_1})+\frac{3}{2} T \partial_r \phi (r_{L_1})^2+\frac{3}{2}
\frac{k_\eff}{m} T \partial_r \phi (r_{L_1})^2+\frac{3}{2} T v_L \partial_r \phi (r_{L_1})^2+ \nonumber \\ 
& & 
\frac{3}{2} T v_L^2 \partial_r \phi (r_{L_1})^2-T \beta  \partial_r \phi (r_{L_1})^2-\frac{k_\eff}{m} T \beta  \partial_r \phi (r_{L_1})^2-T v_L \beta
 \partial_r \phi (r_{L_1})^2-T v_L^2 \beta  \partial_r \phi (r_{L_1})^2+T v_L \gamma  \partial_r \phi (r_{L_1})^2+ \nonumber \\ & & 
2 T v_L^2 \gamma  \partial_r \phi (r_{L_1})^2-\frac{1}{2} T v_L \gamma ^2 \partial_r \phi (r_{L_1})^2+\frac{5}{2} T \partial_r \phi (r_{L_1})^3-3
T \beta  \partial_r \phi (r_{L_1})^3+\frac{1}{2} T^2 \partial_r \phi (r_{L_1})+\frac{3}{2} \frac{k_\eff}{m} T^2 \partial_r \phi (r_{L_1})+ \nonumber \\ & & 
T \partial_r \phi (r_{L_1}) +
\frac{5}{2} \frac{k^2_\eff}{m^2} T^2 \partial_r \phi (r_{L_1})+\frac{3}{2} T^2 v_L \partial_r \phi (r_{L_1})+5 \frac{k_\eff}{m} T^2 v_L \partial_r \phi (r_{L_1})+9
\frac{k^2_\eff}{m^2} T^2 v_L \partial_r \phi (r_{L_1})+ \nonumber \\ & & 
\frac{5}{2} T^2 v_L^2 \partial_r \phi (r_{L_1})+\frac{39}{4} \frac{k_\eff}{m} T^2 v_L^2 \partial_r \phi (r_{L_1})+\frac{11}{4} T^2 v_L^3
\partial_r \phi (r_{L_1})-\frac{3}{2} \frac{k_\eff}{m} T^2 \frac{\omega_a}{m} \partial_r \phi (r_{L_1})- \nonumber \\ & & 
\frac{3}{2} T^2 v_L \frac{\omega_\eff}{m} \partial_r \phi (r_{L_1})+\frac{3}{2} T^2 \partial_r \phi (r_{L_1}) \partial_r \phi (r_{L_1})+\frac{9}{2} \frac{k_\eff}{m}
T^2 \partial_r \phi (r_{L_1}) \partial_r \phi (r_{L_1})+\frac{9}{2} T^2 v_L \partial_r \phi (r_{L_1}) \partial_r \phi (r_{L_1})+ \nonumber \\ & & 
15 \frac{k_\eff}{m} T^2 v_L \partial_r \phi (r_{L_1}) \partial_r \phi (r_{L_1})+\frac{15}{2} T^2 v_L^2 \partial_r \phi (r_{L_1}) \partial_r \phi (r_{L_1})-T^2
\beta  \partial_r \phi (r_{L_1}) \partial_r \phi (r_{L_1})- \nonumber \\ & & 
3 \frac{k_\eff}{m} T^2 \beta  \partial_r \phi (r_{L_1}) \partial_r \phi (r_{L_1})-3 T^2 v_L \beta  \partial_r \phi (r_{L_1}) \partial_r \phi (r_{L_1})-10 \frac{k_\eff}{m}
T^2 v_L \beta  \partial_r \phi (r_{L_1}) \partial_r \phi (r_{L_1})- \nonumber \\ & & 
5 T^2 v_L^2 \beta  \partial_r \phi (r_{L_1}) \partial_r \phi (r_{L_1})-\frac{1}{2} T^2 \gamma  \partial_r \phi (r_{L_1}) \partial_r \phi (r_{L_1})-\frac{3}{2}
\frac{k_\eff}{m} T^2 \gamma  \partial_r \phi (r_{L_1}) \partial_r \phi (r_{L_1})+ \nonumber \\ & & 
\frac{5}{2} T^2 v_L^2 \gamma  \partial_r \phi (r_{L_1}) \partial_r \phi (r_{L_1})+\frac{15}{4} T^2 \partial_r \phi (r_{L_1})^2 \partial_r \phi (r_{L_1})+\frac{45}{4}
T^2 v_L \partial_r \phi (r_{L_1})^2 \partial_r \phi (r_{L_1})- \nonumber \\ & & 
\frac{9}{2} T^2 \beta  \partial_r \phi (r_{L_1})^2 \partial_r \phi (r_{L_1})-\frac{27}{2} T^2 v_L \beta  \partial_r \phi (r_{L_1})^2 \partial_r \phi (r_{L_1})-\frac{3}{2}
T^2 \gamma  \partial_r \phi (r_{L_1})^2 \partial_r \phi (r_{L_1})+T^2 \beta  \gamma  \partial_r \phi (r_{L_1})^2 \partial_r \phi (r_{L_1})+ \nonumber \\ & & 
\frac{3}{4} T^2 \gamma ^2 \partial_r \phi (r_{L_1})^2 \partial_r \phi (r_{L_1})+\frac{1}{2} T^3 \partial_r \phi (r_{L_1})^2+\frac{7}{3} \frac{k_\eff}{m} T^3 \phi
'[r_{L_1}]^2+\frac{7}{3} T^3 v_L \partial_r \phi (r_{L_1})^2+\frac{35}{3} \frac{k_\eff}{m} T^3 v_L \partial_r \phi (r_{L_1})^2+ \nonumber \\ & & 
\frac{35}{6} T^3 v_L^2 \partial_r \phi (r_{L_1})^2-\frac{1}{3} \frac{k_\eff}{m} T^3 \beta  \partial_r \phi (r_{L_1})^2-\frac{1}{3} T^3 v_L \beta
 \partial_r \phi (r_{L_1})^2-\frac{8}{3} \frac{k_\eff}{m} T^3 v_L \beta  \partial_r \phi (r_{L_1})^2- \nonumber \\ & & 
\frac{4}{3} T^3 v_L^2 \beta  \partial_r \phi (r_{L_1})^2-\frac{1}{6} \frac{k_\eff}{m} T^3 \gamma  \partial_r \phi (r_{L_1})^2-\frac{1}{6} T^3 v_L
\gamma  \partial_r \phi (r_{L_1})^2-\frac{4}{3} \frac{k_\eff}{m} T^3 v_L \gamma  \partial_r \phi (r_{L_1})^2- \nonumber \\ & & 
\frac{2}{3} T^3 v_L^2 \gamma  \partial_r \phi (r_{L_1})^2+\frac{5}{2} T^3 \partial_r \phi (r_{L_1}) \partial_r \phi (r_{L_1})^2+\frac{35}{3} T^3 v_L
\partial_r \phi (r_{L_1}) \partial_r \phi (r_{L_1})^2-2 T^3 \beta  \partial_r \phi (r_{L_1}) \partial_r \phi (r_{L_1})^2- \nonumber \\ & & 
\frac{31}{3} T^3 v_L \beta  \partial_r \phi (r_{L_1}) \partial_r \phi (r_{L_1})^2-T^3 \gamma  \partial_r \phi (r_{L_1}) \partial_r \phi (r_{L_1})^2-\frac{17}{6}
T^3 v_L \gamma  \partial_r \phi (r_{L_1}) \partial_r \phi (r_{L_1})^2+ \nonumber \\ & & 
\frac{2}{3} T^3 v_L \beta  \gamma  \partial_r \phi (r_{L_1}) \partial_r \phi (r_{L_1})^2+\frac{1}{2} T^3 v_L \gamma ^2 \partial_r \phi (r_{L_1})
\partial_r \phi (r_{L_1})^2+\frac{1}{6} \frac{k_\eff}{m} T^3 \partial^2_r \phi (r_{L_1})+\frac{2}{3} \frac{k^2_\eff}{m^2} T^3 \partial^2_r \phi (r_{L_1})+ \nonumber \\ & & 
\frac{1}{6} T^3 v_L \partial^2_r \phi (r_{L_1})+\frac{4}{3} \frac{k_\eff}{m} T^3 v_L \partial^2_r \phi (r_{L_1})+\frac{13}{3} \frac{k^2_\eff}{m^2} T^3
v_L \partial^2_r \phi (r_{L_1})+\frac{2}{3} T^3 v_L^2 \partial^2_r \phi (r_{L_1}) - \frac{1}{6} T^3 v_L \frac{\omega_\eff}{m} \partial^2_r \phi (r_{L_1}) + \nonumber \\ & & 
\frac{53}{12} \frac{k_\eff}{m} T^3 v_L^2 \partial^2_r \phi (r_{L_1})+\frac{17}{12} T^3 v_L^3 \partial^2_r \phi (r_{L_1})+2 T^3 v_L^4 \phi
''[r_{L_1}]-\frac{1}{6} \frac{k_\eff}{m} T^3 \frac{\omega_a}{m} \partial^2_r \phi (r_{L_1}) -\frac{4}{3} T^3 v_L^2 \frac{\omega_\eff}{m} \partial^2_r \phi (r_{L_1}) \nonumber \\ & & 
+ \frac{1}{2} T^3 v_L \partial_r \phi (r_{L_1}) \partial^2_r \phi (r_{L_1})+4 \frac{k_\eff}{m} T^3 v_L \partial_r \phi (r_{L_1}) \partial^2_r \phi (r_{L_1})+2
T^3 v_L^2 \partial_r \phi (r_{L_1}) \partial^2_r \phi (r_{L_1})- \nonumber \\ & & 
\frac{1}{3} \frac{k_\eff}{m} T^3 \beta  \partial_r \phi (r_{L_1}) \partial^2_r \phi (r_{L_1})-\frac{1}{3} T^3 v_L \beta  \partial_r \phi (r_{L_1}) \partial^2_r \phi (r_{L_1})-\frac{8}{3}
\frac{k_\eff}{m} T^3 v_L \beta  \partial_r \phi (r_{L_1}) \partial^2_r \phi (r_{L_1})- \nonumber \\ & & 
\frac{4}{3} T^3 v_L^2 \beta  \partial_r \phi (r_{L_1}) \partial^2_r \phi (r_{L_1})-\frac{1}{3} \frac{k_\eff}{m} T^3 \gamma  \partial_r \phi (r_{L_1}) \partial^2_r \phi (r_{L_1})-\frac{1}{6}
T^3 v_L \gamma  \partial_r \phi (r_{L_1}) \partial^2_r \phi (r_{L_1})- \nonumber \\ & & 
\frac{4}{3} \frac{k_\eff}{m} T^3 v_L \gamma  \partial_r \phi (r_{L_1}) \partial^2_r \phi (r_{L_1})+\frac{5}{4} T^3 v_L \partial_r \phi (r_{L_1})^2 \phi
''[r_{L_1}]-\frac{3}{2} T^3 v_L \beta  \partial_r \phi (r_{L_1})^2 \partial^2_r \phi (r_{L_1})- \nonumber \\ & & 
\frac{1}{2} T^3 v_L \gamma  \partial_r \phi (r_{L_1})^2 \partial^2_r \phi (r_{L_1})+\frac{1}{3} T^3 v_L \beta  \gamma  \partial_r \phi (r_{L_1})^2
\partial^2_r \phi (r_{L_1})+\frac{1}{4} T^3 v_L \gamma ^2 \partial_r \phi (r_{L_1})^2 \partial^2_r \phi (r_{L_1})
 \nonumber \\ & & 
+ \mathcal{O}(10^{-34} r_{L_1}) \nonumber
\end{eqnarray}


\begin{thebibliography}{10}
\expandafter\ifx\csname url\endcsname\relax
  \def\url#1{{\tt #1}}\fi
\expandafter\ifx\csname urlprefix\endcsname\relax\def\urlprefix{URL }\fi

\bibitem{Oskay}
Oskay, W.H. et. al. Phys. Rev. Lett. {\bf 97} 020801 (2006).

%\cite{Fischer:2004jt}
\bibitem{Fischer:2004jt}
  M.~Fischer {\it et al.},
  %``New limits on the drift of fundamental constants from laboratory
  %measurements,''
  Phys.\ Rev.\ Lett.\  {\bf 92}, 230802 (2004).
  %%CITATION = PRLTA,92,230802;%%

%\cite{Bize:2003bj}
\bibitem{Bize:2003bj}
  S.~Bize {\it et al.},
  %``Testing the stability of fundamental constants with the Hg-199+ single-ion
  %optical clock,''
  Phys.\ Rev.\ Lett.\  {\bf 90}, 150802 (2003).
  %%CITATION = PRLTA,90,150802;%%

%\cite{Peik:2004qn}
\bibitem{Peik:2004qn}
  E.~Peik, B.~Lipphardt, H.~Schnatz, T.~Schneider, C.~Tamm and S.~G.~Karshenboim,
  %``Limit on the present temporal variation of the fine structure constant,''
  Phys.\ Rev.\ Lett.\  {\bf 93}, 170801 (2004).
  %%CITATION = PRLTA,93,170801;%%

\bibitem{PhysRevLett.78.2046}
T.~L. Gustavson, P.~Bouyer, and M.~A. Kasevich.
%T.~L. Gustavson, {\em et~al.\/}
\newblock Phys. Rev. Lett. {\bf 78}, 2046 (1997).

\bibitem{PhysRevLett.81.971}
M.~J. Snadden, J.~M.~McGuirk, P.~Bouyer, K.~G.~Haritos, and M.~A.~Kasevich.
%M.~J. Snadden, {\em et~al.\/}
\newblock Phys. Rev. Lett. {\bf 81}, 971 (1998).

\bibitem{Biedermann Thesis}
G.~Biedermann,{\it Gravity Tests, Differential Accelerometry and Interleaved Clocks with Cold Atom Interferometers } Thesis, Stanford University (2007).

\bibitem{0026-1394-38-1-4}
A.~Peters, K.~Y. Chung, and S.~Chu.
%A.~Peters, {\em et~al.\/}
\newblock Metrologia {\bf 38}, 25 (2001).

\bibitem{HolgerIsotropy}
H. M\"{u}ller, S. Chiow, S. Herrmann, S. Chu, K.Y. Chung, Phys. Rev. Lett. {\bf 100}, 031101 (2008) [arXiv:0710.3768 [gr-qc]].

%\cite{Dimopoulos:2006nk}
\bibitem{Dimopoulos:2006nk}
  S.~Dimopoulos, P.~W.~Graham, J.~M.~Hogan and M.~A.~Kasevich,
  %``Testing General Relativity with Atom Interferometry,''
  Phys.\ Rev.\ Lett.\  {\bf 98}, 111102 (2007)
  [arXiv:gr-qc/0610047].
  %%CITATION = PRLTA,98,111102;%%

\bibitem{Will}
%C.~M. Will  (2005)
  C.~M.~Will,
  %``The confrontation between general relativity and experiment,''
  Living Rev.\ Rel.\  {\bf 9}, 3 (2005)
  [arXiv:gr-qc/0510072];
  %%CITATION = 00222,9,3;%%
%\newblock gr-qc/0510072.
\newblock {\em Theory and experiment in gravitational physics\/} (Cambridge,
  UK: Univ. Pr., 1993).

\bibitem{Brans:1961sx}
C.~Brans and R.~H. Dicke.
\newblock Phys. Rev. {\bf 124}, 925 (1961).

\bibitem{Damour + Polyakov}
T.~Damour and A.~M. Polyakov.
\newblock Nucl. Phys. {\bf B423}, 532 (1994).

\bibitem{DGP}
G.~R. Dvali, G.~Gabadadze, and M.~Porrati.
%G.~R. Dvali, {\em et~al.\/} 
\newblock Phys. Lett. {\bf B485}, 208 (2000).

\bibitem{Arkani-Hamed 2002}
%N.~Arkani-Hamed, {\em et~al.\/} hep-th/0209227.
N.~Arkani-Hamed, S.~Dimopoulos, G.~Dvali, and G.~Gabadadze hep-th/0209227.

\bibitem{Arkani-Hamed Ghosts}
N.~Arkani-Hamed, H.-C. Cheng, M.~A. Luty, and S.~Mukohyama.
%N.~Arkani-Hamed,  {\em et~al.\/}.
\newblock JHEP {\bf 05}, 074 (2004).

\bibitem{Milgrom 1983}
M.~Milgrom.
\newblock Astrophys. J. {\bf 270}, 365 (1983).

\bibitem{Bekenstein 2004}
J.~D. Bekenstein.
\newblock Phys. Rev. {\bf D70}, 083509 (2004).

%\cite{Dimopoulos:2007cj}
\bibitem{Dimopoulos:2007cj}
  S.~Dimopoulos, P.~W.~Graham, J.~M.~Hogan, M.~A.~Kasevich and S.~Rajendran,
  %``Gravitational Wave Detection with Atom Interferometry,''
  arXiv:0712.1250 [gr-qc].
  %%CITATION = ARXIV:0712.1250;%%

\bibitem{big gravity waves}
  S.~Dimopoulos, P.~W.~Graham, J.~M.~Hogan, M.~A.~Kasevich and S.~Rajendran,
  %``An Atomic Gravitational Wave Interferometric Sensor (AGIS),''
  arXiv:0806.2125 [gr-qc].
  %%CITATION = ARXIV:0806.2125;%%
  
\bibitem{previous2}
K.~Varju and L.~H.~Ryder. Phys. Rev. D {\bf 62}, 024016 (2000).
%S.~Wajima, M.~Kasai and T.~Futamase. Phys. Rev. D {\bf 55} 1964 (1997).

\bibitem{previous3}
S.~Wajima,  {\em et~al.\/} Phys. Rev. D {\bf 55} 1964 (1997).

\bibitem{previous4}
J.~Anandan. Phys. Rev. D {\bf 30} 1615 (1984).

\bibitem{previous5}
J. Audretsch and K.-P. Marzlin, Phys. Rev. A {\bf 50}, 2080 (1994).

\bibitem{previous1}
%C.~J.~Borde, J.-C.~Houard and A.~Karasiewicz gr-qc/0008033.
C.~J.~Borde,  {\em et~al.\/} gr-qc/0008033.

\bibitem{atomicsources}
C. Pethick and H. Smith, {\it Bose-Einstein Condensation in Dilute Gases} (New York: Cambridge Univ. Pr., 2002).

\bibitem{Phillips2002:JPhysB}
J.~H. Denschlag, {\em et~al.\/}.
\newblock J. Phys. B: At. Mol. Opt. Phys. {\bf 35}, 3095 (2002).

\bibitem{PhysRevLett.67.181}
M.~Kasevich and S.~Chu.
\newblock Phys. Rev. Lett. {\bf 67}, 181 (1991).

\bibitem{clocks}
G.~Santarelli, {\em et~al.\/}.
\newblock Phys. Rev. Lett. {\bf 82}, 4619 (1999).
J. McGuirk, et al., Opt. Lett. 26, 364 (2001).
S. Bize, et al., J. Phys. B: At. Mol. Opt. Phys. 38 (2005) S449?S468.

\bibitem{spin_squeezing}
D. Leibfried, et. al. Science {\bf 304}, 1476 (2004).

\bibitem{Tuchman_PRA}
A. K. Tuchman {\em et al.}, Phys. Rev. A {\bf 74}, 053821 (2006).

\bibitem{HolgerLMT}
H. Mueller, S.-w. Chiow, Q. Long, S. Herrmann, S. Chu., arXiv:0712.1990v1.

\bibitem{McGuirk}
J.M.~McGuirk,  {\em et~al.\/} Phys. Rev. Lett. {\bf 85} 4498 (2000).

\bibitem{Chu}
M.~Weitz,  {\em et~al.\/} Phys. Rev. Lett. {\bf 73} 2563 (1994).

\bibitem{EP Varenna}
J.~M.~Hogan, D.~M.~S.~Johnson, and M.~A.~Kasevich, arXiv:0806.3261 [physics.atom-ph].

\bibitem{Fixler}
J. B. Fixler {\em et al.}, Science {\bf 315}, 74 (2007).

\bibitem{CCT:1994}
P.~Storey and C.~Cohen-Tannoudji.
\newblock Journal de Physique II {\bf 4}, 1999 (1994).

\bibitem{Bongs:2006}
K. Bongs, R. Launay, and M. Kasevich, Appl Phys. B 84, 599 (2006).

\bibitem{quantum_calculation}
C.~J.~Borde, Gen. Rel. Grav. {\bf 36}, 475 (2004)

  %\cite{BorisDiamonds}
 \bibitem{BorisDiamonds}
   B.~Dubetsky and M.~A.~Kasevich, 
   %"Atom Interferometer as a selective sensor of rotation or gravity,"
   Phys.\ Rev.\ A {\bf 74},  023615 (2006)
   [arXiv:physics/0604082v4 [physics.atom-ph]]
   %%CITATION=PHRVA,A74,023615;%%

\bibitem{ExactPropagator}
%Propagator and Geometric Phase of a General Time-Dependent Harmonic Oscillator
Jae-Hoo Gweon and Jeong-Ryeol Choi, Journal of the Korean Physical Society, Vol. 42, No. 3, (2003).

\bibitem{Audretsch:PRA47.5}
J\"{u}rgen Audretsch and Karl-Peter Marzlin, Phys. Rev. A {\bf 47}, 4441 (1993).

\bibitem{AntoineFiniteTime}
%Matter wave beam splitters in gravito-inertial and trapping potentials: generalized ttt scheme for atom interferometry
C. Antoine, Appl. Phys. B 84, 585?597 (2006).

\bibitem{JansenThesis}
M. Jansen, {\it Atom interferometry with cold metastable helium.} Ph.D. Thesis, Technische Universiteit Eindhoven (2007).

\bibitem{MolerRaman}
Kathryn Moler, David S. Weiss, Mark Kasevich, and Steven Chu, Phys. Rev. A 45, 342 - 348 (1992).

\bibitem{ShoreBragg}
%Coherent atomic deflection by resonant standing waves
A. F. Bernhardt and B. W. Shore, Phys.\ Rev.\ A {\bf 23}, 1290, (1981).

\bibitem{AllenEberly}
L. Allen and J.H. Eberly, \textit{Optical Resonance and Two-Level Atoms}, Dover Publications, Inc., New York, (1987).


\bibitem{Borde:FallingBeamsplitter}
  Claus Lammerzahl and Christian J.~Borde, Phys. Lett. A {\bf 203}, 59 (1995).


\bibitem{Audretsch:FallingBeamsplitter}
  Karl-Peter Marzlin and J\"{u}rgen Audretsch, Phys. Rev. A {\bf 53}, 1004 (1996).

%\bibitem{Berman}
%P.R.~Berman, Ed. {\em Atom Interferometry\/} (New York: Acad. Pr., 1997).

\bibitem{Weinberg}
S. Weinberg  (1972)
\newblock {\em Gravitation and Cosmology\/} (New York: John Wiley \& Sons, 1972).

\bibitem{Peters Thesis}
A.~Peters, {\it High Precision Gravity Measurements Using Atom Interferometry.} Thesis, Stanford University (1998).

\bibitem{Chu recoil}
D.~S.~Weiss, B.~C.~Young, and S.~Chu,
Appl. Phys. B {\bf 59}, 217 (1994).

\bibitem{Chu recoil2}
H.~Muller, S.-W.~Chiow, Q.~Long, C.~Vo, and S.~Chu, Appl. Phys. B {\bf 84}, 633 (2006).

\bibitem{nanoradian}
M. Pisani and M. Astrua, Appl. Opt. 45, 1725-1729 (2006).


\bibitem{nnbar B shield}
  T.~Bitter {\it et al.},
  %``A large volume magnetic shielding system for the ILL neutron-antineutron oscillation experiment"
  Nucl.\ Instrum.\ Meth.\  A {\bf 309}, 521 (1991).

\bibitem{GibbleNumberDensity}
S. J. J. M. F. Kokkelmans, B. J. Verhaar, K. Gibble, and D. J. Heinzen, Phys. Rev. A {\bf 56}, R4389 (1997).

\bibitem{NumberDensityRbData}
Y. Sortais, S. Bize, C. Nicolas, A. Clairon, C. Salomon, and C. Williams,
\newblock Phys. Rev. Lett. {\bf 85}, 3117 (2000).

\bibitem{Hansch EP exp}
S.~Fray, C.~A.~Diez, T.~W.~Hansch, and M.~Weitz, Phys. Rev. Lett. {\bf 93}, 240404 (2004).

\bibitem{Marion}
H. Marion, et. al., Phys. Rev. Lett. {\bf 90}, 150801 (2003).

%\cite{Adelberger:2003zx}
\bibitem{Adelberger:2003zx}
  E.~G.~Adelberger, B.~R.~Heckel and A.~E.~Nelson,
  %``Tests of the gravitational inverse-square law,''
  Ann.\ Rev.\ Nucl.\ Part.\ Sci.\  {\bf 53}, 77 (2003)
  [arXiv:hep-ph/0307284].
  %%CITATION = ARNUA,53,77;%%

  %\cite{Schlamminger:2007ht}
\bibitem{Schlamminger:2007ht}
  S.~Schlamminger, K.~Y.~Choi, T.~A.~Wagner, J.~H.~Gundlach and E.~G.~Adelberger,
  %``Test of the Equivalence Principle Using a Rotating Torsion Balance,''
  Phys.\ Rev.\ Lett.\  {\bf 100}, 041101 (2008)
  [arXiv:0712.0607].
  %%CITATION = PRLTA,100,041101;%%

%\cite{Kaplan:2000hh}
\bibitem{Kaplan:2000hh}
  D.~B.~Kaplan and M.~B.~Wise,
  %``Couplings of a light dilaton and violations of the equivalence
  %principle,''
  JHEP {\bf 0008}, 037 (2000)
  [arXiv:hep-ph/0008116].
  %%CITATION = JHEPA,0008,037;%%

\bibitem{Audretsch:PRA53.1}
  Karl-Peter Marzlin and J\"{u}rgen Audretsch, Phys. Rev. A {\bf 53}, 312 (1996).

%\cite{Anderson:1998jd}
\bibitem{pioneer}
  J.~D.~Anderson, P.~A.~Laing, E.~L.~Lau, A.~S.~Liu, M.~M.~Nieto and S.~G.~Turyshev,
  %``Indication, from Pioneer 10/11, Galileo, and Ulysses Data, of an Apparent
  %Anomalous, Weak, Long-Range Accelerattion,''
  Phys.\ Rev.\ Lett.\  {\bf 81}, 2858 (1998)
  [arXiv:gr-qc/9808081].
  %%CITATION = PRLTA,81,2858;%%

\bibitem{McVittie}
G.~C.~McVittie, Mon.~Not.~R.~Astron.~Soc. 933, 325 (1933).

%\cite{Lammerzahl:2001qr}
\bibitem{Lammerzahl:2001qr}
  C.~Lammerzahl, C.~W.~F.~Everitt and F.~W.~Hehl,
  %``Gyros, Clocks, Interferometers...: Testing Relativistic Gravity In Space.
  %Proceedings, 220th We-Heraeus Seminar, Bad Honnef, Germany, August 21-27,
  %1999,''
%\href{http://www.slac.stanford.edu/spires/find/hep/www?irn=6466249}{SPIRES entry}
{\it Prepared for 220th WE-Heraeus Seminar on Gyros, Clocks, and Interferometers: Testing General Relativity in Space, Bad Honnef, Germany, 22-27
Aug 1999},
(Springer-Verlag, New York, 2001)

%\cite{Ciufolini:2006up}
\bibitem{Ciufolini:2006up}
  I.~Ciufolini and E.~Pavlis,
  %``On the Measurement of the Lense-Thirring effect Using the Nodes of the
  %LAGEOS Satellites in reply to On the reliability of the so-far performed
  %tests for measuring the Lense-Thirring effect with the LAGEOS satellites by
  %L. Iorio,''
  New Astron.\  {\bf 10}, 636 (2005)
  [arXiv:gr-qc/0601015].
  %%CITATION = NEWAS,10,636;%%

%\cite{Iorio:2007nn}
\bibitem{Iorio:2007nn}
  L.~Iorio,
  %``High-precision measurement of frame-dragging with the Mars Global Surveyor
  %spacecraft in the gravitational field of Mars,''
  arXiv:gr-qc/0701042.
  %%CITATION = GR-QC/0701042;%%

  %\cite{Nordtvedt:1988vt}
\bibitem{Nordtvedt:1988vt}
  K.~Nordtvedt,
  %``EXISTENCE OF THE GRAVITOMAGNETIC INTERACTION,''
  Int.\ J.\ Theor.\ Phys.\  {\bf 27}, 1395 (1988).
  %%CITATION = IJTPB,27,1395;%%

%\cite{AngoninWillaime:2003vp}
\bibitem{AngoninWillaime:2003vp}
  M.~C.~Angonin-Willaime, X.~Ovido and P.~Tourrenc,
  %``Gravitational perturbations on local experiments in a satellite : The
  %dragging of inertial frame in the HYPER project,''
  Gen.\ Rel.\ Grav.\  {\bf 36}, 411 (2004)
  [arXiv:gr-qc/0310021].
  %%CITATION = GRGVA,36,411;%%

\bibitem{Warburton}
R.~J.~Warburton and J.~M.~Goodkind, Ap.\ J.\ {\bf 208}, 881 (1976).

%\bibitem{Rapol}
%U.~D.~Rapol, A. Wasan, and V. Natarajan,
%"Loading of a Rb magneto-optic trap from a getter source"
%Phys.\ Rev.\ A {\bf 64}, 023402 (2001).

\end{thebibliography}
\end{document}